\documentclass[fleqn,usenatbib]{mnras}

\usepackage{newtxtext,newtxmath}
\usepackage[T1]{fontenc}

\DeclareRobustCommand{\VAN}[3]{#2}
\let\VANthebibliography\thebibliography
\def\thebibliography{\DeclareRobustCommand{\VAN}[3]{##3}\VANthebibliography}

\usepackage{graphicx}	% Including figure files
\usepackage{amsmath}	% Advanced maths commands

\newcommand{\msun}{\mathrm{M}_\odot}

\title[Supernova Gravitational Waves]{Gravitational Waves from Core-Collapse Supernovae: Dependence on the Progenitor Star, Rotation Rate and Nuclear Equation of State }

\author[Powell et al.]{
Jade Powell,$^{1,2}$\thanks{E-mail: dr.jade.powell@gmail.com}
Bailey Sykes,$^{3}$
Bernhard M\"uller$^{3}$\thanks{E-mail: bernhard.mueller@monash.edu}
\\
$^{1}$ Centre for Astrophysics and Supercomputing, Swinburne University of Technology, Hawthorn, VIC 3122, Australia.\\
$^{2}$ ARC Centre of Excellence for Gravitational Wave Discovery (OzGrav), Melbourne, VIC 3122, Australia. \\
$^{3}$ School of Physics and Astronomy, Monash University, Clayton, VIC 3800, Australia. 
}

\pubyear{\the\year{}}

\begin{document}
\label{firstpage}
\pagerange{\pageref{firstpage}--\pageref{lastpage}}
\maketitle

\begin{abstract}
 We simulate 137 axisymmetric core-collapse supernova explosions to systematically investigate the impact of different progenitor properties, rotation rates, and equations of state on the supernova gravitational-wave emission. We use 15 different progenitor stars, with masses ranging from $9.71\,\mathrm{M}_{\odot}$ to $36.61\,\mathrm{M}_{\odot}$, three equations of state, and three rotation rates, with durations extending up to 5.5\,s after bounce. We find substantial differences in the gravitational-wave emission between equations of state, with the CMF equation of state producing lower gravitational-wave amplitudes and a longer low frequency mode due to the standing accretion shock instability that remains visible even at $\sim5$\,s post bounce. Rapid rotation also significantly alters the gravitational-wave signal, with more visible modes in the gravitational-wave emission, and lower energies due to later shock revival times. We include the gravitational-wave emission due to neutrino memory, and show it can significantly improve the detectability of core-collapse supernovae for observatories with improved low frequency sensitivity. We propose a recalibration of universal relations for the high frequency mode, using fits to our waveforms, that reduces the fit error at late times in the gravitational-wave signal. Finally, we investigate, for the first time, the impact of random seed perturbations on the gravitational-wave emission. We find that stochastic perturbations can produce dramatic variations in gravitational-wave amplitude, and can alter the signal-to-noise ratio of a detection by as much as 62\% for models close to the boundary between failed and successful shock revival.   
\end{abstract}


%%%%%%%%%%%%%%%%%%%%%%%%%%%%%%%%%%%%%%%%%%%%%%%%%%
%%%%%%%%%%%%%%%%% BODY OF PAPER %%%%%%%%%%%%%%%%%%
%%%%%%%%%%%%%%%%%%%%%%%%%%%%%%%%%%%%%%%%%%%%%%%%%%
\section{Introduction}

The current network of ground-based gravitational-wave detectors LIGO \citep{aLIGO_15}, Virgo \citep{Virgo_15} and KAGRA \citep{Kagra_20} have discovered hundreds of gravitational-wave signals from the mergers of black holes or neutron stars in binary systems \citep{gwtc5}. As the sensitivity of this global detector network continues to improve, it will become increasingly possible to observe other sources of gravitational-wave bursts \citep{powell_25}. Among the most promising of these new gravitational-wave sources are core-collapse supernovae (CCSNe) \citep{abdikamalov_20, burrows_21, mezzacappa_24, mueller_26}. 

CCSNe are the explosions of stars with zero-age-main-sequence (ZAMS) masses of at least $9\,\mathrm{M}_{\odot}$. Massive stars undergo thermonuclear burning until their iron core becomes massive enough that it can no longer support itself and undergoes gravitational collapse. The core then bounces and launches a powerful shock wave. Due to photodisintegration, and to a lesser extent neutrino losses, the shockwave starts to lose energy at about 200\,km. The shock wave then needs a further source of energy to power a full CCSN explosion. Typical CCSNe are thought to power the revival of the shock by absorbing some of the energy from the emitted neutrinos \citep{janka_17}. In more extreme CCSNe, the shock revival can be powered by rotation and strong magnetic fields \citep{leblanc_70, moesta_18, obergaulinger_21, bugli_21, powell_23, mueller_24}. CCSNe are regularly observed electromagnetically, however a detection of gravitational waves or neutrinos will be needed to confirm the central engine behind CCSN explosions. The gravitational-wave emission from CCSNe is predicted through complex  hydrodynamical simulations \citep{bruenn_09, mueller_10, takiwaki_14, just_15, oconnor_18, skinner_19}. Although rapid progress has been made in advancing simulation codes, there are still a lot of uncertainties in CCSN simulation results due to the complex, and sometimes unknown, astrophysics involved in the explosion. However, multiple groups are now in agreement about several features of the gravitational-wave emission. 

One of the most well understood features are the gravitational waves emitted during rotating core-bounce \citep{dimmelmeier_08, scheidegger_08, abdikamalov_14, richers_17, kuroda_14}. When the iron core collapses, a proto-neutron star (PNS) forms at the centre of the massive star. If the star is rotating, centrifugal support results in a slower collapse at the equator than at the poles, which results in the PNS having an oblate asymmetric deformation at formation. The faster the star is rotating, the larger the PNS deformation. This results in a burst of gravitational-wave emission during the core-bounce, with an amplitude proportional to the rotation, followed by a ringdown phase that lasts approximately 10-20\,ms. As the deformation is axisymmetric, the time series spike at core-bounce is only observed in the $h_{+}$ polarisation of the gravitational-wave emission. The gravitational-wave emission from rotating core-bounce has been simulated extensively in previous work, 
and is understood well enough for the development of templated gravitational-wave search and parameter estimation tools \citep{engels_14, edwards_14, edwards_21, mitra_24, pastor_24}. For non-rotating progenitors, it is expected that there will be no gravitational-wave emission during the core-bounce phase.  

Shortly after bounce, if the shock wave produces a negative entropy gradient, it can result in prompt convection that occurs for a period less than 50\,ms \citep{bruenn_94, mueller_13}. Prompt convection results in a burst of gravitational waves at frequencies of $\sim 100$\,Hz, in the most sensitive frequency band of current gravitational-wave observatories. Some previous 3D CCSN simulations have shown that prompt convection can be the highest amplitude component of the gravitational-wave emission \citep{powell_20, mezzacappa_23, cusinato_25}, although the gravitational-wave amplitude due to prompt convection is also known to be somewhat stochastic. In some models, the gravitational waves from prompt convection occur in all observer directions, while in others, only in observer angles towards the poles.  
The LS220 equation of state (EoS) \citep{lattimer_91}, which has been popular in CCSNe simulations in recent years, has been known to result in the development of weaker prompt convection \citep{richers_17}. This may have resulted in an underestimation of the CCSN gravitational-wave energy and detectability of some 3D models. 

About $\sim 100$\,ms after the prompt convection phase, a lower frequency mode may start to develop due to oscillations in the shock wave known as the standing accretion shock instability (SASI) \citep{blondin_04, blondin_06, foglizzo_07}. The gravitational-wave frequency of the SASI is dependent on the properties of the shock radius \citep{mueller_14, foglizzo_26}. The SASI mode usually starts at frequencies below 100\,Hz, and will continue to slowly increase in frequency with time if the shockwave is not revived. Therefore, a long SASI signature in the gravitational-wave emission can be an indicator of a failed explosion and rapid black hole formation \citep{powell_25b}. After shock revival, the SASI mode disappears quickly from the gravitational-wave emission. The SASI also has a strong imprint on the CCSN neutrino signal \citep{walk_20}, and therefore is a promising aspect for a multi-messenger search \citep{drago_23}. 

The dominant feature in CCSN gravitational-wave emission is expected to be PNS g-modes or f-modes \citep{murphy_08, mueller_13, oconnor_18b, pan_18, powell_19,  andresen_19, radice_19, powell_20, vartanyan_23, mezzacappa_23}. The gravitational-wave frequency of these modes increases as the PNS gains mass and shrinks in radius \citep{torres_forne_19, sotani_21}. The emission starts at a few hundred Hz, at $\sim 150$\,ms after bounce, and typically rises to somewhere between 800\,Hz and 2500\,Hz. In modern 3D simulations of neutrino-driven explosions, there is still some disagreement between different simulation codes about the gravitational-wave amplitude of the high frequency modes. This leads to some uncertainties in the rates of observations for CCSNe in next generation gravitational-wave observatories. The majority of simulations show that most of the gravitational-wave energy comes from the first second after bounce, but the full signal can last for several seconds \citep{vartanyan_23}. If a black hole forms rapidly, there will be an abrupt end to the gravitational waves from the PNS modes \citep{powell_21, burrows_25}. The measured black hole formation time, and predictions of the size of the PNS from the frequency of the gravitational-wave emission, may tell us about the hot nuclear EoS \citep{wolfe_23, murphy_24, powell_25b, powell_25c}. 

In recent years, gravitational-wave signals from simulations that include magnetic fields and/or rotation, and extend far beyond the core-bounce phase, have become available  \citep{andresen_19, powell_20, jardine_22, bugli_23, powell_23, powell_24, sykes_26}. The impact of rotation and magnetic fields on the later parts of the gravitational-wave signal are currently not well understood. Both strong rotation and strong magnetic fields are known to produce smaller PNS masses, which will alter the gravitational-wave frequency of the PNS modes. If the magnetic fields and rotation power the shock revival, then the explosion will be significantly more energetic, and result in much higher amplitude gravitational waves and larger detection horizons. Having full gravitational-wave signals for magneto-rotational explosions is essential for us to be able to determine the explosion mechanism from the next nearby CCSN \citep{powell_17, powell_24b}. However, the majority of available magneto-rotational explosion mechanism gravitational waveforms are produced by simulations that do not include general relativity. This is likely artificially increasing the difference between neutrino-driven and magneto-rotational explosion mechanism waveforms. Therefore, to be ready to infer the explosion mechanism from the next nearby CCSN, we need gravitational waveforms from long duration magneto-rotational explosion mechanism simulations that include general relativity. Some efforts have already been made in this direction \citep{kuroda_21, shankar_23, shibagaki_24, schnauck_25}, however only the first 200\,ms to 500\,ms of the gravitational-wave signal is captured in these simulations, and they do not include sophisticated neutrino transport, which also creates significant differences to the gravitational-wave emission. 

Gravitational-wave memory from the asymmetric emission of matter and the asymmetric emission of neutrinos is another aspect of the gravitational-wave emission that we are starting to understand better now that longer duration simulations of CCSNe are available. Previous work has shown extremely high gravitational-wave amplitudes at frequencies below the current ground based gravitational-wave detectors frequency band \citep{vartanyan_20, choi_24, powell_24, richardson_24}. 
A better understanding of this aspect of the gravitational-wave emission is important for developing the science case for next generation ground and space-based observatories. 

Despite the increase in our understanding of gravitational waves from CCSNe, current CCSN gravitational-wave searches make no assumption about the morphology of the gravitational-wave signal \citep{szczepanczyk_21}. In the current advanced gravitational-wave detectors observing runs, targeted searches were performed for CCSNe within 20\,Mpc \citep{lvk_O1O2_SN, szczepanczyk_24, lvk_SN2023ixf}. All-sky gravitational-wave searches are also performed, and may detect gravitational waves from failed or obscured CCSNe \citep{lvk_allsky_O3, lvk_allsky_O4a}. However, significant improvements in the sensitivity of these searches to CCSN signals are still needed. 
Advances in understanding of the gravitational-wave emission from CCSNe has enabled the development of universal relations that describe the relationship between the gravitational-wave frequency and the properties of the PNS \citep{torres_forne_19, sotani_21}. These universal relations, and our better understanding of the gravitational-wave signal, are being used to develop new phenomenological models for CCSN gravitational-wave emission, new gravitational-wave searches, and new methods for astrophysical interpretation of the signal \citep{astone_18, bizouard_21, bruel_23, powell_22, powell_25b, cerda_duran_25}. However, these universal relations may not be accurate for CCSNe with magnetic fields and rotation, or at later times in the signal when the emission is weaker. 

To fully understand CCSN gravitational-wave emission, a larger number of simulations are still needed that represent the entire CCSN progenitor parameter space. Longer duration simulations are also needed to better understand the gravitational-wave emission at late times, and to investigate further the low frequency gravitational-wave memory. The gravitational-wave emission in 3D simulations of up to $\sim 6$\,s post bounce has been studied in the work of \citet{vartanyan_23} and \citet{choi_24}. However, they only include one EoS, and only non-rotating models, and do not include full general relativity. It is also important to investigate the gravitational-wave emission at late times from multiple simulation codes.
Therefore, to expand our understanding of CCSN gravitational-wave emission, we perform 137 2D simulations that include 15 different masses, 3 EoS, and 3 different rotation rates. Performing the simulations in 2D allows us to continue the simulations for up to 5.5\,s post bounce, allowing us to cover a large area of the CCSN parameter space at late times where the gravitational-wave signal is more poorly understood. 
We find significant differences in gravitational-wave emission between different EoS and rotation rates. Due to lack of shock revival, the CMF EoS produces lower amplitude gravitational-wave emission and SASI modes that are visible even at $\sim 5$\,s post bounce. We find rapid rotation leads to later shock revival times, and a larger number of visible modes in the gravitational-wave emission. The long duration of our simulations also allows us to more systematically study the gravitational waves due to neutrino and matter memory. 

CCSNe are stochastic in nature, but the impact of the stochasticity on their gravitational-wave emission has not been investigated beyond the first few hundred milli-seconds. The authors in \citet{cardall_15} perform 160 simulations with a highly simplified model, to examine the effects of stochasticity on the SASI, but  do not investigate the impacts on gravitational-wave emission. In \citet{powell_25c}, we showed that different random seed perturbations can have a large impact on the explosion dynamics of models close to the boundary between failed and successful shock revival. As the change in explosion dynamics will also have a significant impact on the gravitational-wave emission, this motivates including the impact of stochasticity in our gravitational-wave analysis. 
We repeat several simulations with different random velocity perturbations, to investigate the stochasticity of the gravitational-wave emission. We found significant differences in the gravitational-wave amplitudes and detectability for some models close to the explosion boundary. Finally, the large number of long duration simulations presented here enables us to derive an improved universal relation between the gravitational-wave frequency and the properties of the PNS, that reduces the error at late times in fits to our signal predictions. 
 
This paper is structured as follows: In Section \ref{sec:sim}, we describe the numerical methods and progenitor models included in this work. In Section \ref{sec:dynamics}, we give an overview of the explosion dynamics and remnant properties that are relevant for the gravitational-wave emission. In Section \ref{sec:gws}, the main features of the gravitational-wave emission are described. In Section \ref{sec:uni_relations}, we fit current universal relations to our gravitational-wave emission, and include an improved relation that reduces errors in the fits to our models at late times. In Section \ref{sec:low_freq}, we calculate the low frequency gravitational waves from asymmetric emissions of matter and neutrinos. In Section \ref{sec:detection}, we determine the detection prospects of our models, and discuss the signal features most likely to be observed from the next nearby CCSN. In Section \ref{sec:perturbations}, we investigate the impact of seed perturbations on the gravitational-wave emission. The conclusions are given in Section \ref{sec:conclusion}.

%%%%%%%%%%%%%%%%%%%%%%%%%%%%%%%%%%%%%%%%%%%%%%%%%%%%%%%
\section{Simulation Set Up and Numerical Methods}
\label{sec:sim}

We perform our simulations in 2D with the general relativistic CCSN simulation code CoCoNuT-FMT \citep{dimmelemeier_02, mueller_15}. CoCoNuT-FMT uses the xCFC scheme for Einstein's equations \citep{cordero_09}, the HLLC Riemann solver for the equations of hydrodynamics \citep{gurski_04}, and the fast multi-group neutrino transport (FMT) scheme \citep{mueller_15}. The gravitational-wave emission is calculated using the quadrupole formula \citep{finn_90}. We use spherical coordinates with a spacial resolution of 550 x 144 in radius $r$ and polar angle $\theta$. 
We use 15 different single star, solar metallicity, progenitor star models from \cite{mueller_16}, obtained with the stellar evolution code KEPLER \citep{heger_10, weaver_78}. The progenitor ZAMS masses range from $9.71\,\mathrm{M}_{\odot}$ to $36.61\,\mathrm{M}_{\odot}$. 
To explore the gravitational-wave dependence on the EoS we use three different EoS at high densities. They are the CMF EoS from \citet{Motornenko_20} and the SFHo and SFHx EoS from \citet{Steiner_13}. 
At low densities we use an EoS accounting for photons, electrons, positrons and an ideal gas of nuclei with a flashing treatment for nuclear reactions \citep{rampp_02}. For the CMF EoS, we use only non-rotating models. For the SFHo and SFHx EoS, we use three different rotation rates, non-rotating, slowly rotating and rapidly rotating. We use the same rotation profile that has been used in previous works \citep{mueller_e_04, andresen_19, summa_18}. In the inner 1747\,km, we use a rotation rate of 0.5 rad/s for the rapidly rotating models, and 0.1 rad/s for the slowly rotating models. Beyond this radius, the rotation decays as $r^{-3/2}$. We do not include magnetic fields.

%%%%%%%%%%%%%%%%%%%%%%%%%%%%%%%%%%%%%%%%%%%%%%%%%%%%%%%%
%%%%%%%%%%%%%%%%%%%%%%%%%%%%%%%%%%%%%%%%%%%%%%%%%%%%%%%%
\section{Explosion Dynamics \& Remnant Properties}
\label{sec:dynamics}

In this section, we give a brief overview of the aspects of the explosion dynamics that are most relevant for the development of the gravitational-wave emission. 

%%%%%%%%%% NO ROTATION 
\subsection{No Rotation}

\begin{figure*}
\includegraphics[width=0.65\columnwidth]{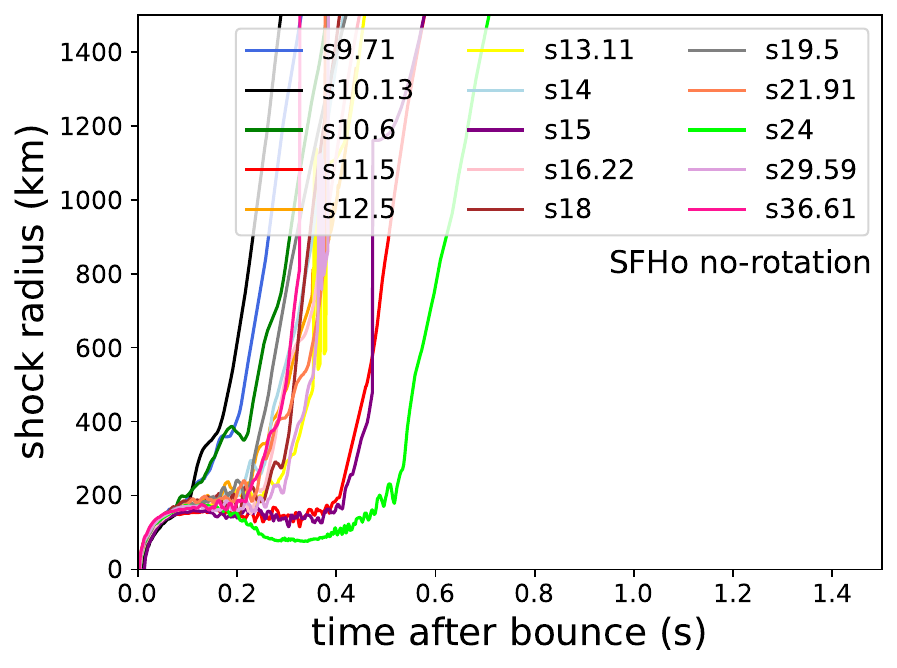}
\includegraphics[width=0.65\columnwidth]{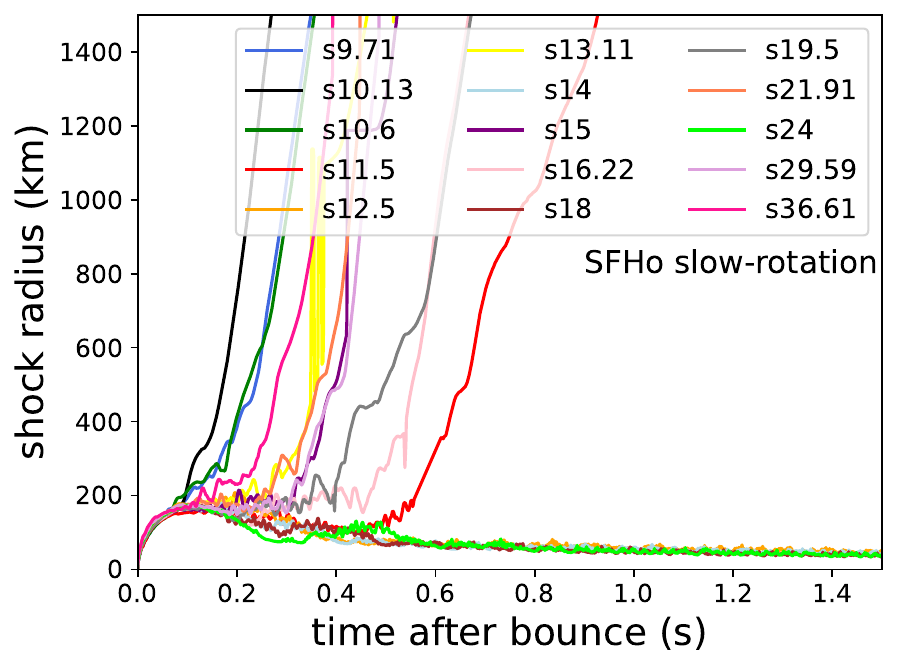}
\includegraphics[width=0.65\columnwidth]{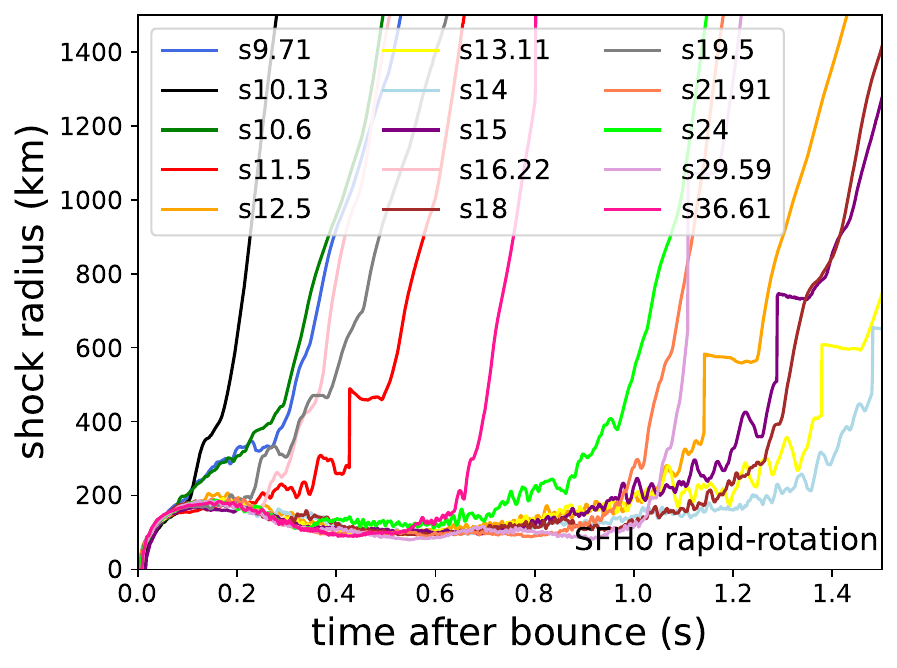}
\includegraphics[width=0.65\columnwidth]{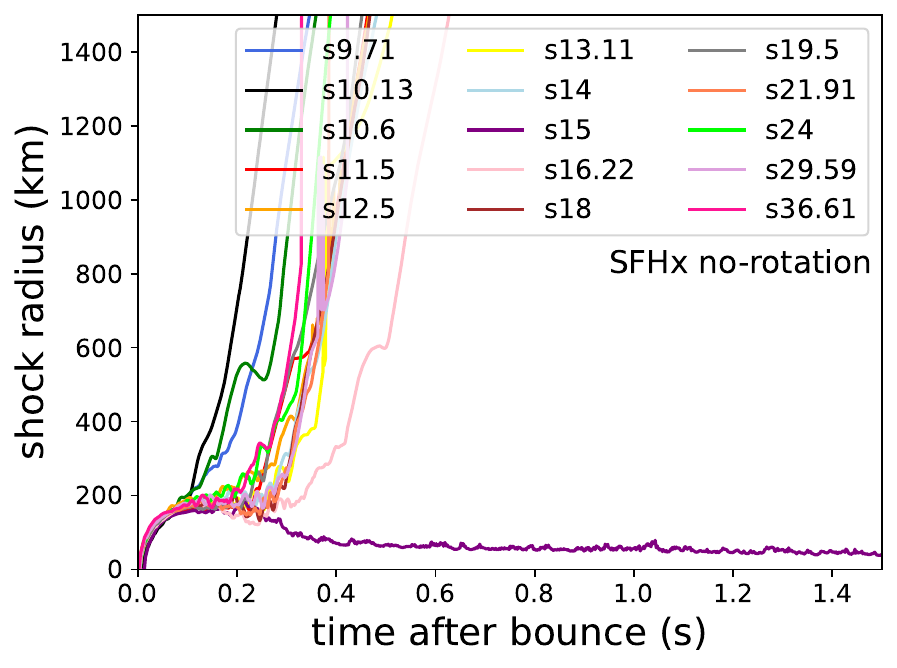}
\includegraphics[width=0.65\columnwidth]{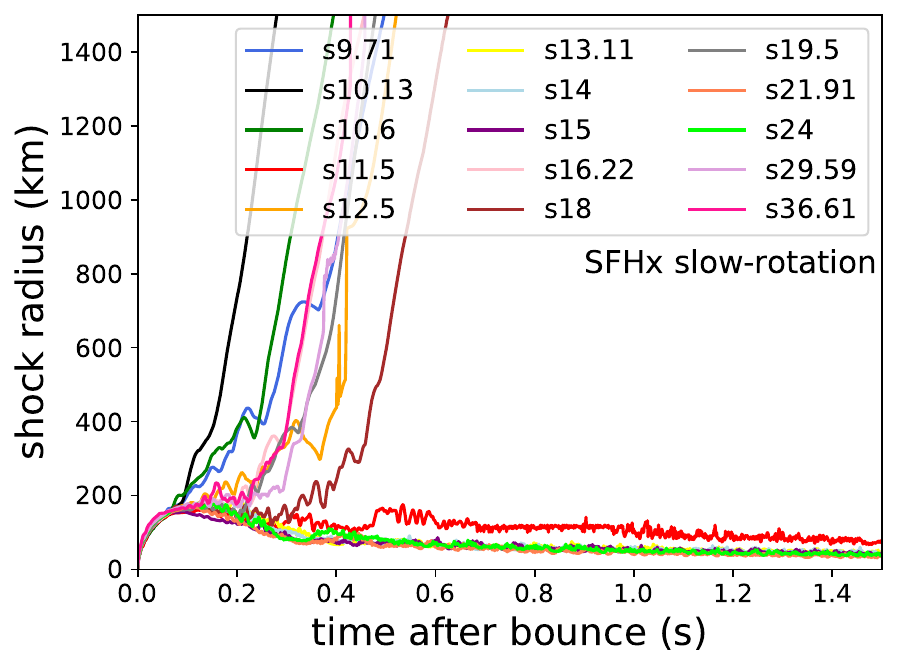}
\includegraphics[width=0.65\columnwidth]{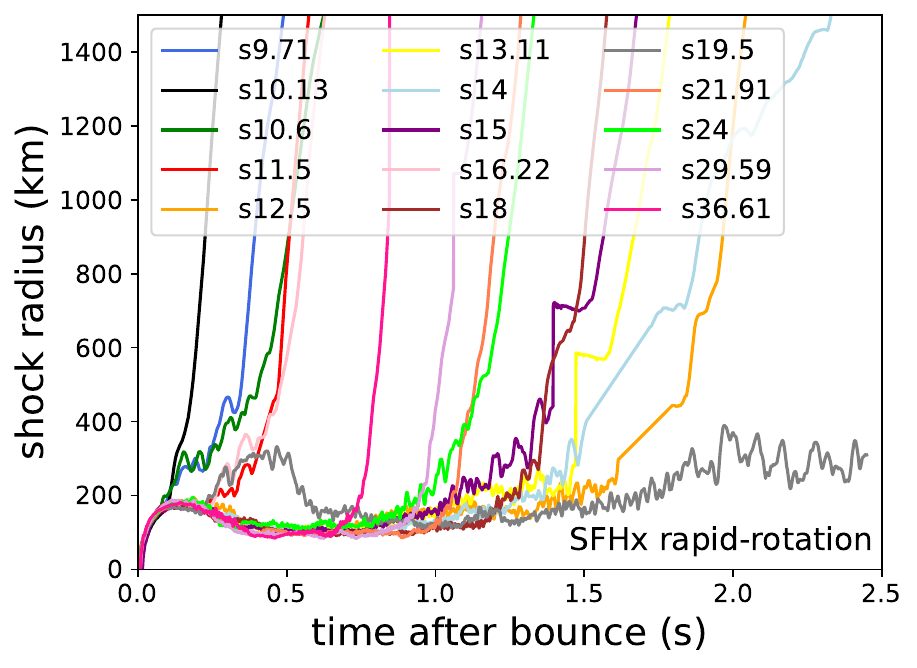}
\caption{The average shock radius for the SFHo models (top), and the SFHx models (bottom). From left to right are the non-rotating models, the slowly rotating models and the rapidly rotating models. We observe later shock revival in models with rotation. We do not show the non-rotating CMF models as only one of the models, s29.59, undergoes shock revival at $\sim 1$\,s post bounce.  }
\label{fig:shock_radius}
\end{figure*}

\begin{figure}
\includegraphics[width=\columnwidth]{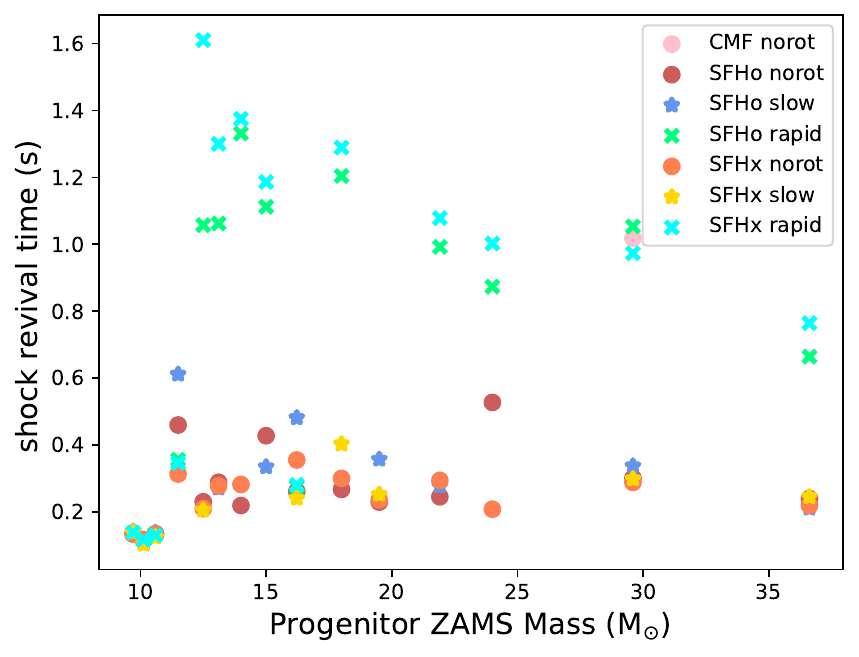}
\caption{The shock revival time, defined as the time when the shock reaches 250\,km, for all models. The lowest mass models all undergo shock revival at a similar time. For the higher mass models, the shock revival is significantly delayed in the rapidly rotating models. The one exploding CMF model has a significantly delayed explosion time in comparison with the other non-rotating models. }
\label{fig:shock_mass}
\end{figure}

The angle-averaged shock radius of our models is shown in Figure \ref{fig:shock_radius}, and the shock revival time for all models is shown in Figure~\ref{fig:shock_mass}. The majority of the CMF models do not undergo shock revival before the end of the simulation time. This is due to a larger PNS radius in the CMF models, which leads to reduced levels of neutrino heating and neutrino luminosity \citep{powell_25c}. The exception is model s29.95, which undergoes shock revival $\sim 1$\,s after bounce. For the SFHo and SFHx models, almost all models undergo rapid shock revival, with the exception of s15 SFHx. There are no significant differences observed between the shock revival times for SFHo and SFHx. There are a few models, such as s11.5 and s24, where the shock revival times vary by a few hundred milli-seconds between the different EoS, but this is likely somewhat stochastic. 

\begin{figure}
\includegraphics[width=\columnwidth]{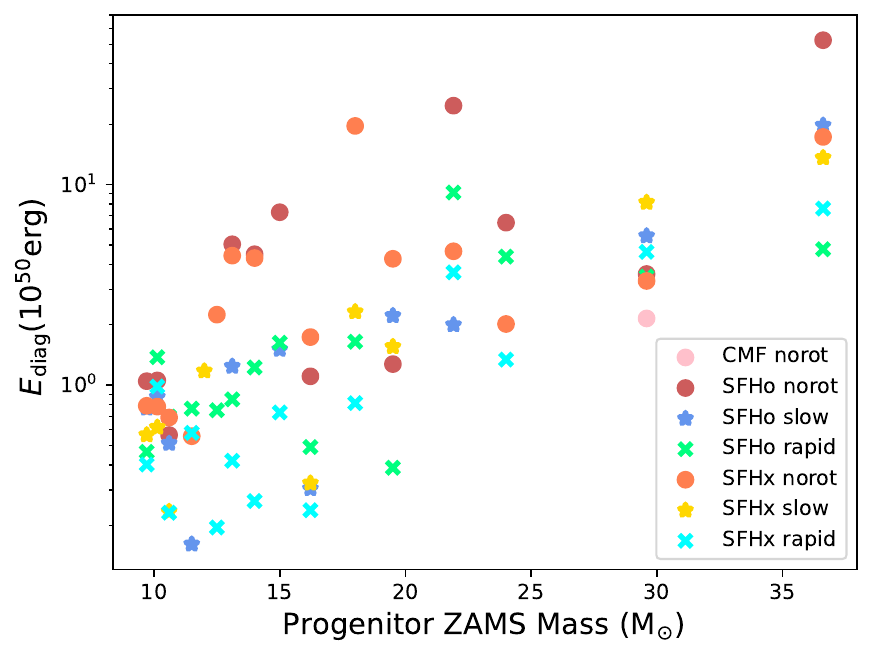}
\caption{The final explosion energies for all exploding models and their ZAMS mass. The models show no significant difference in explosion energy between the different equations of state. The rotating models have lower final explosion energies. This is in part due to the later shock revival times, and shorter simulation durations.}
\label{fig:energy}
\end{figure}

The final diagnostic explosion energy \citep{mueller_12} for all models is shown in Figure \ref{fig:energy}. As the 2D simulations here are long duration, most have asymptoted before the end of the simulation time. We find no noticeable trends in the explosion energy between the different EoS. We find that there is a general increasing trend between the progenitor ZAMS masses and the final diagnostic explosion energies. Some of the higher mass models, such as s18, s21.91 and s36.61, are highly energetic, and reach over $10^{51}$\,erg before the end of the simulation time. 

\begin{figure}
\includegraphics[width=\columnwidth]{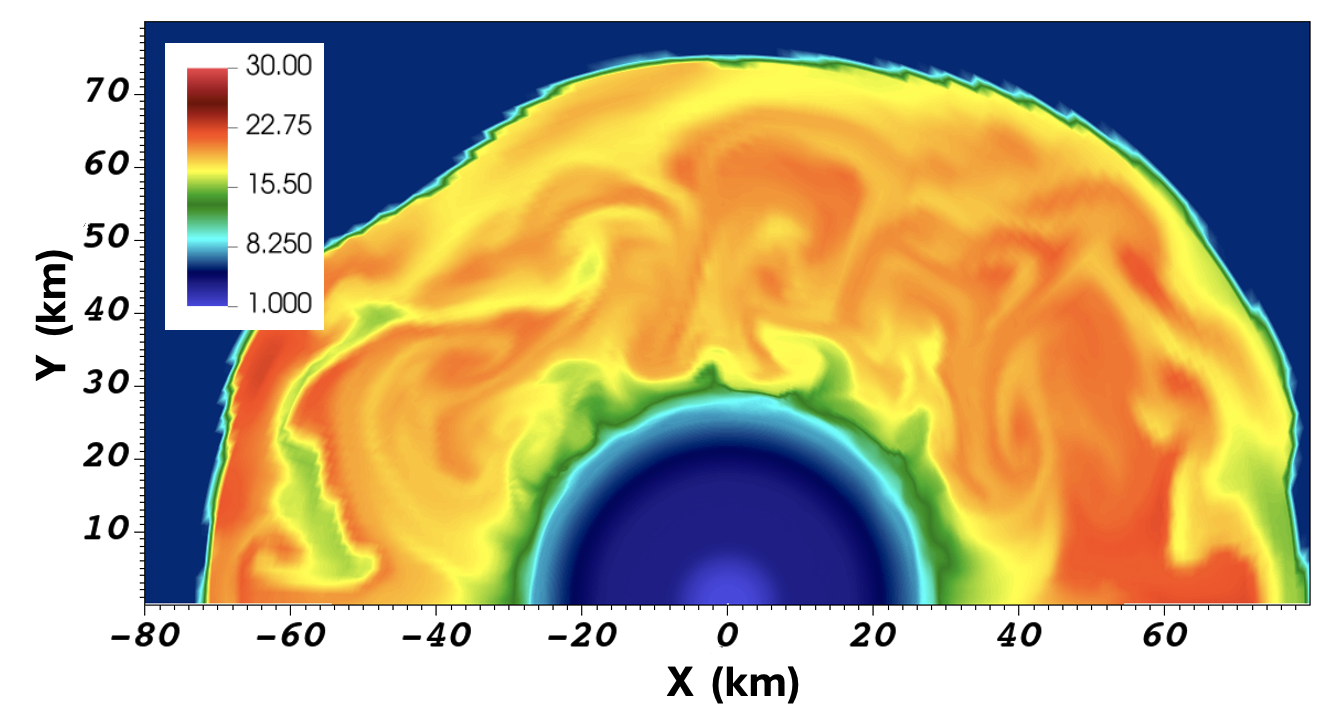}
\includegraphics[width=\columnwidth]{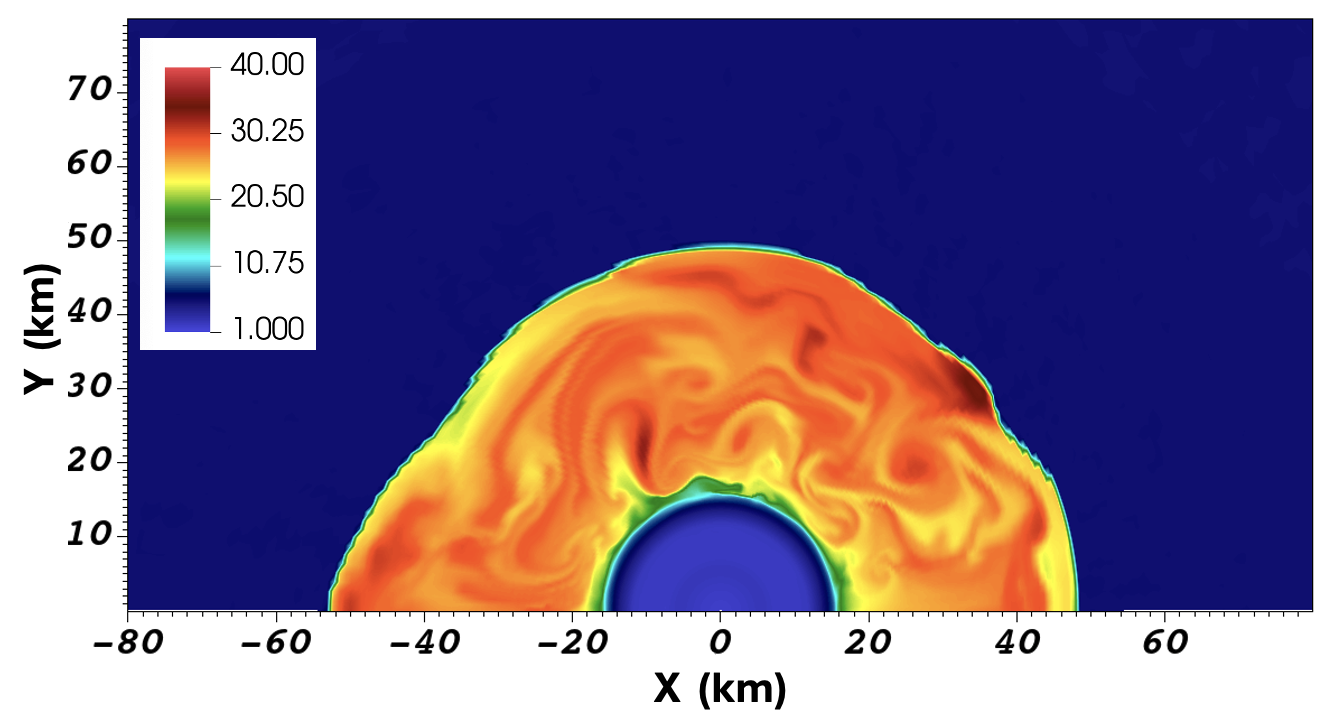}
\includegraphics[width=\columnwidth]{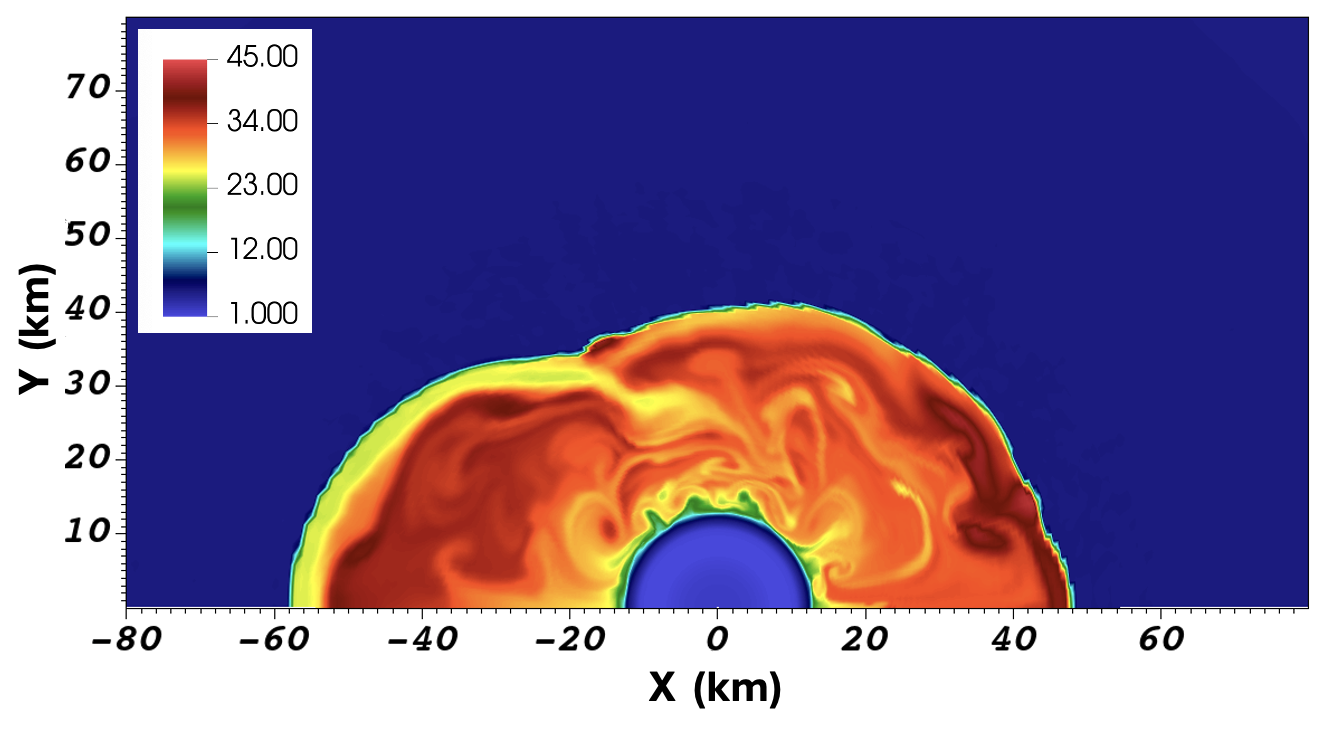}
\caption{The entropy (in $k_\mathrm{B}/\mathrm{nucleon}$) of model s10.13 with the CMF EoS at 0.5\,s (top), 2\,s (middle), and 4\,s after bounce. The CMF models exhibit strong SASI even at several seconds post bounce.  }
\label{fig:visit_cmf}
\end{figure}

\begin{figure}
\includegraphics[width=\columnwidth]{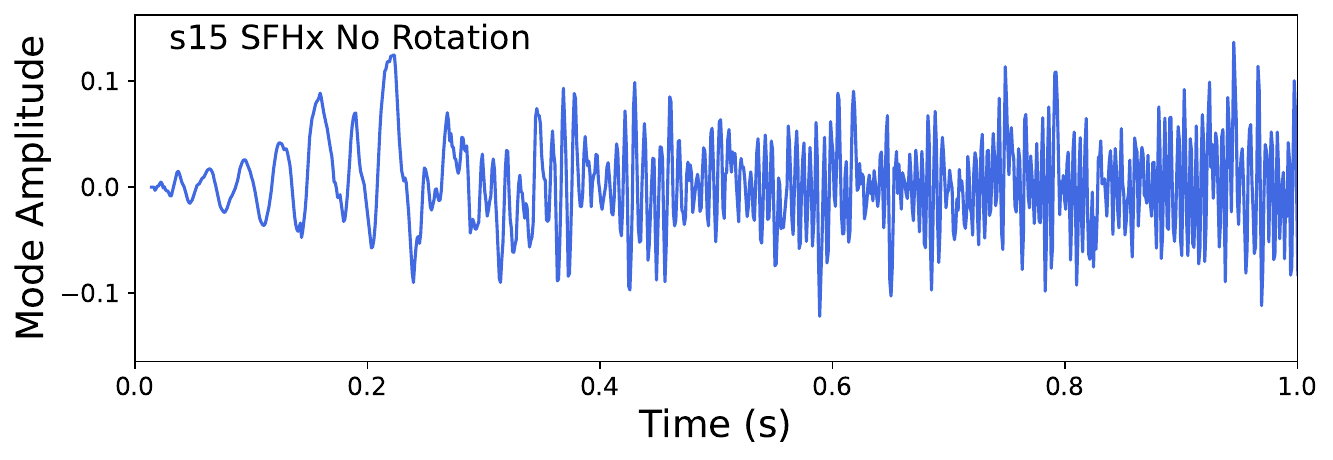}
\includegraphics[width=\columnwidth]{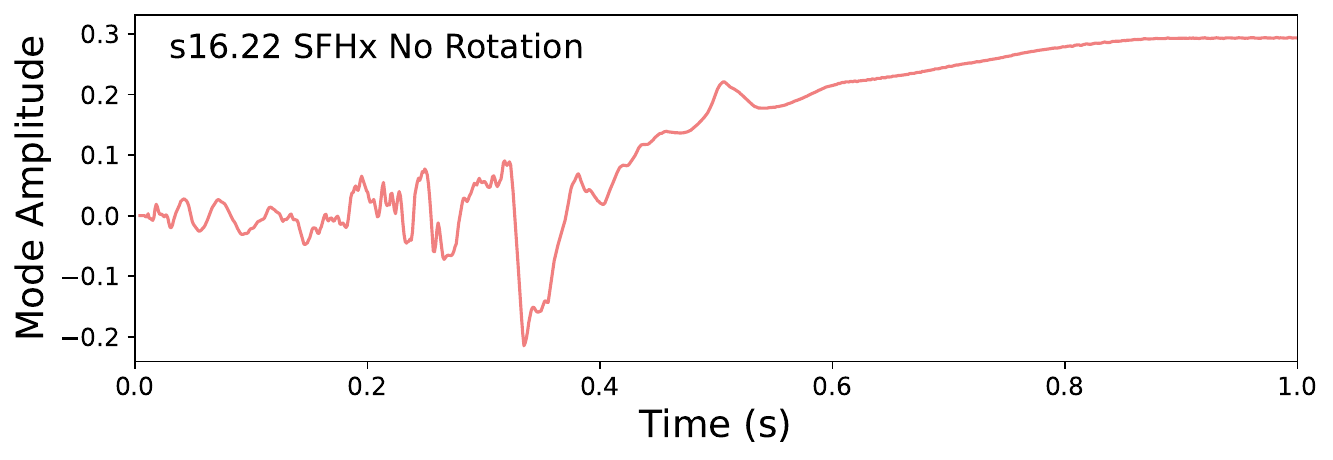}
\includegraphics[width=\columnwidth]{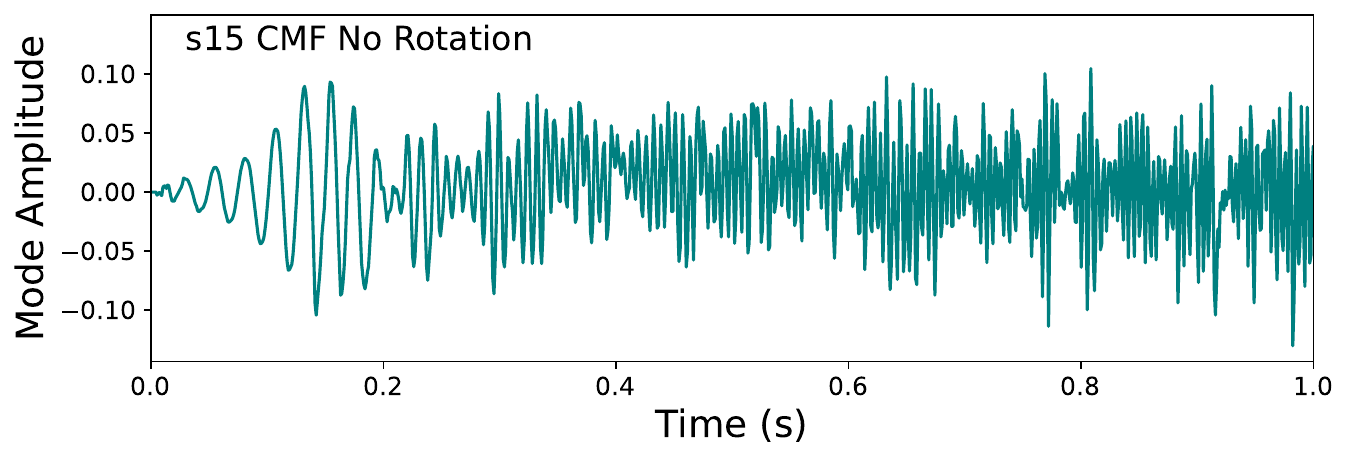}
\caption{Normalised dipole coefficients ($l=1$) of the angle-dependent shock position for the first second after bounce. The top panel is the s15 SFHx which fails to explode. The middle panel is the s16.22 SFHx model that undergoes successful shock revival. The bottom panel is the s15 CMF model which also fails to revive the shock. }
\label{fig:sasi_modes}
\end{figure}

Shortly after bounce, prompt convection occurs in all models, for all EoS. After the prompt convection phase ends, SASI activity begins in multiple models. 
To show some examples of the SASI activity, we include slices of the entropy for a representative CMF model in Figure \ref{fig:visit_cmf}, and in Figure \ref{fig:sasi_modes} we show the normalised dipole coefficients $c_1$ of the angle-dependent shock position for the non-exploding s15 CMF model, the non-exploding s15 SFHx model, and the s16.22 SFHx model that explodes shortly after bounce.  
In 2D, the normalised coefficients $c_l$ for the expansion of the angle-dependent shock position $r_\mathrm{sh}$ in polar coordinates into Legendre polynomials $P_l$ are given by
\begin{equation}
    c_{l}=
    \frac{\int P_l(\cos\theta) r_\mathrm{sh}(\theta)\,\mathrm{d}\cos\theta}{\int P_0(\cos\theta) r_\mathrm{sh}(\theta)\,\mathrm{d}\cos\theta}.
\end{equation}
The non-exploding CMF models, and s15 SFHx, all show significant SASI activity up to the end of the simulation time. The exploding non-rotating models do not have time to build up significant SASI activity before the shock revival.

\begin{figure*}
\includegraphics[width=0.50\columnwidth]{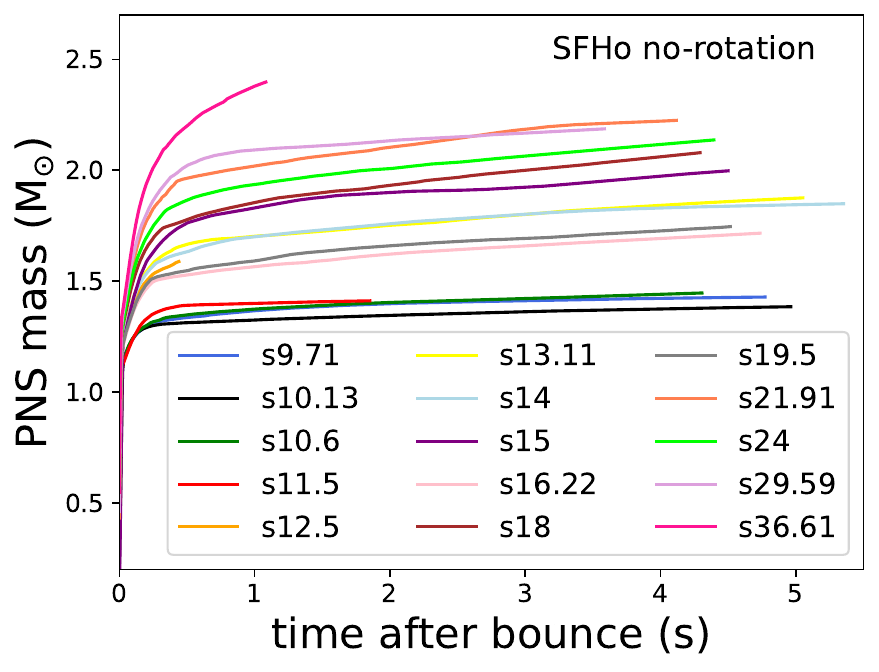}
\includegraphics[width=0.50\columnwidth]{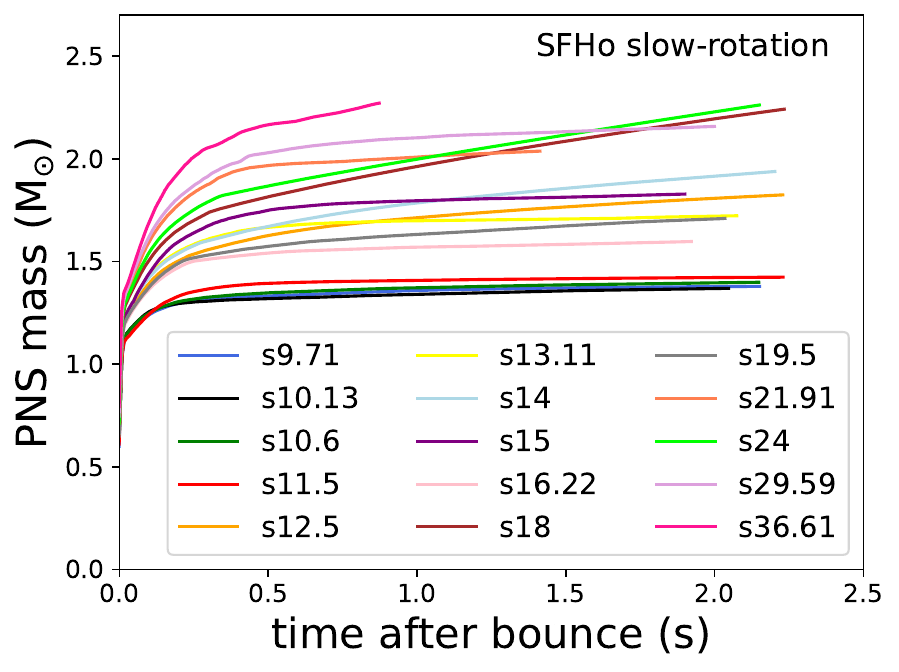}
\includegraphics[width=0.50\columnwidth]{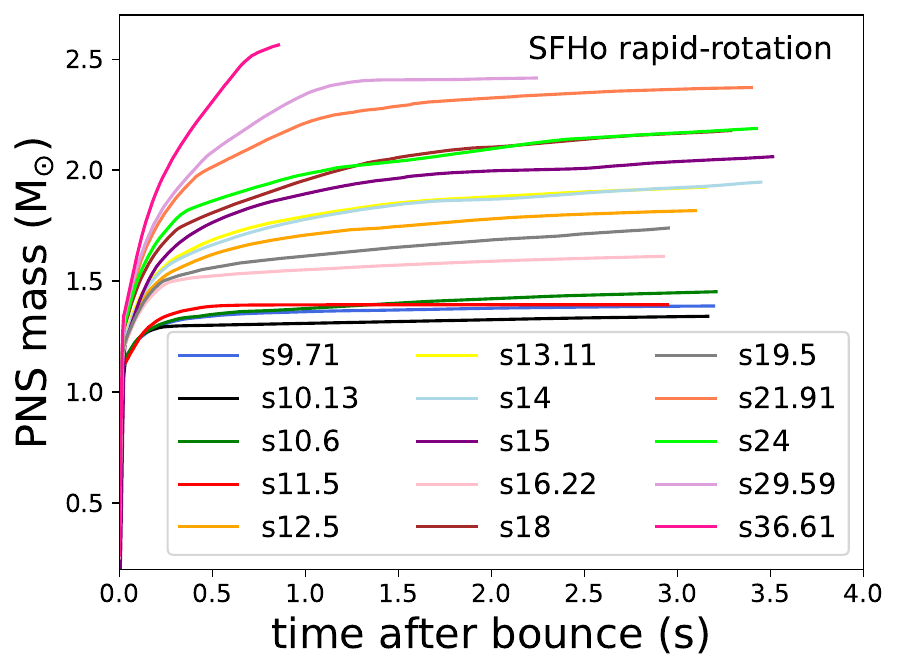}
\includegraphics[width=0.50\columnwidth]{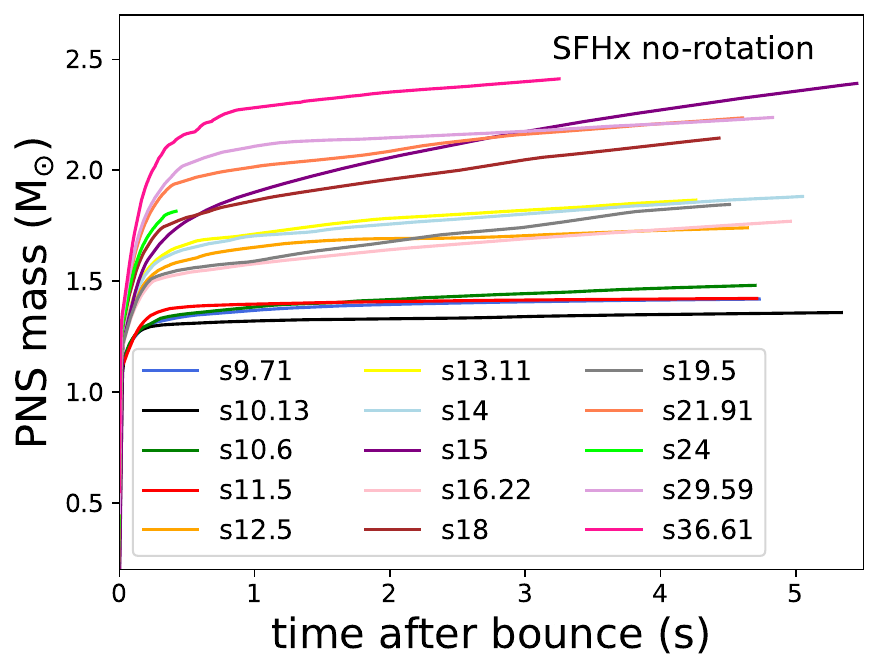}
\includegraphics[width=0.50\columnwidth]{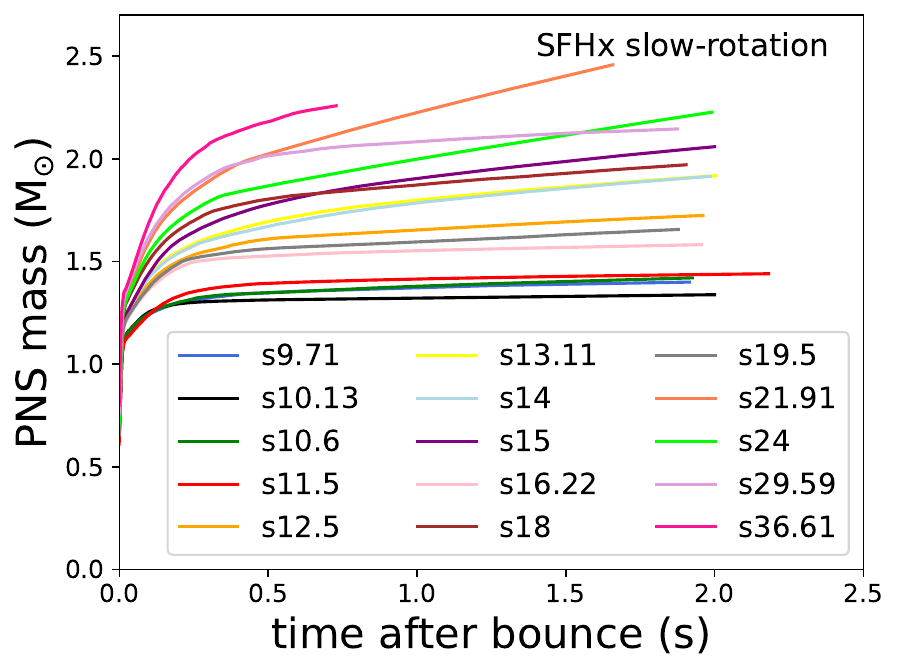}
\includegraphics[width=0.50\columnwidth]{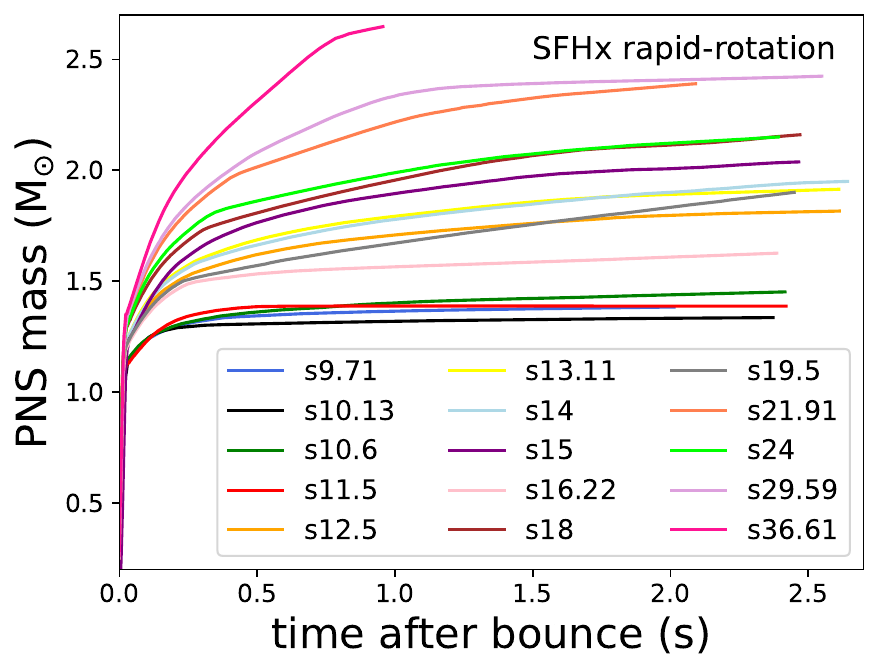}
\includegraphics[width=0.50\columnwidth]{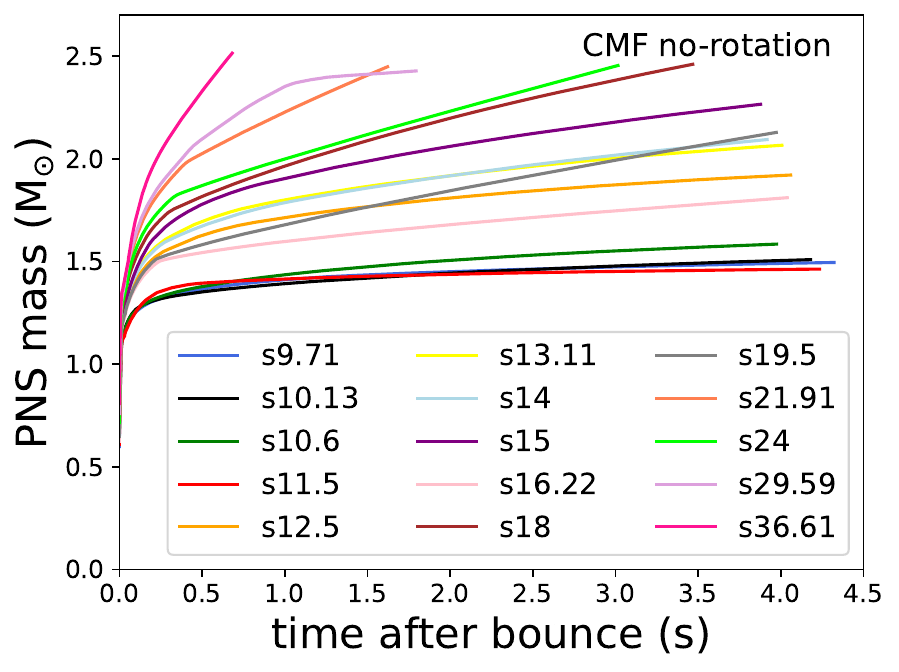}
\includegraphics[width=0.50\columnwidth]{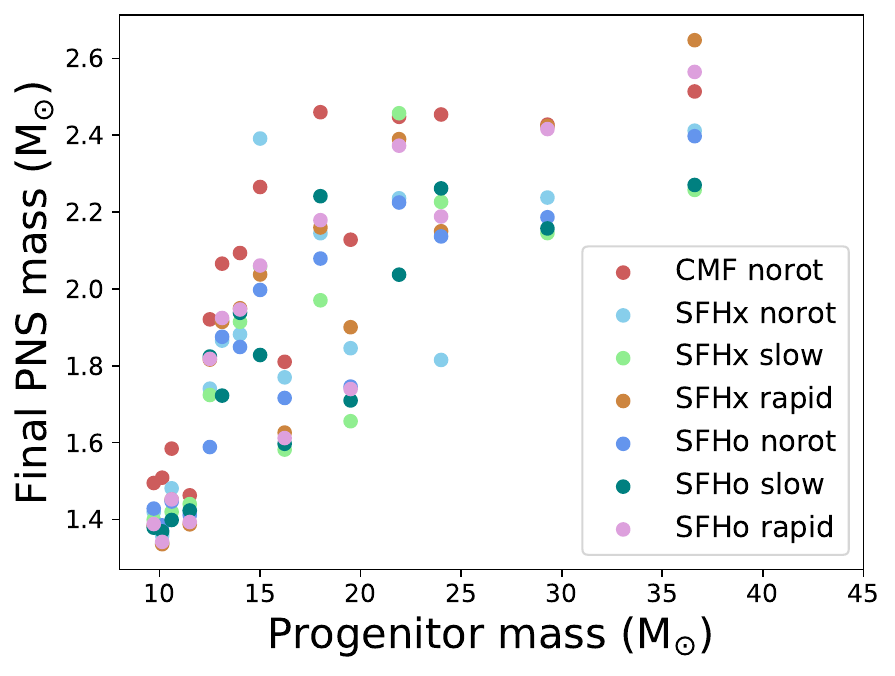}
\caption{The time evolution of the baryonic mass of the PNS for all models. Top row from left to right is SFHo no rotation, SFHo slow rotation, SFHo rapid rotation and SFHx slow rotation. The bottom row from left to right is SFHx slow rotation, SFHx rapid rotation, CMF no rotation, and the final PNS mass as a function of progenitor mass. The final PNS mass, and time to black hole formation, varies significantly for different EoS and rotation rates, which imprints into the gravitational-wave and neutrino emission. 
}
\label{fig:pns_mass}
\end{figure*}

In Figure~\ref{fig:pns_mass}, we show the evolution of the PNS mass for all of the models. The SFHo s36.61 model undergoes shock revival but still rapidly forms a black hole within 1.5\,s post bounce. The corresponding SFHx model also forms a black hole, but several seconds later. These two models add to the growing sample of simulations of so-called BHSNe, where black hole formation occurs shortly after shock revival \citep{powell_21, eggenberger_26}. The non-exploding s15 SFHx model forms a black hole more than 5\,s post bounce. Five of the CMF models, s36.61, s29.59, s24, s21.91 and s18, all formed a black hole before the end of the simulation time. The CMF s36.61 model forms a black hole quicker than the other EoS, at $\sim0.7$\,s, as the lack of explosion results in a higher mass accretion rate. The CMF models generally have significantly larger PNS masses due to the lack of shock revival. For the majority of models, the final PNS mass is similar between SFHx and SFHo, with SFHx usually producing slightly larger masses. The exception is the s15 model, where the SFHx model is significantly larger, due to shock revival only occurring for the SHFo model.

% PNS mass

%%%%%%%%%%%%%%%% SLOW ROTATION 
\subsection{Slow Rotation}

The angle-average shock radius for the slowly rotating models is shown in Figure \ref{fig:shock_radius}. No shock revival is observed in models s11.5, s13.11, s14, s15, s21.91 and s24 for SFHx, and models s12.5, s14, s18 and s24 for SFHo. This is a significantly larger number of non-explosions in comparison to the non-rotating models. The lowest mass models, for both EoS, explode very rapidly and effectively without ever retracting, similar to the analogous non-rotating and rapidly-rotating cases. Some of the larger mass models have a longer time between bounce and shock revival than in the non-rotating case. Some examples are the s16.22 and s11.5 models for SFHo, and model s18 for SFHx.   

The slow-rotating models show a smaller final diagnostic explosion energy than non-rotating models in Figure \ref{fig:energy}. However, some of this difference is due to a shorter average simulation time for the slowly rotating models, which are $\sim 2$\,s duration, compared to the $\sim 5$\,s duration of non-rotating models. Even when evaluated at the same post-bounce time, there is a trend towards lower energy explosions in the slowly rotating models. This is most noticeable in the highest mass models, for example the slowly rotating s36.61 and s21.91 SFHo models both have a factor of $\sim3$ less explosion energy than their non-rotating counterparts. 

As with the non-rotating case, the slowly rotating models experience a burst of prompt convection after bounce. The SASI also develops in a greater number of models, as there are more models that either do not undergo shock revival, or explode later.  

The evolution of the PNS mass for slowly rotating models is also shown in Figure \ref{fig:pns_mass}. The s36.61 SFHo model formed a black hole at 0.87\,s post bounce, about 200\,ms earlier than the non-rotating model. 
Large differences in PNS mass, with respect to the non-rotators, is mainly caused by differences in shock revival. 
Due to the failure to revive the shock, the SFHo s24 and s18 slowly rotating models are significantly more massive, and very close to black hole formation by the end of the simulation. The non-rotating s15 SHFo model is more massive than the slowly rotating model due to the lack of shock revival in the non-rotating case. For slowly rotating models that explode at a similar time to the non-rotating models, smaller final PNS masses are observed. For example, SFHo s10.6 has a final mass of $1.446\,\mathrm{M}_{\odot}$ without rotation, and $1.398\,\mathrm{M}_{\odot}$ with slow rotation. 

For the SFHx models, s36.61 and s21.91 both form black holes before the end of the simulation time, at 0.73\,s and 1.66\,s, respectively. The black hole formation for SFHx s36.61 was significantly earlier than for the non-rotating model, but a similar time to the SFHo slowly rotating model. The other non-exploding models are rapidly accreting mass onto the PNS, and would have formed black holes a few seconds later if the simulations were longer duration. Similar to the SFHo slow rotating models, when shock revival occurs at a similar time, the SFHx slow rotators also show a general trend towards smaller final PNS masses compared to the non-rotating case. 

%%%%%%%%%%%%%%%%%% RAPID ROTATION
\subsection{Rapid Rotation}

The shock radius for all rapidly rotating models are shown in Figure \ref{fig:shock_radius}. The SFHx s19.5 model is the only model that does not undergo shock revival before the end of the simulation time, and the ratio of heating timescales shows it is unlikely to explode at later times. 
% does not show signs of being likely to explode at later times. 
The smallest models (s11.5 and below) still rapidly undergo shock revival. However, all the other larger models undergo shock revival significantly later than the non-rotating and slowly rotating models, between 1\,s and 1.5\,s after core bounce. 

\begin{figure}
\includegraphics[width=\columnwidth]{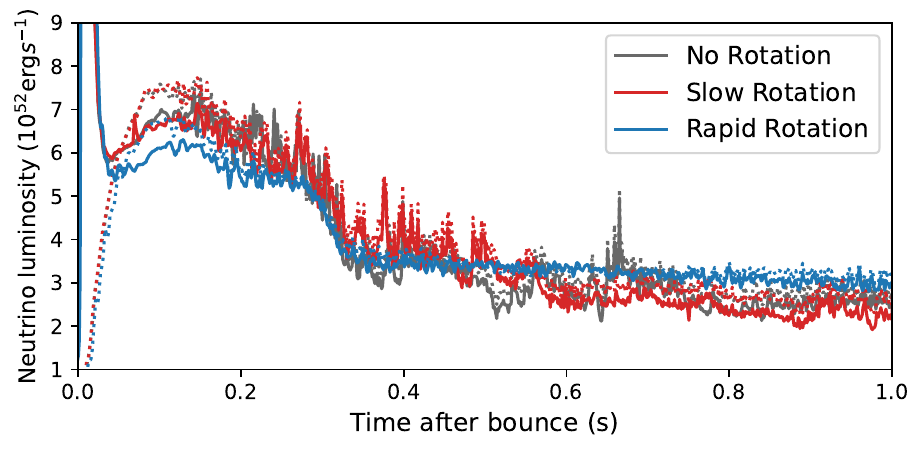}
\includegraphics[width=\columnwidth]{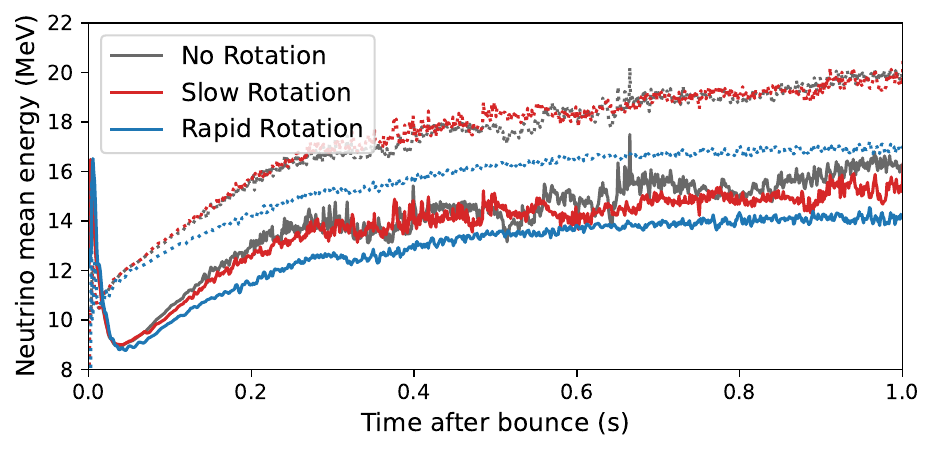}
\includegraphics[width=\columnwidth]{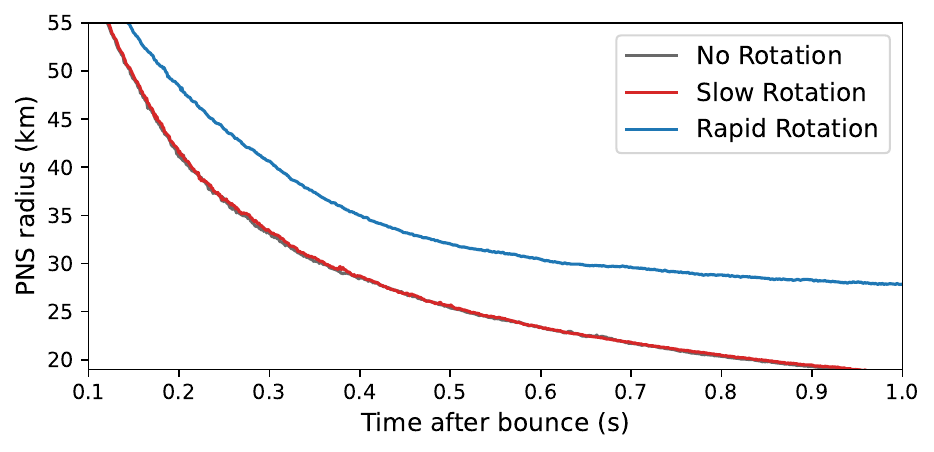}
\caption{The electron-flavor neutrino luminosity (top), neutrino mean energy (middle), and PNS radius (bottom) for the s18 SFHx model. In the top panel, the solid line is $\nu_\mathrm{e}$ and the dotted is $\bar{\nu}_\mathrm{e}$. The rapidly rotating models have a larger PNS radius, and lower levels of neutrino heating, leading to later shock revival times. }
\label{fig:s24_neutrino}
\end{figure}

In Figure \ref{fig:s24_neutrino}, we show the electron-flavor neutrino luminosity, neutrino mean energy, and mean PNS radius for an example of one model, s18 SFHx, that has much later shock revival when rapid rotation is included. The rapidly rotating models have a significantly larger mean PNS radius and gain radius, lower neutrino mean energy, and lower neutrino luminosity during the first few hundred milli-seconds. This leads to less neutrino heating in the gain region, and therefore later shock revival times for the rapid rotators. 

The diagnostic explosion energies at the end of the simulation times are shown in Figure \ref{fig:energy}. The rapidly rotating SFHo models show higher final explosion energies than the rapidly rotating SFHx models as they were run for a longer duration. For the lowest mass models, where shock revival occurs on similar timescales for both rotating and non-rotating models, the final energy is higher for the rapidly rotating models. The higher mass rapidly rotating models do not reach energies as high as the non-rotating models. The lower energies for the higher mass models are likely due to the later shock revival times and the shorter total simulation time, which mean the energies are not converged to their asymptotic value. 

\begin{figure}
\includegraphics[width=\columnwidth]{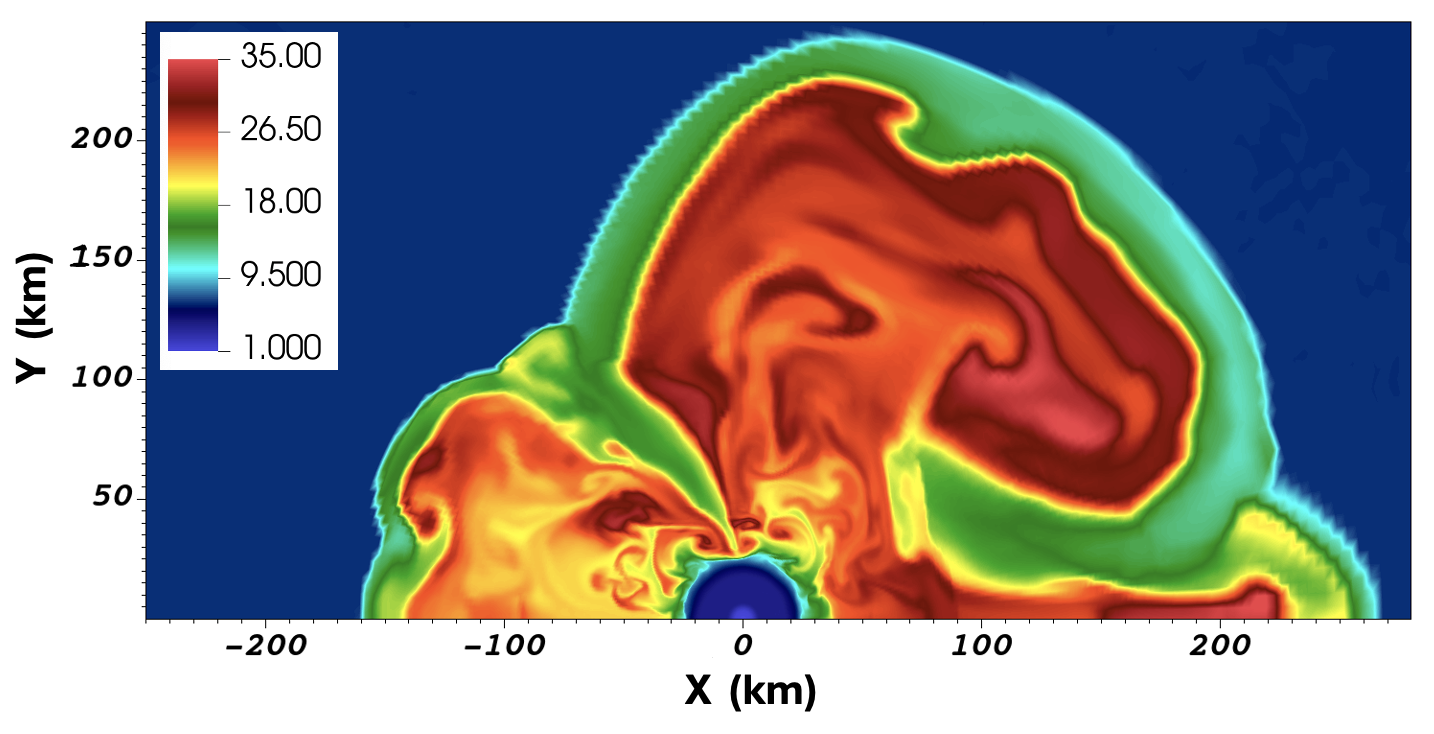}
\includegraphics[width=\columnwidth]{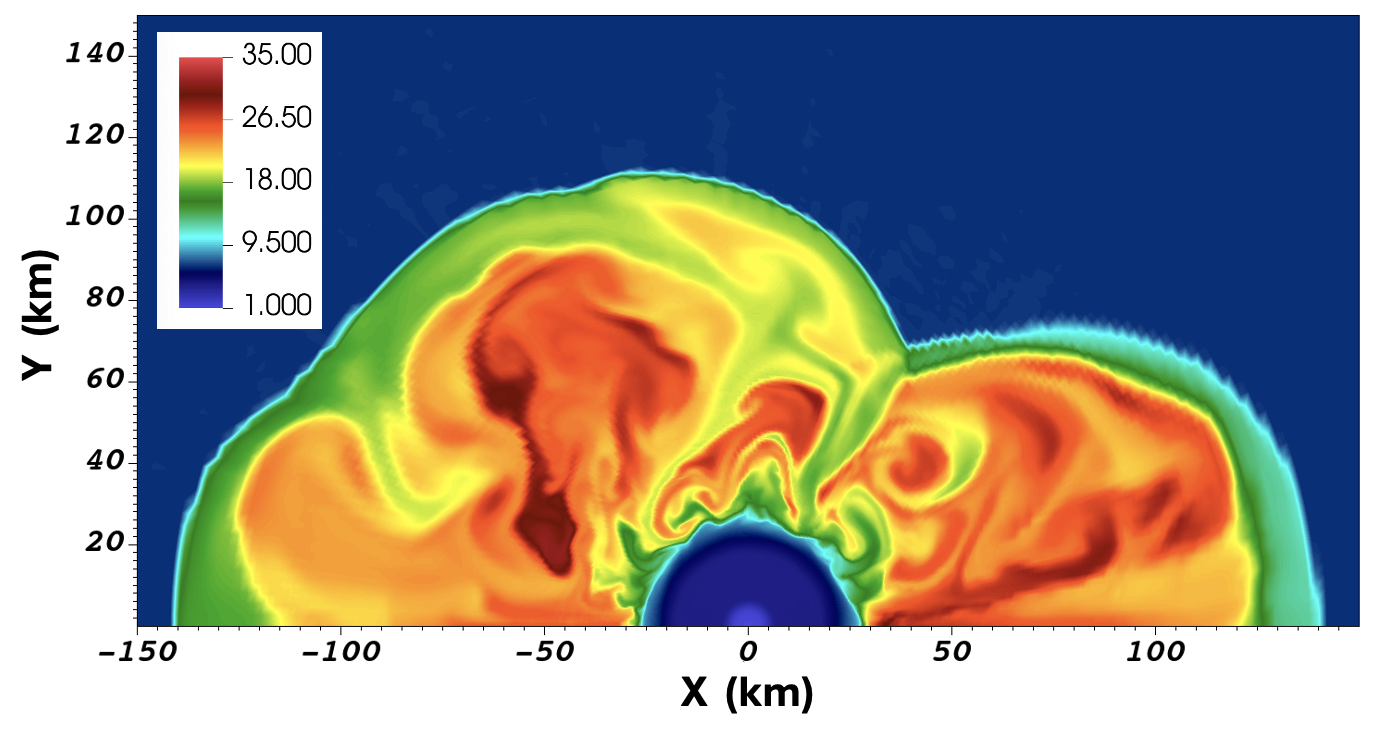}
\includegraphics[width=\columnwidth]{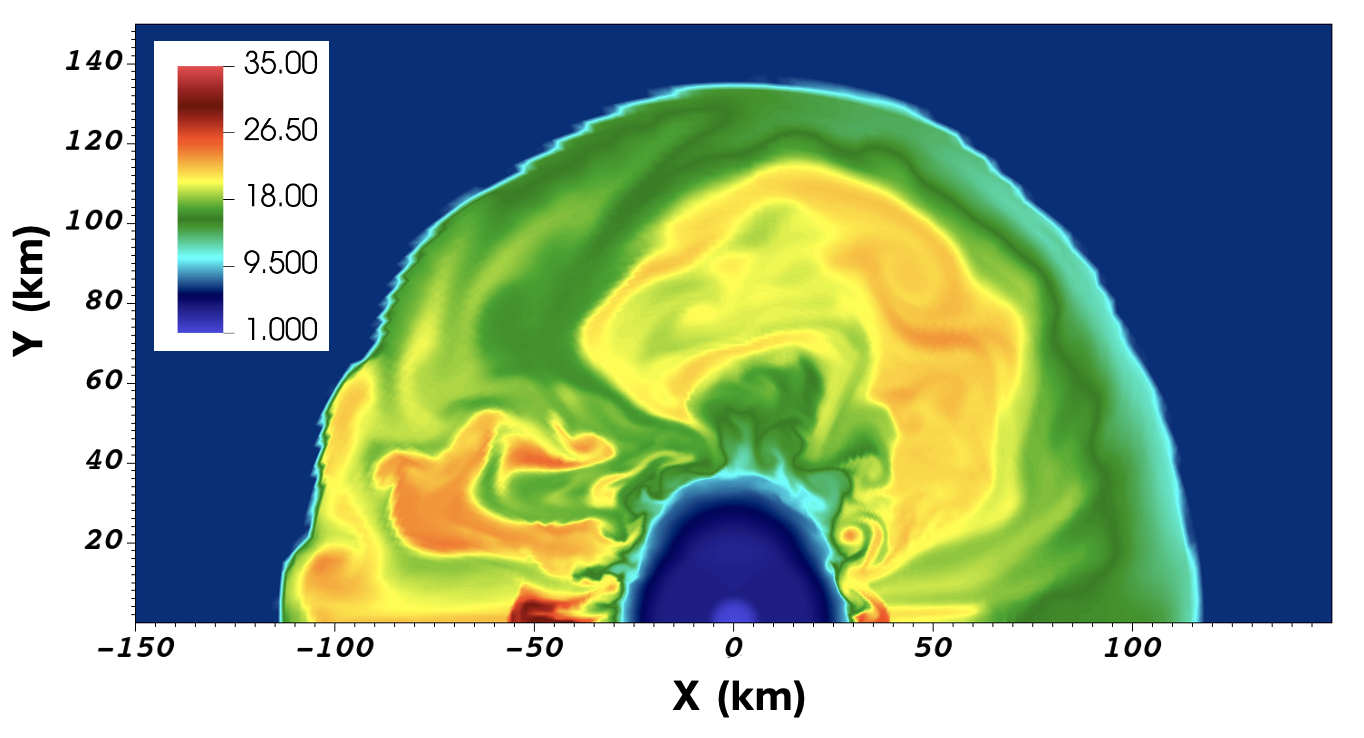}
\caption{Entropy in units of $k_\mathrm{B}/\mathrm{nucleon}$ at 0.5\,s post bounce for the s24 SFHo model. The top is non-rotating, the middle is slowly rotating, and the bottom is rapidly rotating. The rapidly rotating models show SASI activity for a longer duration due to later shock revival times.  }
\label{fig:visit_s24}
\end{figure}

After the initial core bounce, the rapidly rotating models also have a period of strong prompt convection. They then show SASI activity up to the shock revival time. As the high mass models explode at much later times than the non-rotating models, the SASI persists for a much longer time than in the non-rotating and the exploding slow rotation cases. We show a comparison of the entropy between models with different rotation rates in Figure \ref{fig:visit_s24}. For the s24 SFHo model, only the non-rotating model has undergone shock revival at 0.5\,s post bounce, and the slow and rapidly rotating models are still showing SASI activity. 

The PNS masses for the rapidly rotating models are shown in Figure \ref{fig:pns_mass}. The s36.61 models, for both SFHo and SFHx, form black holes at 0.85\,s and 0.95\,s respectively. The rotation allows the models to reach much higher final masses, with $2.65\,\mathrm{M}_{\odot}$ for s36.61 SFHx, and $2.56\,\mathrm{M}_{\odot}$ for s36.61 SFHo. For other masses, for example s24, higher PNS masses are observed in the slowly rotating models due to the lack of shock revival. In some of the smaller models for both EoS, like s10.13 and s11.5, the rapidly rotating models produce the smallest final PNS masses.

%%%%%%%%%%%%%%%%%%%%%%%%%%%%%%%%%%%%%%%%%%%%%%%%%%%%%%%%%
%%%%%%%%%%%%%%%%%%%%%%%%%%%%%%%%%%%%%%%%%%%%%%%%%%%%%%%%%
\section{The Gravitational-Wave Emission}
\label{sec:gws}

In this section, we describe the main features of the gravitational-wave emission, and how they differ between the different masses, EoS and rotation rates. 

%%%%%%%%%%%%%%%%%%%%%%%%%%%%%%%%%%%%%%%%%%
\subsection{No Rotation}

\begin{figure}
\includegraphics[width=\columnwidth]{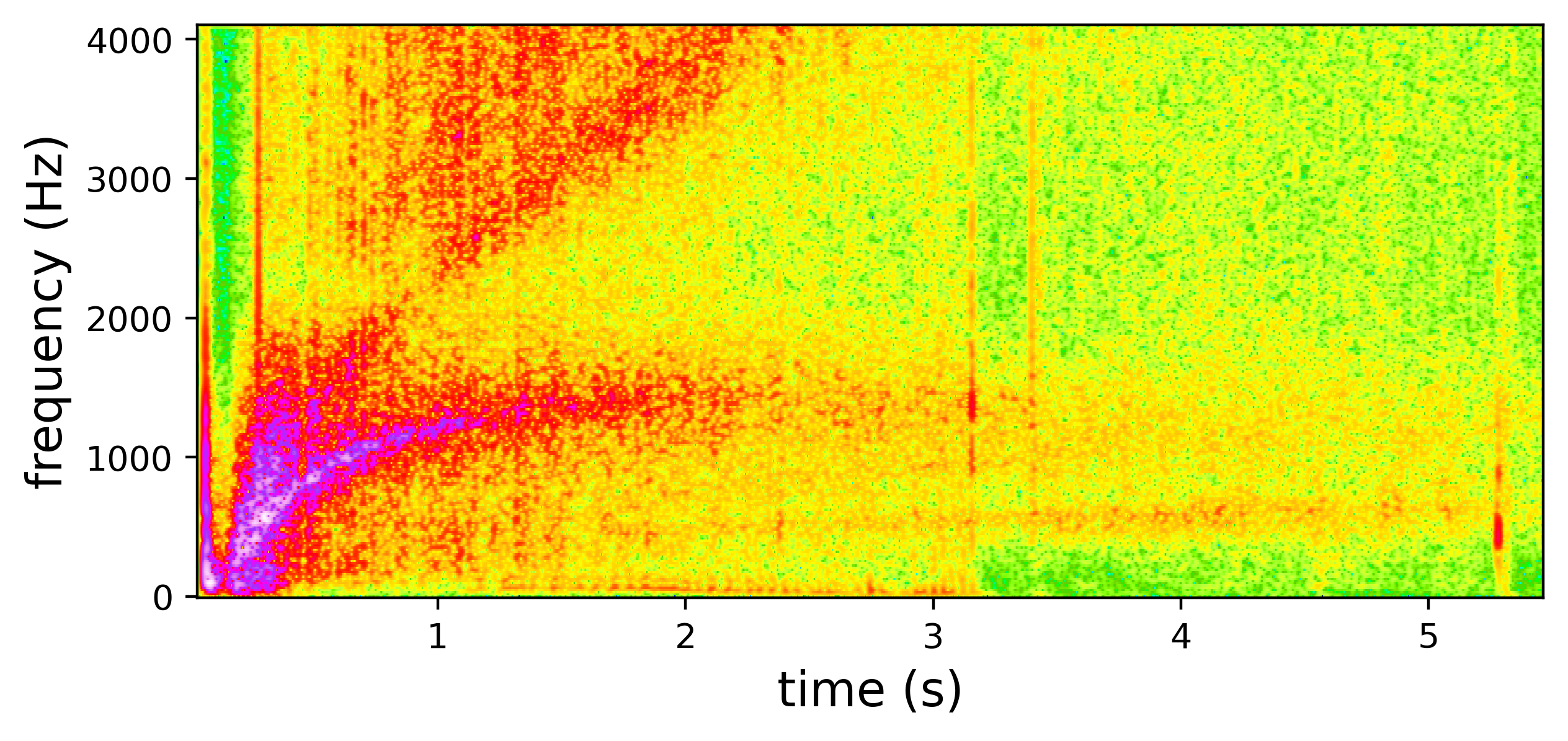}
\includegraphics[width=\columnwidth]{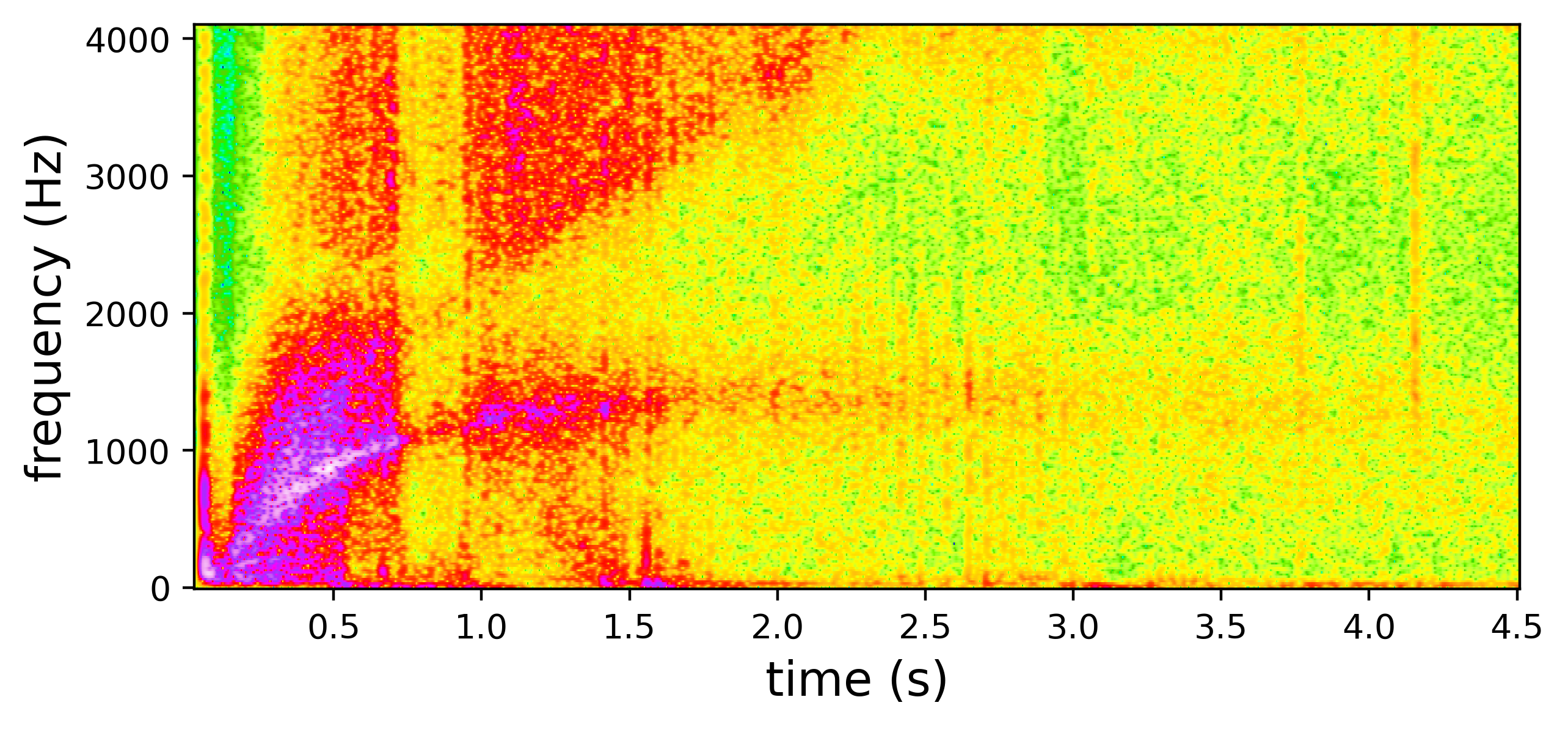}
\includegraphics[width=\columnwidth]{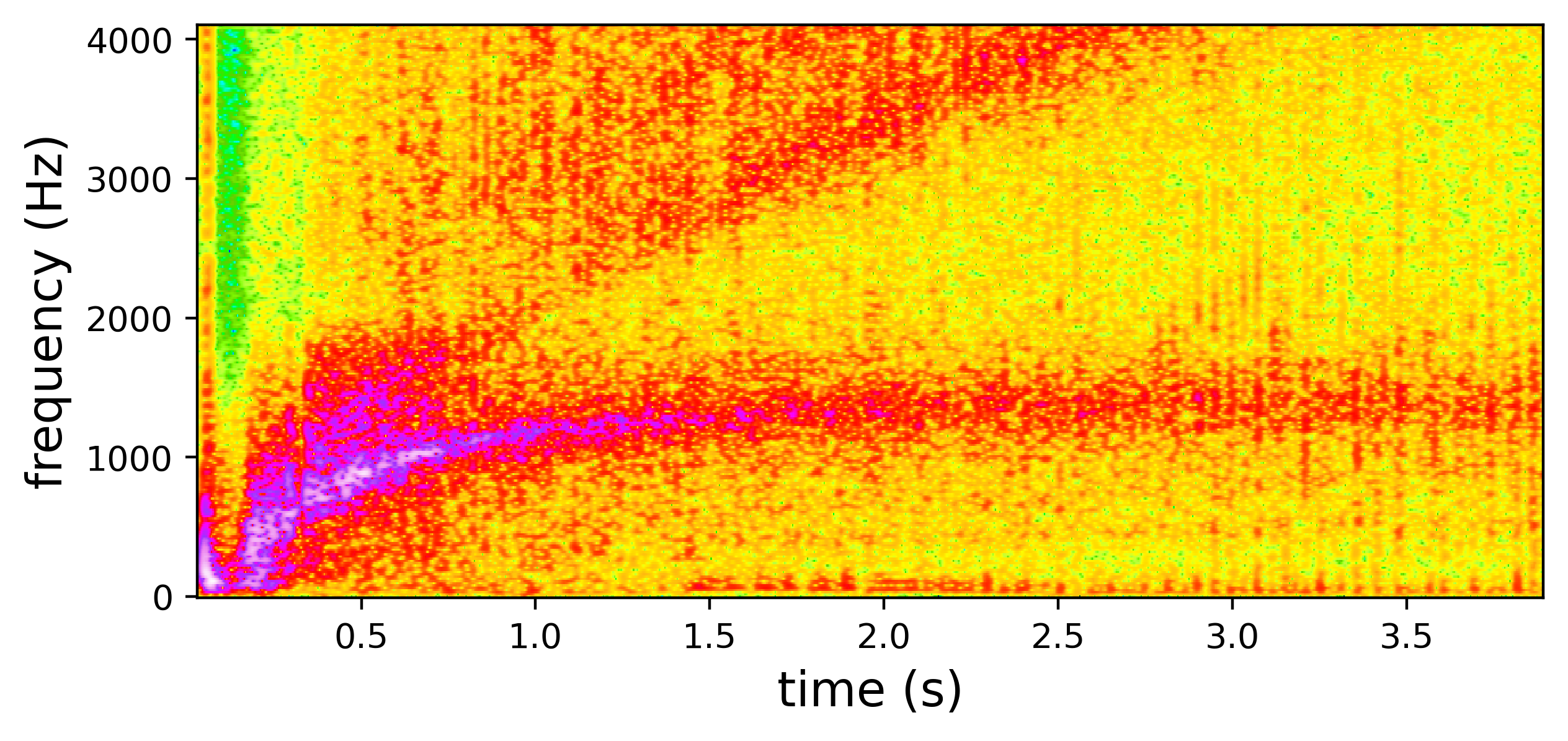}
\caption{Normalised amplitude spectrograms of the gravitational-wave emission for three representative non-rotating models. The top panel is model s15 SFHx, the middle panel is s15 SFHo, and the bottom is s15 CMF. As well as the dominant emission mode, a second higher frequency mode is observed that extends above 4000\,Hz.}
\label{fig:norot_gws}
\end{figure}

\begin{figure}
\includegraphics[width=\columnwidth]{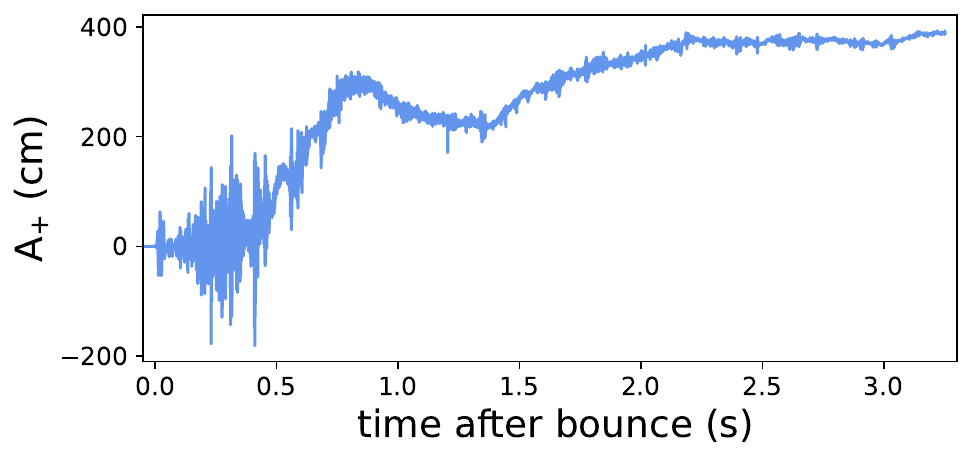}
\includegraphics[width=\columnwidth]{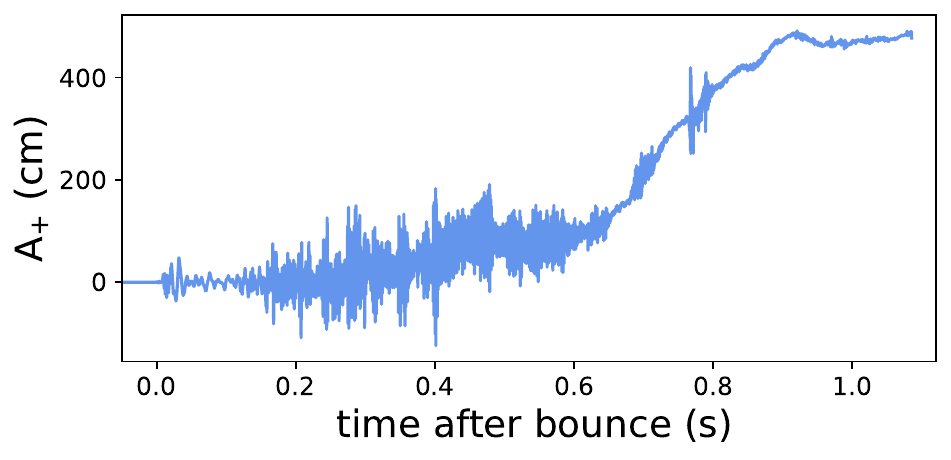}
\includegraphics[width=\columnwidth]{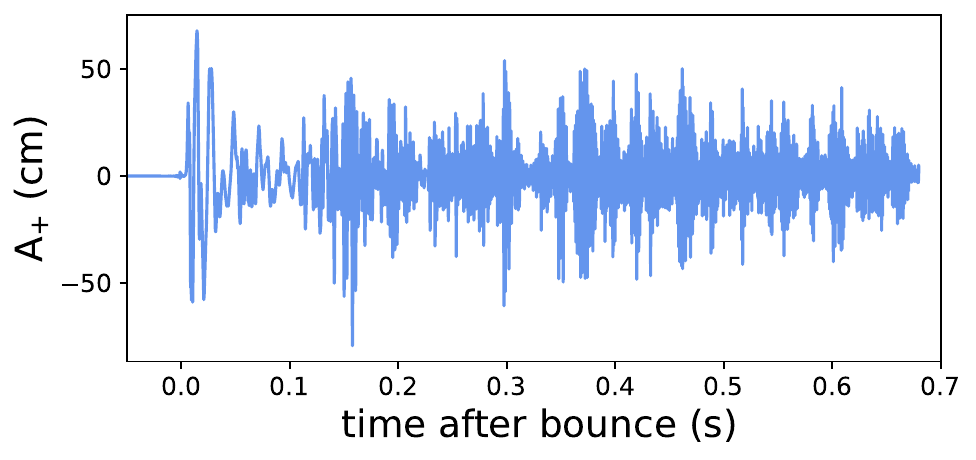}
\caption{Examples of the time-series gravitational-wave emission from the non-rotating models. The bottom is s36.61 CMF, the middle is s36.61 SFHo, and the top is s36.61 SFHx. The different signal durations are due to different black hole formation times. The SFHo and SFHx models are higher amplitude, and show matter memory, which is absent from the CMF model due to the lack of shock revival. }
\label{fig:norot_timeseries}
\end{figure}

Example spectrograms of the non-rotating gravitational-wave signals are shown in Figure \ref{fig:norot_gws}, and example time-series are shown in Figure \ref{fig:norot_timeseries}. 
All models show high amplitude gravitational-wave emission from prompt convection shortly after the core bounce, at frequencies  $\sim 100-150$\,Hz. The amplitude of the prompt convection varies between 39\,cm and 127\,cm for the SFHx models,  40\,cm and 115\,cm for the SFHo models, and 29\,cm and 82\,cm for the CMF models. The amplitude of the prompt convection can be somewhat stochastic, however the maximum amplitude of the prompt convection for the CMF EoS is significantly lower than for the other two EoS. The overall gravitational-wave amplitude of each model is overestimated because the simulations are axisymmetric \citep{andresen_19}. However, the relative differences between the amplitudes should remain the same if the models were in 3D.  

All of the models have an f/g-mode that begins shortly after the end of the prompt convection phase. This high frequency mode is clearly visible for the first 1.5\,s after bounce, and rises in frequency to $\sim 1300$\,Hz. In general, the larger mass models have higher amplitude f/g-mode emission, with the s36.61 SFHo model reaching amplitudes of 210\,cm, and the s36.61 SFHx model reaching amplitudes of 182\,cm. The CMF models have much lower amplitude g/f-mode emission, reaching a maximum of 63\,cm for model s29.59. This is likely mainly due to the lack of shock revival in the CMF models. Previous work has shown that there is often a high frequency ``haze'' of gravitational-wave emission above the dominant mode \citep{vartanyan_23, andresen_26}. In Figure \ref{fig:norot_gws}, it appears that the ``haze'' during the first 1\,s is the start of a secondary higher frequency PNS mode. This mode reaches frequencies of up to 4000\,Hz at 3\,s post bounce. This higher frequency mode also appears to show a mode crossing at around 2200\,Hz. 

\begin{figure}
\includegraphics[width=\columnwidth]{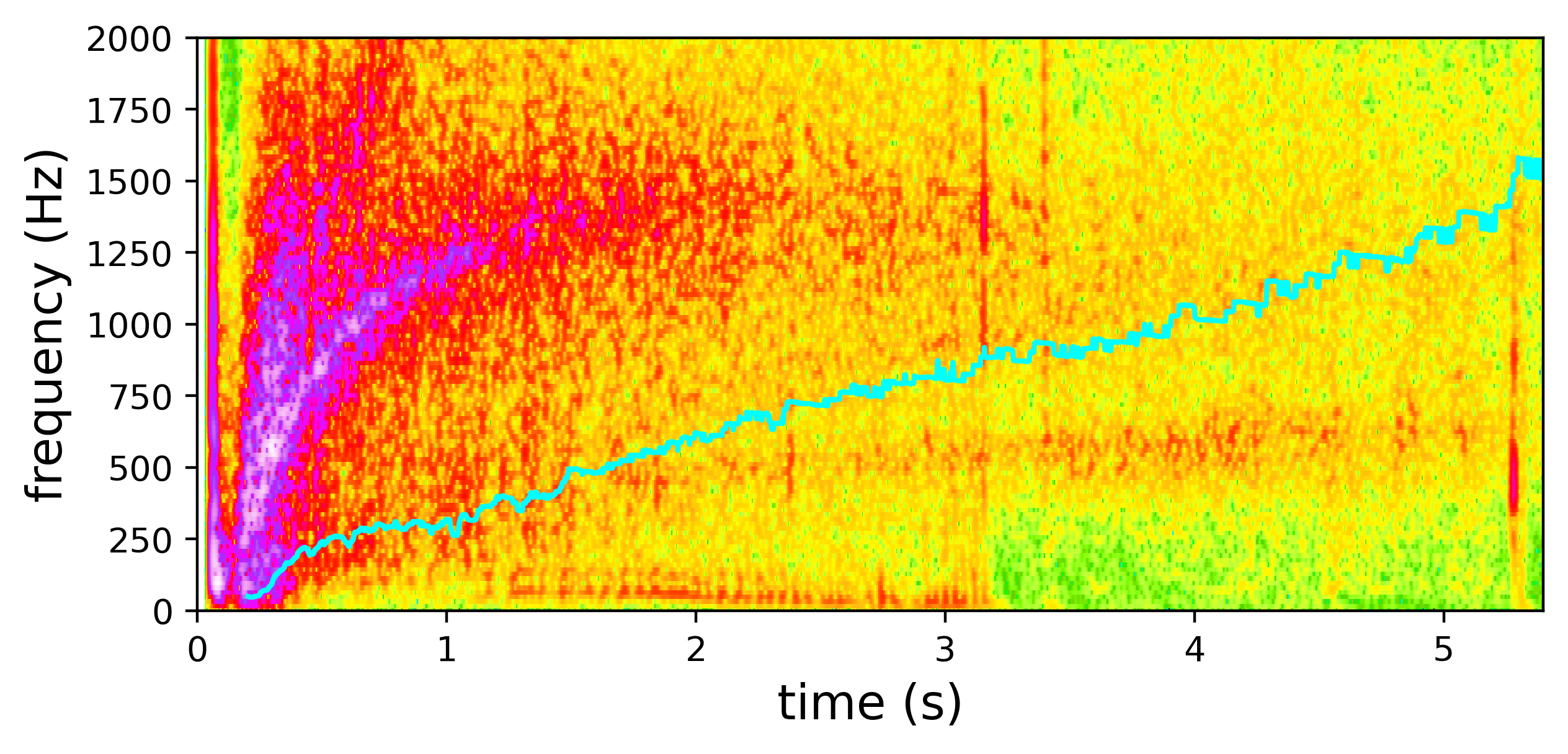}
\includegraphics[width=\columnwidth]{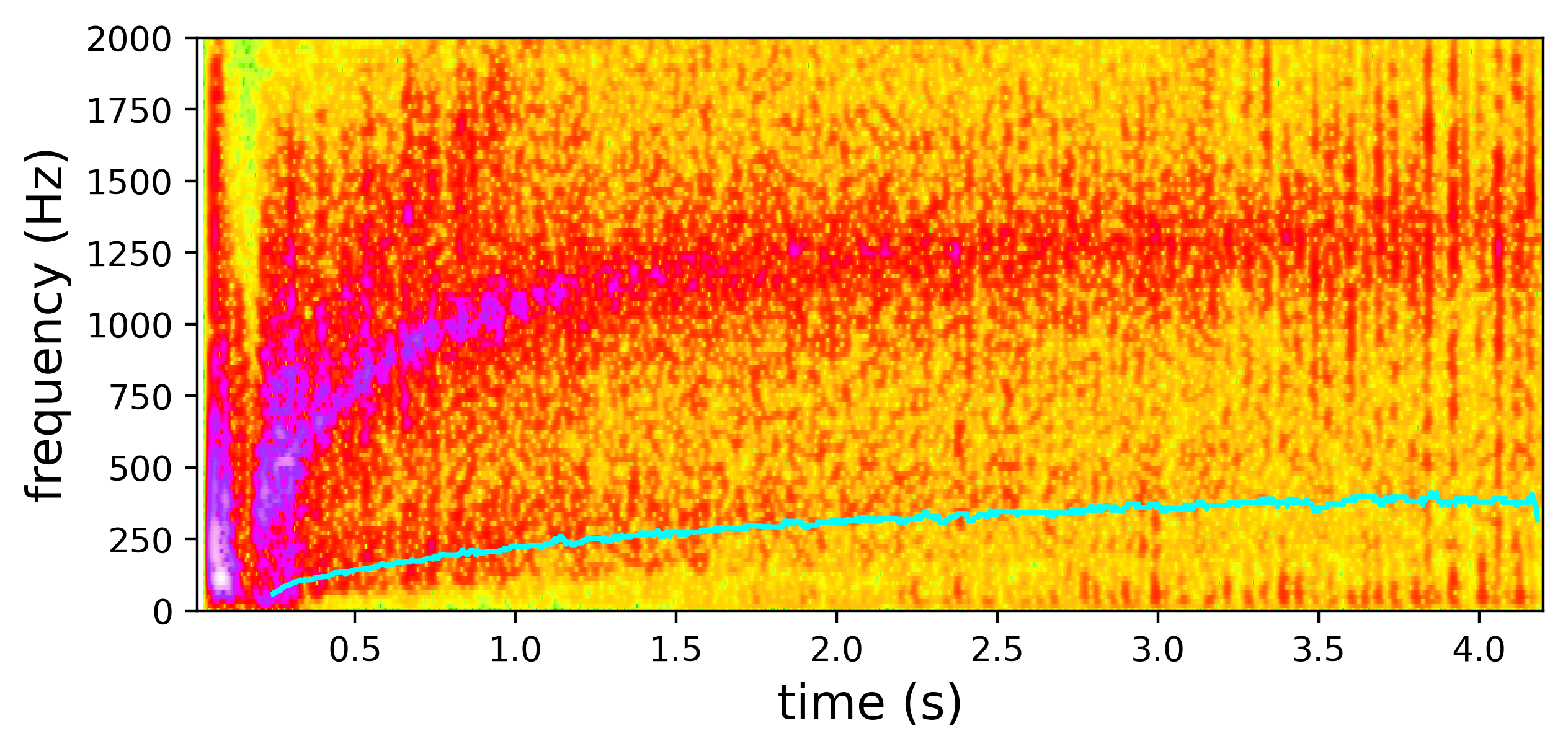}
\caption{The gravitational-wave signals for the non-rotating models s15 SFHx (top), CMF s10.13 model (bottom). The blue line shows the gravitational-wave frequency for the SASI predicted by Equation \ref{eqn:sasi}.}
\label{fig:sasi_norot}
\end{figure}

Most of the models show low frequency gravitational-wave emission due to the SASI before the time of shock revival. As the s15 SFHx model does not undergo shock revival, it has SASI activity right up until the end of the simulation at 5.45\,s.  
The same is observed in the non-exploding CMF models. However, it is difficult to see the lower frequency mode from the SASI in the gravitational-wave emission from the higher mass CMF models, as the energy in the spectrograms becomes so dominated by the higher frequency mode. Previous work has shown that the gravitational-wave emission from the SASI can be approximated by,
\begin{equation}
f_{\mathrm{SASI}} =  \frac{2}{19\,\mathrm{ms}} \left( \frac{R_{sh}}{100\,\mathrm{km}} \right)^{-3/2} \left[ \ln \left( \frac{R_\mathrm{sh}}{R_{\mathrm{PNS}}} \right) \right]^{-1},
\label{eqn:sasi}
\end{equation}
where $R_\mathrm{sh}$ is the average shock radius, and $R_{\mathrm{PNS}}$ is the radius of the PNS \citep{mueller_14}. As found in previous work \citep{andresen_17, powell_21}, the SASI mode is expected to be at $2 \times f_{\mathrm{SASI}} $ in the spectrograms due to frequency doubling. 
Some examples of the gravitational-wave frequency predicted by this equation are given in Figure \ref{fig:sasi_norot}. For the CMF models, we find that this equation is a good description for the lower frequency SASI mode in the gravitational-wave signal. The s15 SFHx model has a mode at $\sim 500$\,Hz that is visible even after 5\,s. However, Figure \ref{fig:sasi_norot} shows that the $f_{\mathrm{SASI}}$ relation is not a good fit for this mode. It may still be due to the SASI, as the mode could be at the SASI base frequency, instead of having the frequency doubling that we see for the SASI modes in the other models. Another possibility is that this is a higher order g-mode. 

%%%%%%%%%%%%%%%%%%%%%%%%%%%%%%%%%%%%%%%%%
\subsection{Slow Rotation}

\begin{figure}
\includegraphics[width=\columnwidth]{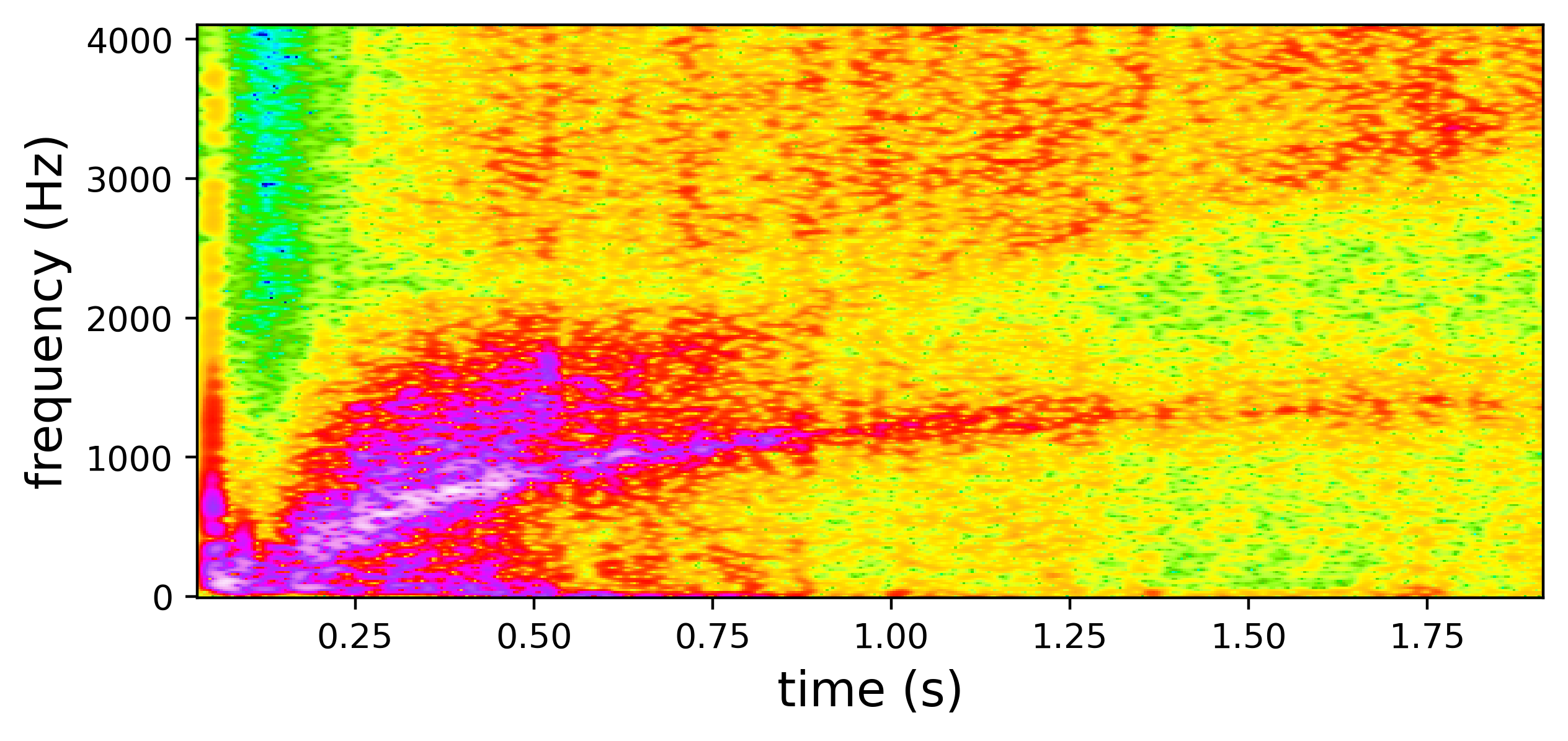}
\includegraphics[width=\columnwidth]{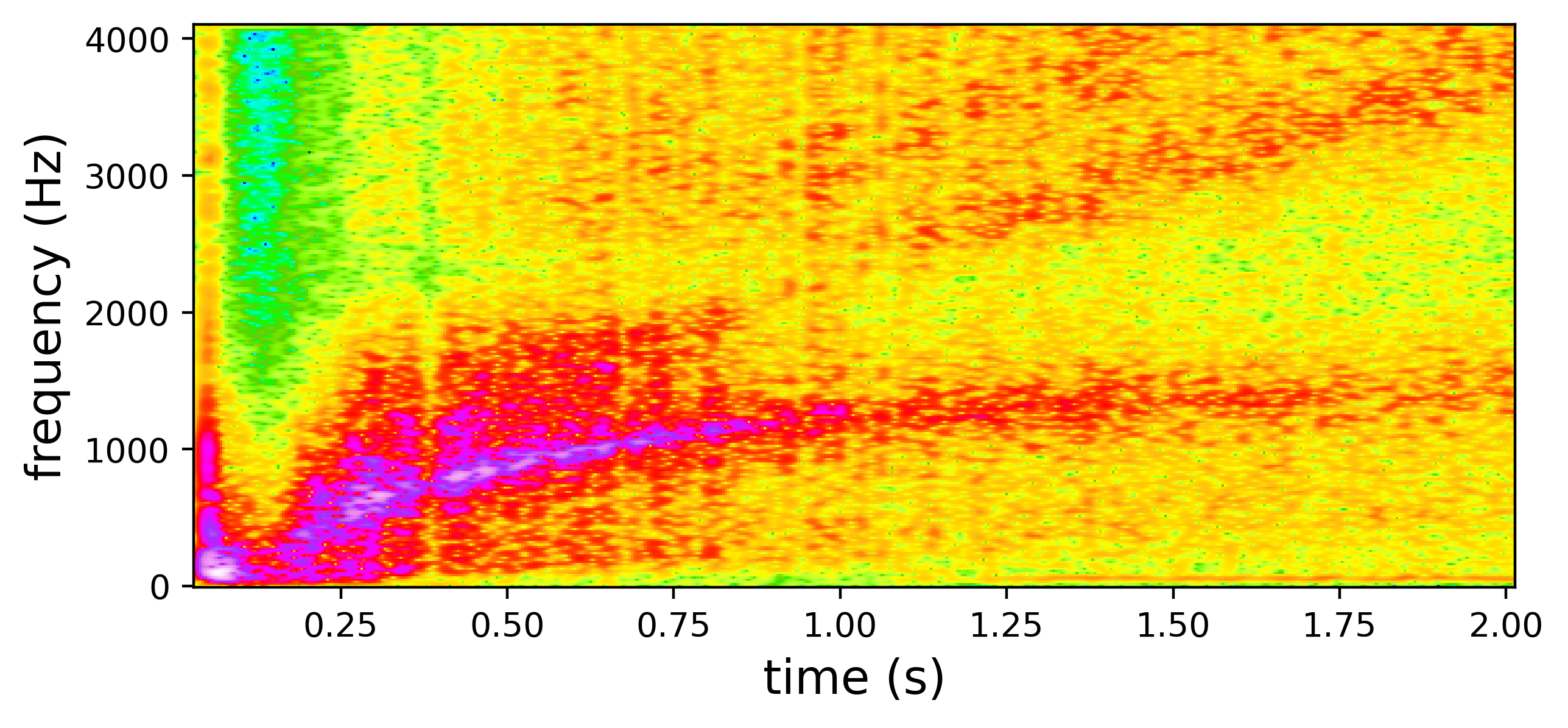}
\caption{Examples of the gravitational-wave emission from the slowly rotating models in the time-frequency domain. The top panel shows the s15 SFHo model. The bottom panel shows the s15 SFHx model. Both models show the dominant PNS mode and a weaker higher frequency mode that reaches up to 4000\,Hz by the end of the simulation time. }
\label{fig:srot_gws}
\end{figure}

\begin{figure}
\includegraphics[width=\columnwidth]{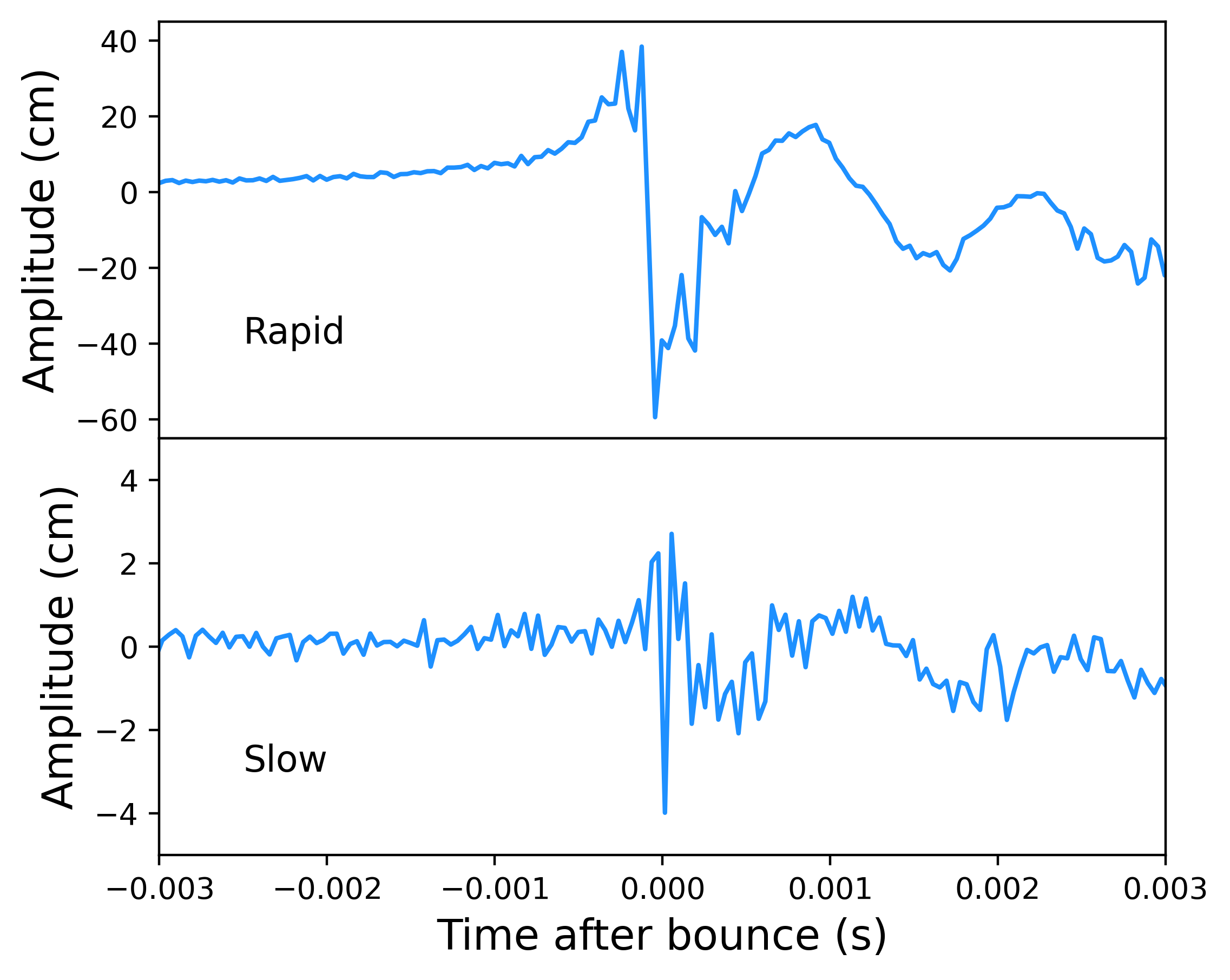}
\caption{The rotating bounce signal for the s24 SFHo models. Top is the rapidly rotating model, and the bottom is the slowly rotating model. The bounce signal is significantly larger for more rapidly rotating models. }
\label{fig:bounce}
\end{figure}

The slowly rotating models start with a small spike in the time series at core bounce. An example is shown in the bottom panel of Figure \ref{fig:bounce}. The core bounce signal has very low amplitude in comparison to later parts of the gravitational-wave signal. However, the later parts of the signal have artificially increased amplitudes due to the 2D set-up, whereas the core-bounce amplitude is known to not vary between 2D and 3D \citep{scheidegger_08}. The amplitude varies between $\sim 2-7$\,cm for the different masses. 

Example spectrograms of the slowly rotating models are shown in Figure \ref{fig:srot_gws}. Similar to the non-rotating models, the slowly rotating models have a short burst of gravitational waves from prompt convection shortly after the bounce. The amplitude of the prompt convection signal varies between 48\,cm and 113\,cm for SHFo, and between 25\,cm and 98\,cm for SFHx. For the SFHo models, this is a similar amplitude to the non-rotating models, while the SFHx amplitudes are a bit smaller. 

The models then develop dominant f/g-modes, with similar frequencies to the non-rotating case. There is again also a secondary mode that reaches about 4000\,Hz after two seconds. There is a trend towards higher gravitational-wave amplitudes in higher mass models. The highest amplitudes are found in the s29.59 models, which reach 128\,cm for SFHx and 190\,cm for SFHo. In general, the slowly rotating models have slightly less gravitational-wave energy than the non-rotating models, due to slightly later shock revival times, and less energetic explosions. 

\begin{figure}
\includegraphics[width=\columnwidth]{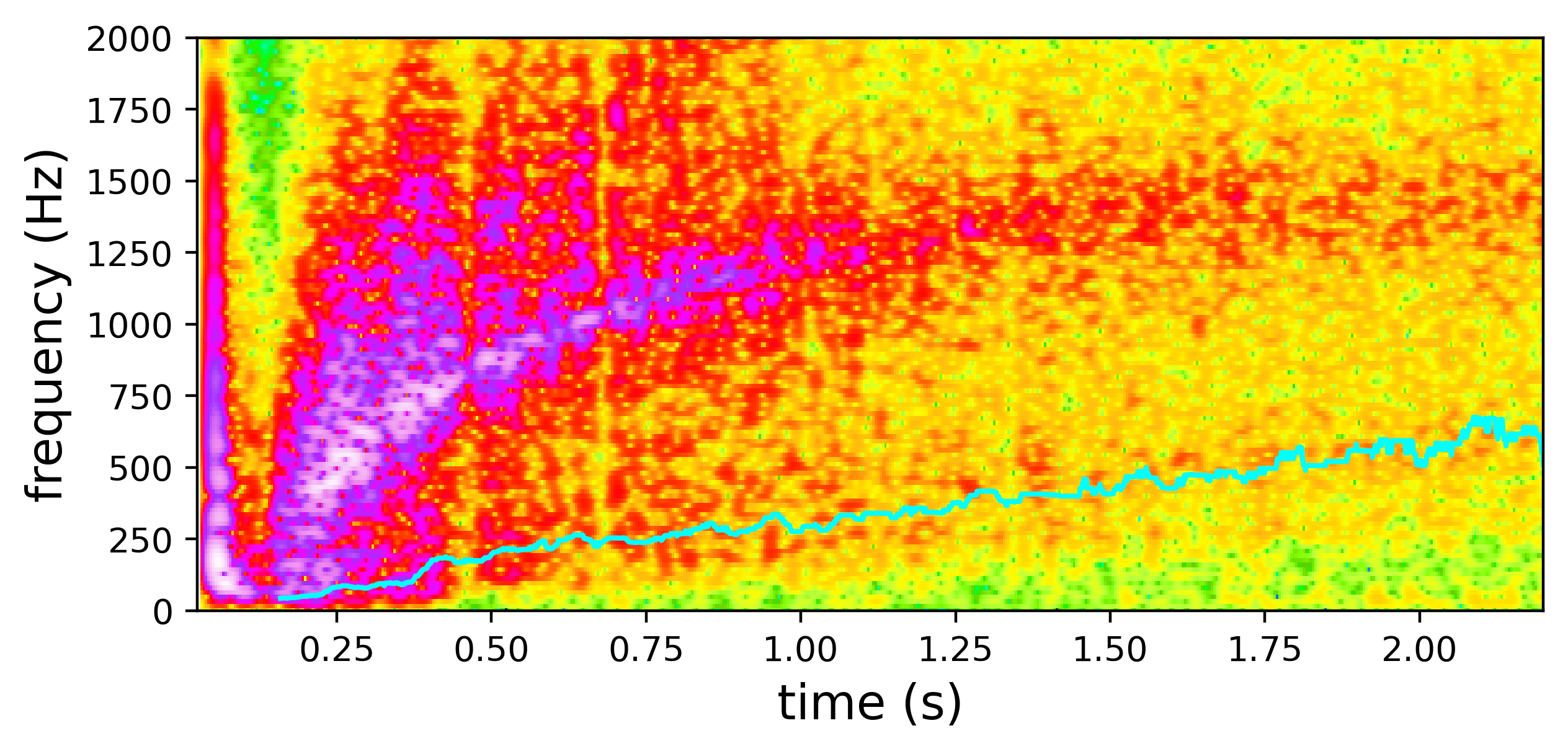}
\includegraphics[width=\columnwidth]{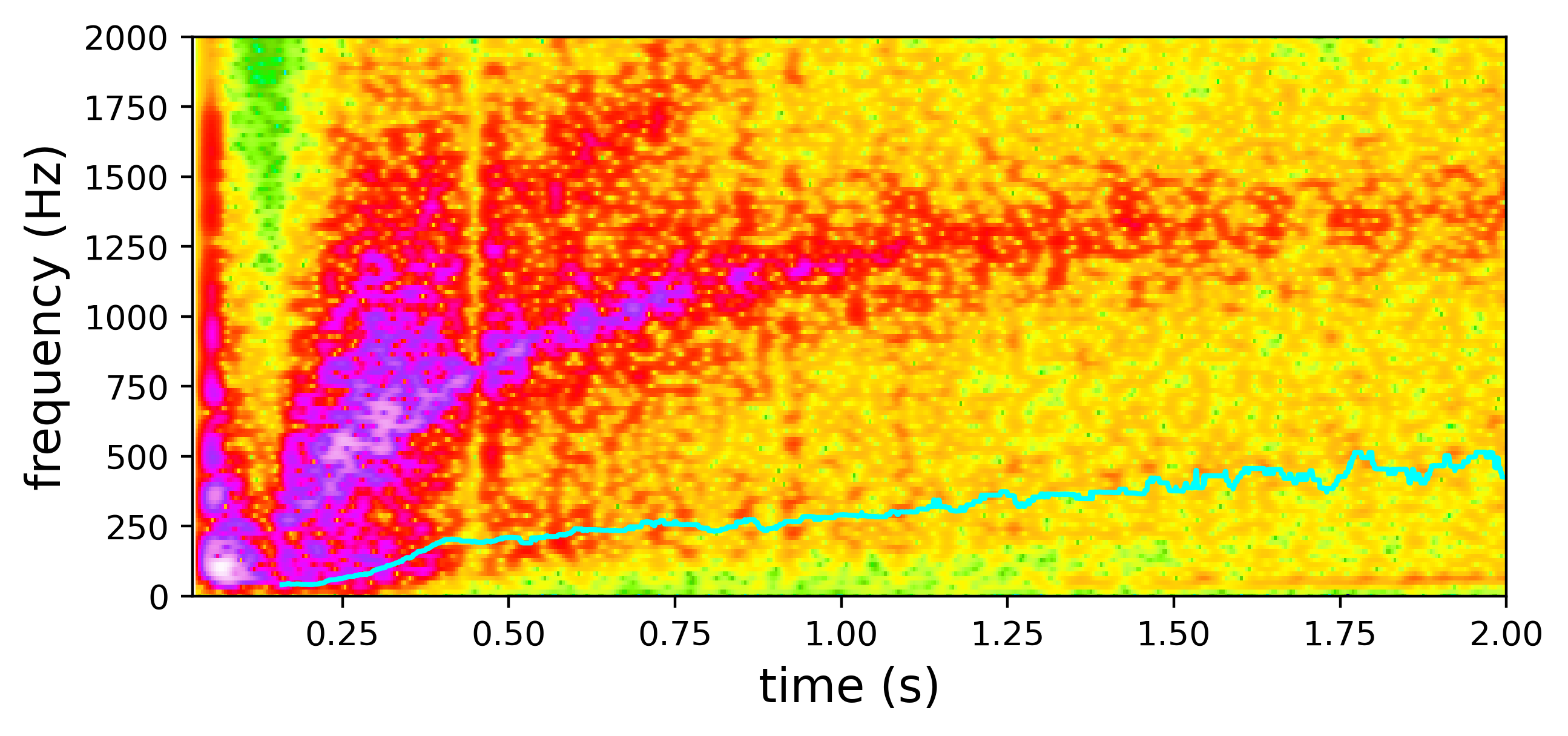}
\caption{The top panel is a spectrogram of the slowly rotating model s14 SFHo, and bottom is the slowly rotating model s13.11 SFHx. Both of these models do not undergo shock revival. The blue line is the frequency predicted by equation \eqref{eqn:sasi}. We find this equation is still a good description of the gravitational-wave emission when rotation is slow.}
\label{fig:sasi_srot}
\end{figure}

SASI is also observed in the gravitational-wave emission of the slowly rotating models. Two examples are shown in Figure \ref{fig:sasi_srot}. The gravitational-wave frequency of the SASI is still well described by Equation \eqref{eqn:sasi}. SASI is seen in a larger number of slowly rotating models due to a larger number of non-exploding models, and later shock revival in the higher mass models.  

%%%%%%%%%%%%%%%%%%%%%%%%%%%%%%%%%%%%%%%%%%%%
\subsection{Rapid Rotation}

\begin{figure}
\includegraphics[width=\columnwidth]{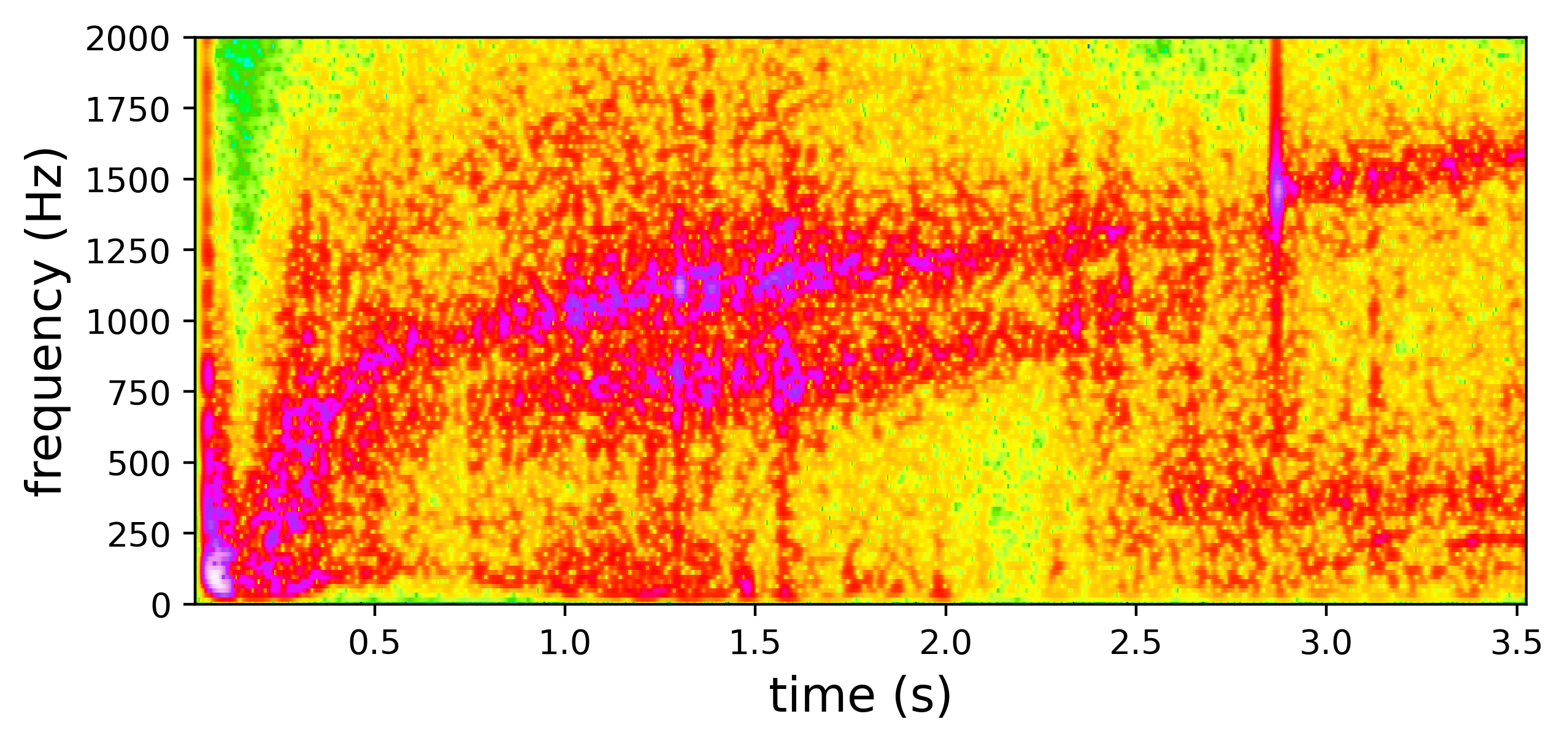}
\includegraphics[width=\columnwidth]{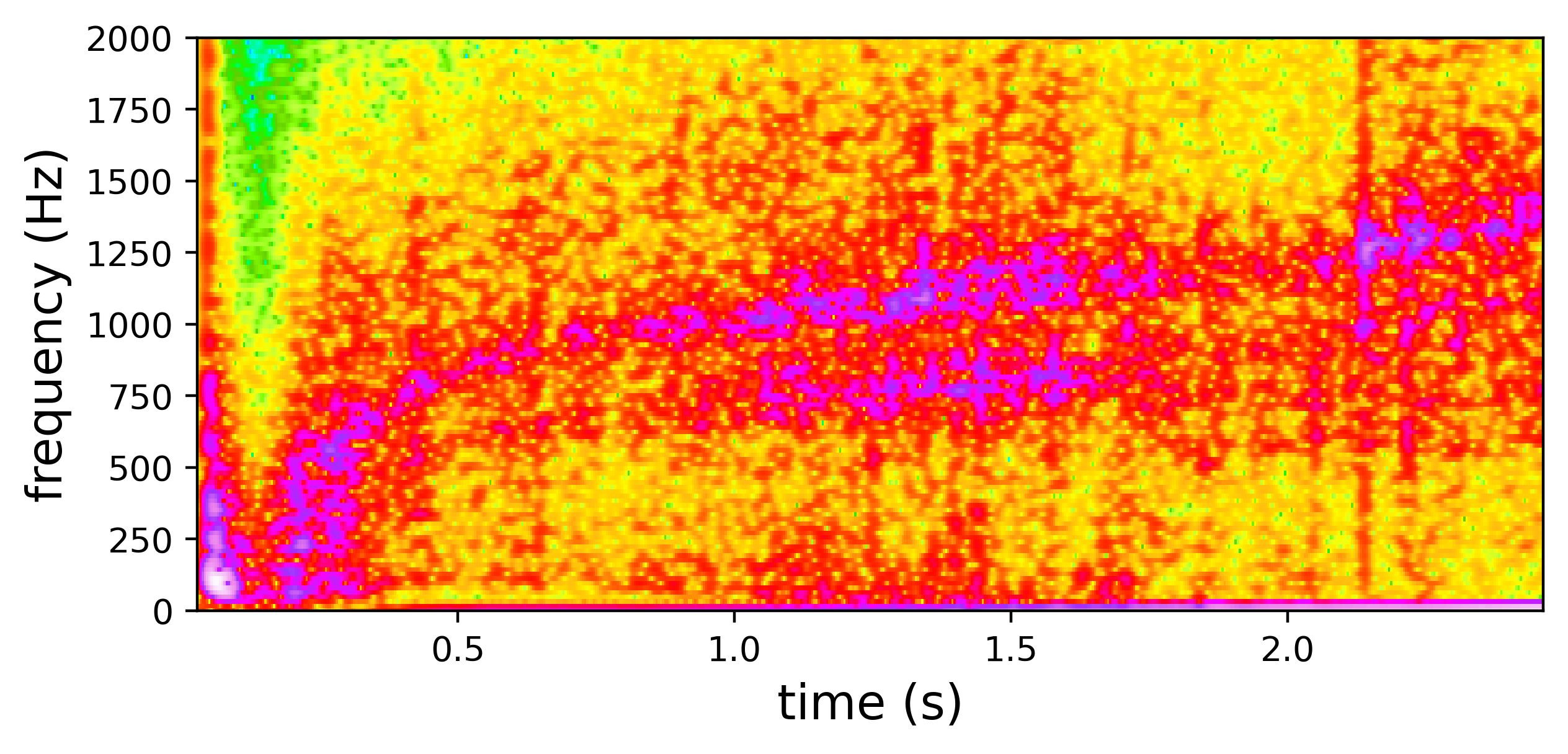}
\includegraphics[width=\columnwidth]{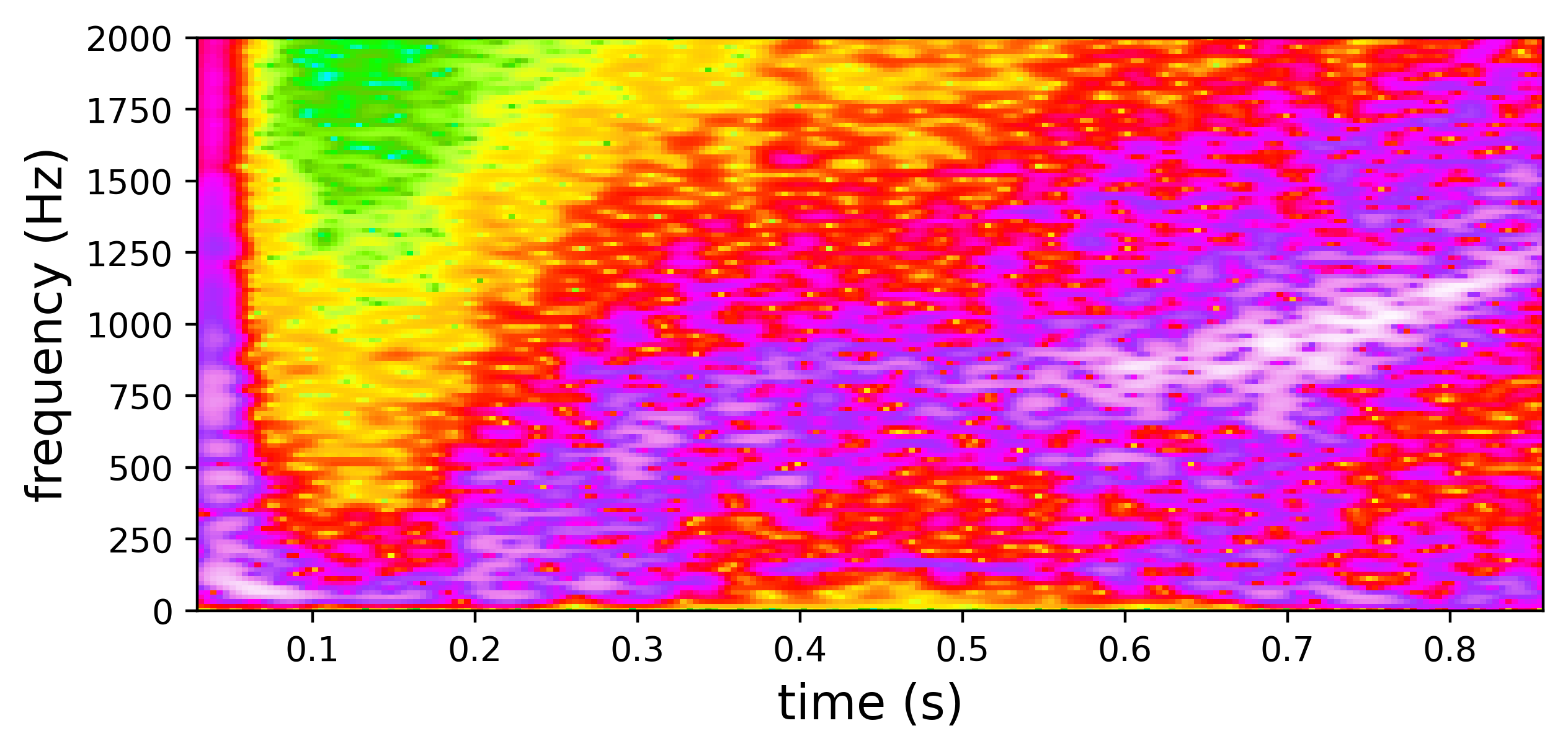}
\caption{Examples of the gravitational-wave emission from the rapidly rotating models. The top panel shows model s15 SFHo, the middle panel shows s15 SFHx, and the bottom panel is s36.61 SFHo. The models rise steeply in frequency for the first few hundred milliseconds and then increase more slowly towards the end of the simulation time.}
\label{fig:rapid_gws}
\end{figure}

The gravitational-wave emission for the rapidly rotating models starts with a spike in the time series at core bounce. The bounce signal is shown in the top panel of Figure \ref{fig:bounce}. The bounce signal is about 10 times larger than for the slowly rotating models, which is expected. There is also some variation of bounce amplitude with mass. This was also found by \citet{dimmelmeier_08}. The bounce signal varies between 10\,cm and 145\,cm for the SFHo models, and 10\,cm and 140\,cm for the SFHx models. 

Example spectrograms of the rapidly rotating models are shown in Figure \ref{fig:rapid_gws}. The rapidly rotating models also have strong prompt convection shortly after bounce. For the SFHx models, the amplitude of the prompt convection varies between 48\,cm and 133\,cm. As with the slowly rotating models, the prompt convection for the SFHo models is a little higher, varying between 50\,cm and 155\,cm. The maximum amplitude of the prompt convection is higher in the rapidly rotating models. The frequency of the prompt convection peaks at $\sim 100$\,Hz for all models. 

The rapidly rotating models also show the typical dominant rising f/g-mode, and reach a higher frequency by the end of the simulation time than the non-rotating models. The lower mass models reach frequencies between $\sim 1500 -1600$\,Hz. The higher mass models reach frequencies as high at $\sim 1700$\,Hz for SFHx models, and $\sim 2000$\,Hz for SFHo models. 
However, for both EoS, the slope of the f/g-mode is different than what we see for slow and non-rotating models. The slope is very steep for the first $0.5\,\mathrm{s}$, and then becomes shallower for the remainder of the signal duration, but continues to rise in frequency. An exception is the s36.61 model, shown in Figure \ref{fig:rapid_gws}, whose high frequency mode becomes flat at around 800\,Hz before continuing to steeply rise before black hole formation. After the f/g-mode slope becomes more shallow, a secondary high frequency mode is visible in the models above $15\,M_{\odot}$. The models above $20\,M_{\odot}$ have f/g-mode amplitudes that become larger than those from prompt convection. The majority of the SFHx models have a spike in the time series at $\sim 2.1$\,s, which is smaller in amplitude than the prompt convection, but comparable in amplitude to the f/g-mode in the first 1\,s after bounce. The SFHo EoS models also have the small spike, but at a later time at 2.55\,s after core-bounce. 
In only the s29.59 model, the spike becomes huge, with an amplitude of 452\,cm for SFHx, and 161\,cm for SFHo. The s36.61 model also reaches amplitudes of 200\,cm for SFHx, and 344\,cm for SFHo shortly before black hole formation.  

\begin{figure}
\includegraphics[width=\columnwidth]{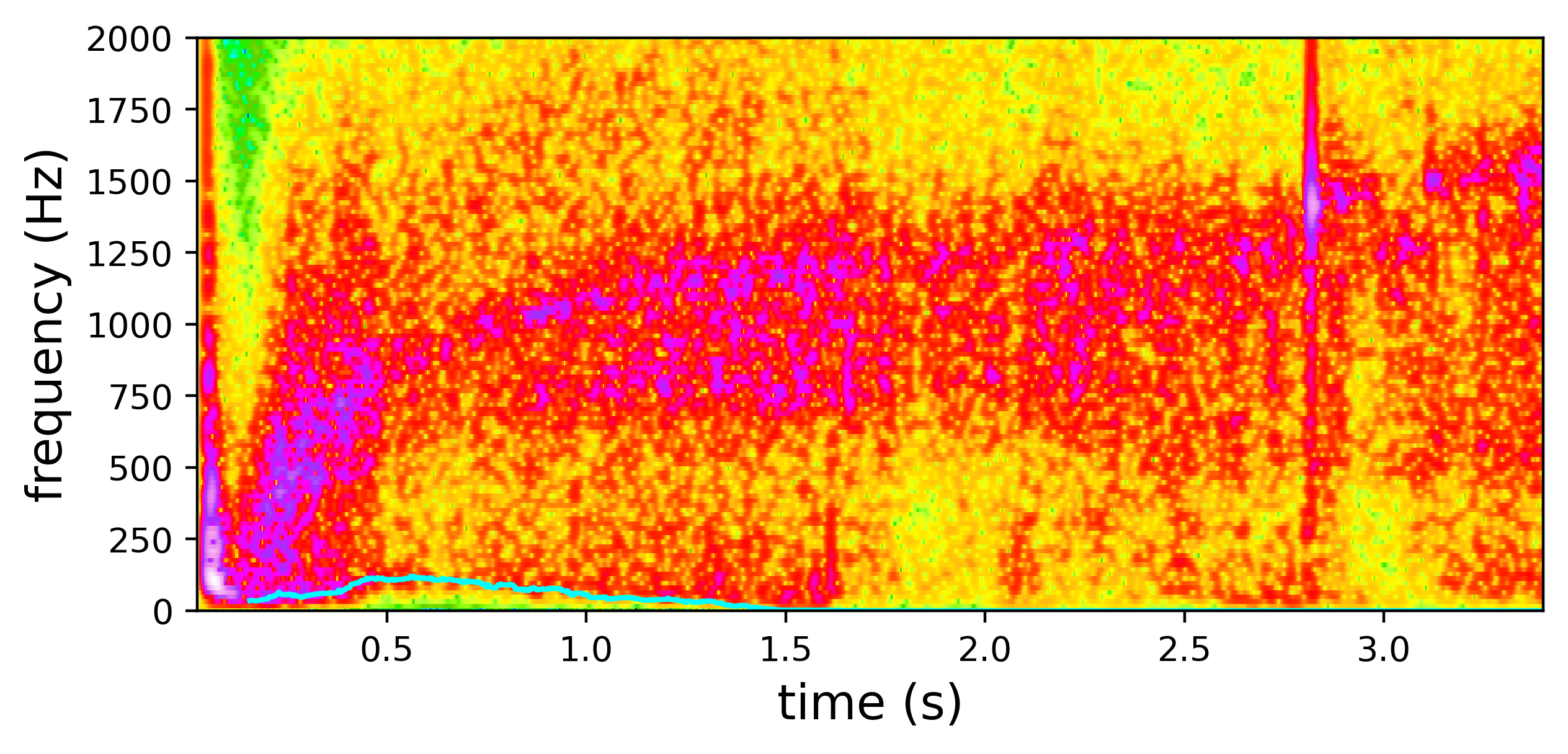}
\caption{A spectrogram of the rapidly rotating model s14 SFHo. The blue line is the SASI frequency predicted by Equation \eqref{eqn:sasi}. The SASI is much lower frequency in the rapidly rotating models, however rapid rotation has additional effects on the SASI in 3D, such as spiral SASI modes which would impact the results.  }
\label{fig:sasi_rapid}
\end{figure}

The rapidly rotating models also have gravitational-wave emission due to the SASI before the shock revival time. However, the SASI looks significantly different from the non-rotating and slowly rotating case. An example is shown in Figure \ref{fig:sasi_rapid}. The frequency of the SASI mode is much lower, only reaching a maximum $\sim70$ Hz at 0.5\,s post bounce. The SASI mode frequency also does not continue to rise in frequency with time, and instead starts to decrease in frequency after 0.5\,s post bounce. However, SASI is known to be impacted by effects that are only present in 3D, such as the spiral SASI \citep{foglizzo_26}. Therefore, we need simulations of gravitational waves in 3D with rapid rotation to fully understand the impact of the SASI on the gravitational-wave emission. Our previous 3D simulations with rapid rotation underwent shock revival too rapidly for the development of significant SASI activity \citep{powell_23, powell_24}.

%%%%%%%%%%%%%%%%%%%%%%%%%%%%%%%%%%%%%%%%%%%
%%%%%%%%%%%%%%%%%%%%%%%%%%%%%%%%%%%%%%%%%%%
\section{Universal Relations}
\label{sec:uni_relations}

\begin{figure}
\includegraphics[width=\columnwidth]{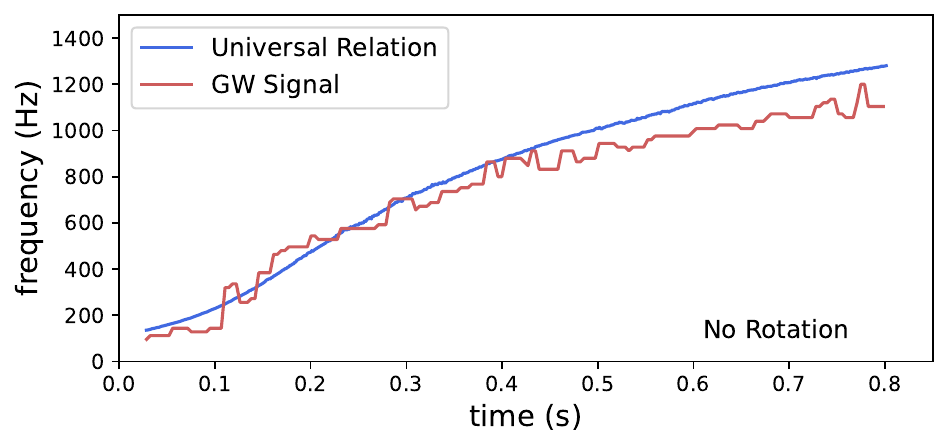}
\includegraphics[width=\columnwidth]{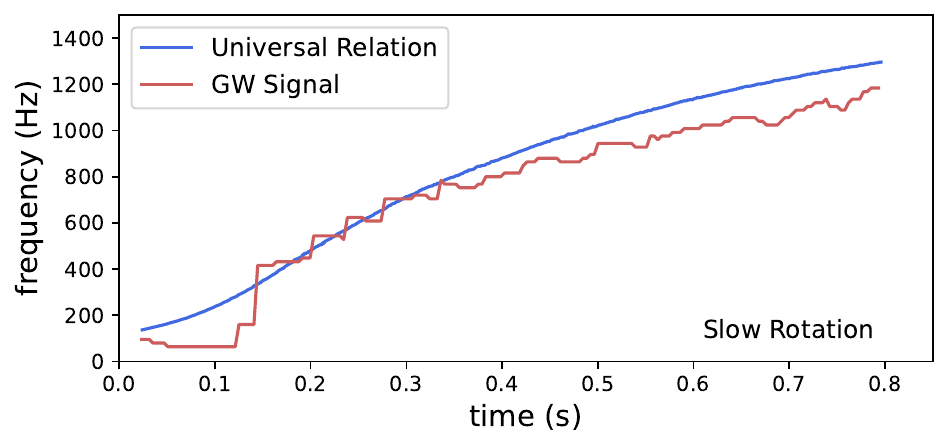}
\includegraphics[width=\columnwidth]{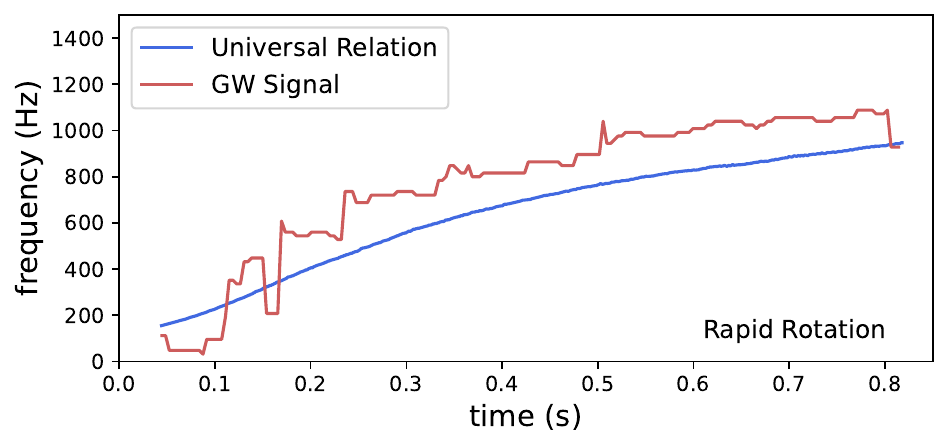}
\caption{The lines labelled "GW signal" show the time-frequency bins with the highest gravitational-wave energies. The "Universal Relation" line shows the time-frequency predicted by the $^2\mathrm{g}_2$ universal relation in \citet{torres_forne_19}. The top panel is the non-rotating s14 SFHo model. The middle is the slowly rotating s14 SFHo model. The bottom is the s12.5 SFHo rapidly rotating model. The non and slowly rotating models fit the universal relations well up until $\sim 0.4$\,s post bounce. Then the universal relation starts to under-predict the frequency up to a maximum of 200\,Hz at 0.8\,s. For the rapidly rotating models, the universal relations consistently over predict the gravitational-wave frequency. 
}
\label{fig:uni_relation}
\end{figure}

In recent years, universal relations that describe the relationship between the gravitational-wave frequency and the size of the PNS have been used to develop data analysis tools for the extraction of PNS properties from a real CCSN gravitational-wave detection \citep{bizouard_21, powell_22, bruel_23, powell_25b}. Therefore, it is important to test the accuracy of these relations in order to understand the errors on any astrophysical inference of PNS properties made from the next nearby CCSN. 
Thus, we investigate how well our gravitational-wave signals obey the universal relations given in \citet{torres_forne_19}. For the $^2\mathrm{g}_2$ mode, which best tracks the high-frequency signal, they found the approximate relation
\begin{equation}
    f_\mathrm{high}/\mathrm{Hz}=
    5.88\times 10^5 \times x -8.62\times 10^7\times x^2
+ 4.67\times 10^9 \times x^3
\end{equation}
for the mode frequency $f_\mathrm{high}$, where
$x=(M/\msun)/(R/\mathrm{km})^2$.
We plot the $^2\mathrm{g}_2$ mode, and compare it to our gravitational-wave signals. To this end, we identify the frequency bin of maximum power in the spectrograms for each time slice.
Some example results are shown in Figure \ref{fig:uni_relation}. For the non-rotating, and slowly rotating models, we find a good fit to the universal relations for the first 400\,ms where the gravitational-wave amplitude is maximal. After this time, the universal relation starts to under-predict the frequency, reaching an error of $\sim 200$\,Hz by 0.8\,s post-bounce. This error continues to grow several seconds post bounce. For the rapidly rotating models, the universal relations do not fit so well, even at early times when the emission is highest. They consistently over-predict the gravitational-wave frequency by $\sim 200$\,Hz for the full duration of the gravitational-wave emission. Poorer fits to universal relations in rapidly rotating models is consistent with what we observed in our prior 3D models \citep{powell_23, powell_24}. At later times, when the gravitational-wave amplitude is lower, the waveform reconstruction will also have large error bars on the gravitational-wave frequency, resulting in further difficulties in measuring the size of the PNS \citep{powell_22}.

\begin{figure*}
    \centering
    \includegraphics[width=0.48\linewidth]{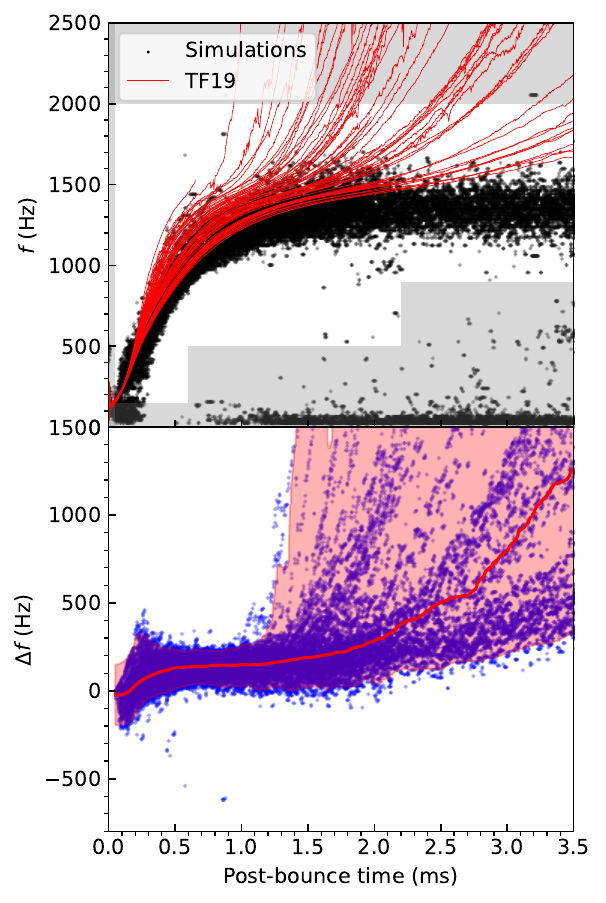}
    \hfill
    \includegraphics[width=0.48\linewidth]{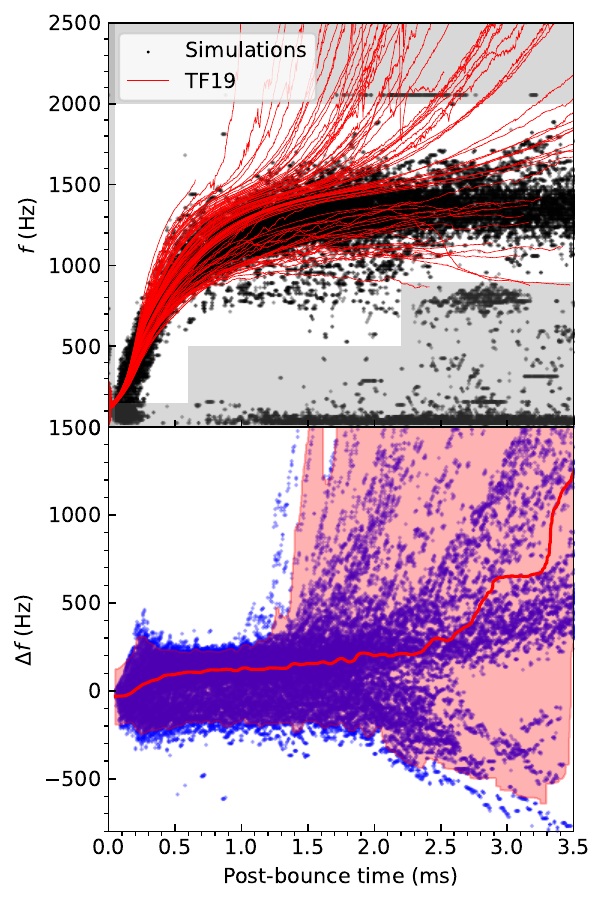}
    \caption{Top: Time-dependent frequency with maximum power (black dots) for the non-rotating and slowly rotating models (left) and for all models (right) compared to the universal relation from 
    \citet[][TF19, red curves]{torres_forne_19}. 
    Data points in the grey-shaded areas are deemed to not belong to the high-frequency signal, and are excluded from the calculation of errors and the new fit.
    Bottom: The corresponding fit errors (blue dots), the average error (red curve) and the 95\% coverage band (red shaded area) for the universal relation.}
    \label{fig:error_band}
\end{figure*}

To analyse the fit of the universal relation more quantitatively beyond these illustrative examples, we compute fit residuals for all models. The data underlying the calculation of fit results are shown
in Figure~\ref{fig:error_band}. The top panels
show data points for the time-dependent maxima of the spectra and the universal relations for two subsets of simulations: one containing non-rotating and slowly rotating models, and another containing the entire set of simulations (i.e., including rapidly rotating models). We exclude data points at low and very high frequencies that are due to other features in the spectrograms and are not associated with the high-frequency band. 
The fit residuals $\Delta f$ for all data points are depicted in the bottom panels.
We also calculate $95\%$ coverage regions using the  rolling $2.5\%$ and $97.5\%$ quantiles of the time-ordered residuals. For more smoothly varying rolling quantiles, the data points are weighted with a shifting 
triangular weight window over 1000 data points.

The error bands reveal a broader asymmetric error range when including the rapidly rotating models, which is partly due to the appearance of additional modes in the frequency range of interest. Because of the additional modes, the search for the bin with maximum power sometimes does not select the main mode.
The figure also confirms the deterioration in accuracy at late times. Furthermore, they indicate a tendency for the universal relation of \citet{torres_forne_19} to slightly overpredict the mode frequencies in our models.

We therefore propose a recalibration of the universal relation based on our current 2D data. In keeping with
previous work, we consider only a dependence on the PNS mass and radius, $M$ and $R$, but slightly generalise the form of the universal relation to
\begin{align}
    f_\mathrm{high}&=
    A\times 
    \left(\frac{M}{\msun}\right)^{\alpha}
    \left(\frac{R}{10\,\mathrm{km}}\right)^{-\beta}
    \left(1-\frac{2 G M}{R c^2}\right)^{\gamma}
    +\epsilon,\\
    \epsilon& \sim\mathcal{N}\left(0,\sqrt{\sigma^2+f_\mathrm{r} \sigma_\mathrm{r}^2}\right),
\end{align}
where the normally-distributed error have a variance  $\sigma$ for slow or no rotation ($f_\mathrm{r}$=0), and an additional variance $\sigma_\mathrm{r}^2$ is added for rapid rotation ($f_\mathrm{r}$=1). Different from \citet{torres_forne_19}, we include a relativistic factor, $1-2GM/Rc^2$, which casts the relation into a form more similar to that of \citet{mueller_13}. Deviations from the scaling with $M/R^2$ are accounted for by allowing general power-law exponents $\alpha$, $\beta$ and $\gamma$, which reflect change in the PNS surface temperature and structure that correlate with mass and radius. This approximately accounts for the mean electron antineutrino energy term in \citet{mueller_13}. The model is fitted to the data for the first $1.5\,\mathrm{s}$ with \textsc{Numpyro} \citep{phan_19} using uniform priors. 

\begin{figure}
    \centering
    \includegraphics[width=\linewidth]{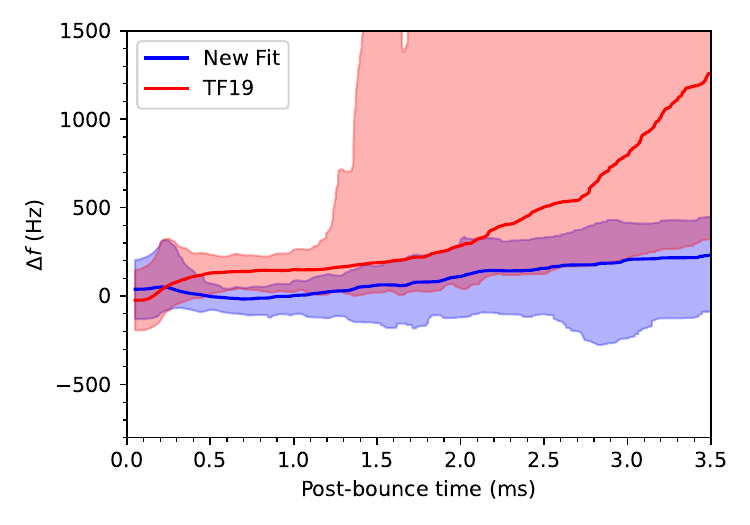}\\
    \includegraphics[width=\linewidth]{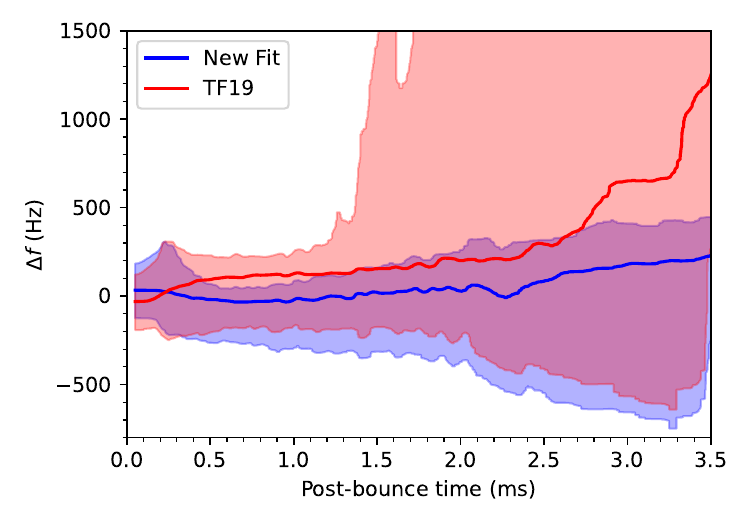}
    \caption{Comparison of the average error (solid curves) and 95\% coverage band (shaded areas) for the new fit from Equation~(\ref{eq:new_fit}) (blue) and the universal relation from TF19. The top panel only includes non-rotating and slowly rotating models, whereas the bottom panel includes all models.}
    \label{fig:new_fit}
\end{figure}

We find that the data are best described by the fit
\begin{equation}
\label{eq:new_fit}
    f_\mathrm{high}=
    3134.64 \,\mathrm{Hz}
    \left(\frac{M}{\msun}\right)^{0.81}
    \left(\frac{R}{10\,\mathrm{km}}\right)^{-1.43}
    \left(1-\frac{2 G M}{R c^2}\right)^{1.54}.
\end{equation}
Error bands for the new fit are shown and compared to the universal relation of \citet{torres_forne_19} in Figure~\ref{fig:new_fit}. Even though the new fit is only based on the first $1.5\,\mathrm{s}$ of data, the fit residuals remain significantly smaller than before even at late times. In the region of greatest accuracy, the new fit reduces the error to about $100\,\mathrm{Hz}$ at the $95\%$ confidence level. For the set including rapid rotation, the fit remains significantly worse than for the set of slow or non-rotating models. In the future, aggregating data from more simulations and controlling for other systematic errors (e.g., due to monopole gravity) may further improve the fit  or at least provide reliable information on uncertainties in the estimation of PNS parameters from gravitational-wave signals.

%%%%%%%%%%%%%%%%%%%%%%%%%%%%%%%%%%%%%%%%%%%%%%%%%%%%%%%%%%
%%%%%%%%%%%%%%%%%%%%%%%%%%%%%%%%%%%%%%%%%%%%%%%%%%%%%%%%%%
\section{Gravitational Wave Memory}
\label{sec:low_freq}

In this section, we discuss the properties of the gravitational-wave memory from both asymmetric ejection of matter and the asymmetric emission of neutrinos. The long duration of our simulations allows us to investigate the low frequency gravitational-wave emission in more detail than in our previous shorter duration 3D simulations.  

%%%%%%%%%%%%%%%%%%%%%%%%%%%%%%%%%%%%
\subsection{Matter Memory }

% No rotation
Almost all of the non-rotating models show at least some low frequency emission. Some examples are shown in Figure \ref{fig:norot_timeseries}. In the lowest mass models, the matter memory amplitudes reach a maximum of 20\,cm. The matter memory starts to become much larger in models above $12\,M_{\odot}$. For the non-rotating models with the SFHx EoS, the largest amplitudes are found in model s36.61 at 392\,cm and model s18 at 314\,cm. The non-rotating SFHo models reach even higher amplitudes with 595\,cm for model s18, 680\,cm for model s21.91, and 493\,cm for model s36.61. For the CMF models, the majority do not have any matter memory due to the lack of shock revival. The one exploding CMF model, s29.59, does have matter memory, which reaches 12\,cm by the end of the simulation time. The matter memory for this model does not have much time to grow, as there was only 0.79\,s between the shock revival time and the time of black hole formation. 

% Slow rotation
For the slowly rotating models, there are significantly fewer low frequency gravitational-wave signals from matter memory due to the prevalence of failed explosions. The small number of models that do show some low frequency emission vary in amplitude between 10\,cm and 30\,cm. The exceptions are the SFHx s29.59 model, which reaches amplitudes of 80\,cm, and the SFHo s36.61 model that reaches 70\,cm. The later shock revival, and the much shorter duration of the slowly rotating models, does not give as much time for the low frequency gravitational-wave amplitude to grow.  

% Rapid rotation
For the rapidly rotating SFHx models, we again only see the matter memory in the highest mass models. The largest amplitude is from the s29.91 model at 60\,cm. For the rapidly rotating SFHo models, we see matter memory in most of the models, however the lower mass models only reach amplitudes of 10-20\,cm. The largest is the s29.91 model, which reaches amplitudes of 50\,cm by the end of the simulation. The highest mass, rapidly rotating models, have the latest shock revival time and the models are also several seconds shorter duration than the non-rotating models, which gives less time for the low frequency emission to grow to large amplitudes. 

Although we clearly see some differences in matter memory amplitude between different progenitors, EoS, and rotation rates, determining these parameters from a single CCSN detection would likely be extremely difficult, as there is a lot of overlap in ranges of amplitudes for different source parameters. Models in 3D have also shown that the amplitude of the memory signal also varies with the orientation of the source \citep{choi_24, powell_24}. Furthermore, this aspect of the signal is at frequencies too low for current gravitational-wave detectors, and will require next generation observatories \citep{et_paper, berti_26}.

%%%%%%%%%%%%%%%%%%%%%%%%%%%%%%%%%%%%%%%%%%%%%%
\subsection{Neutrino Memory}
\label{subsec:neu_memory}

\begin{figure}
\includegraphics[width=\columnwidth]{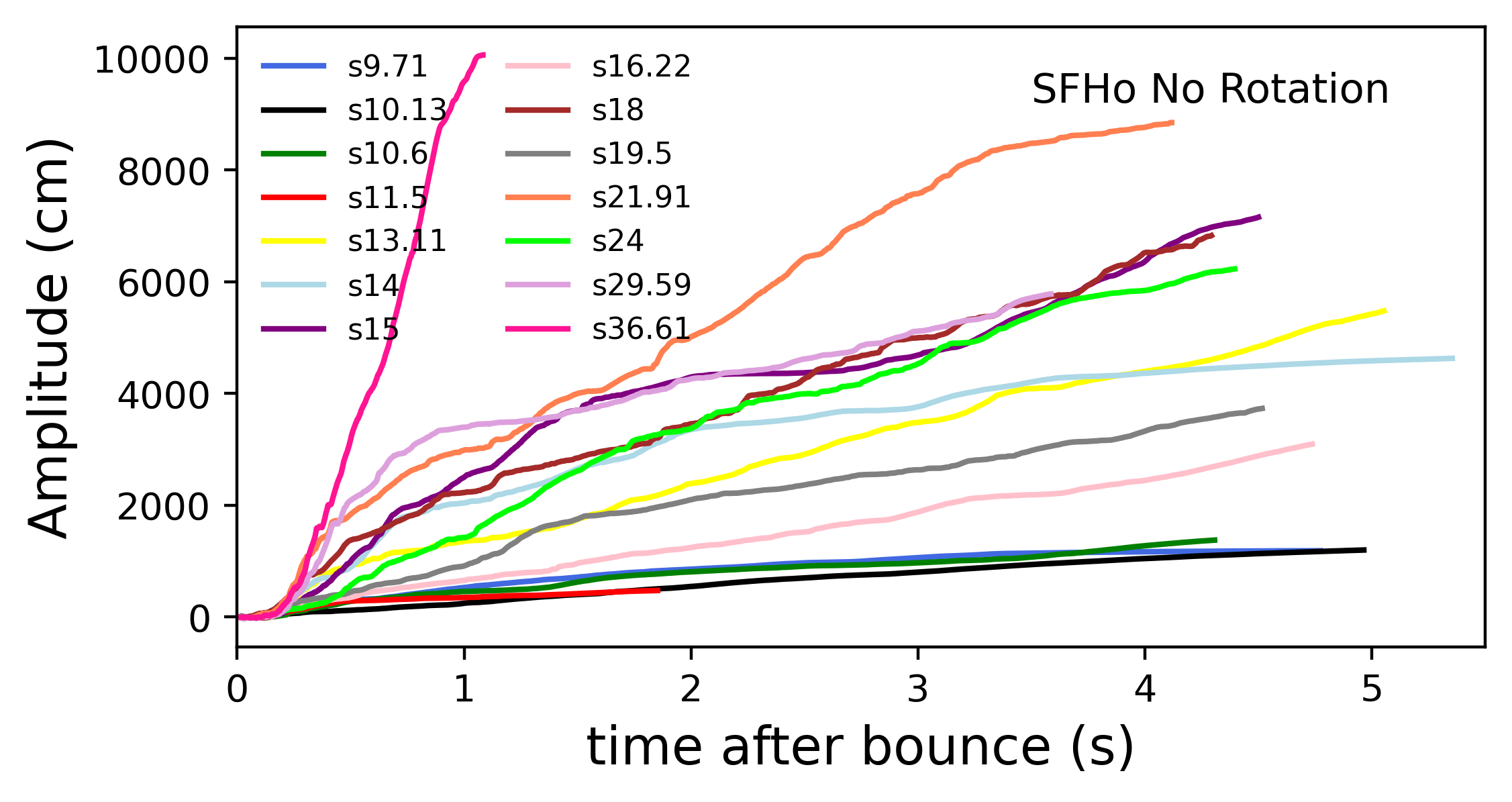}
\includegraphics[width=\columnwidth]{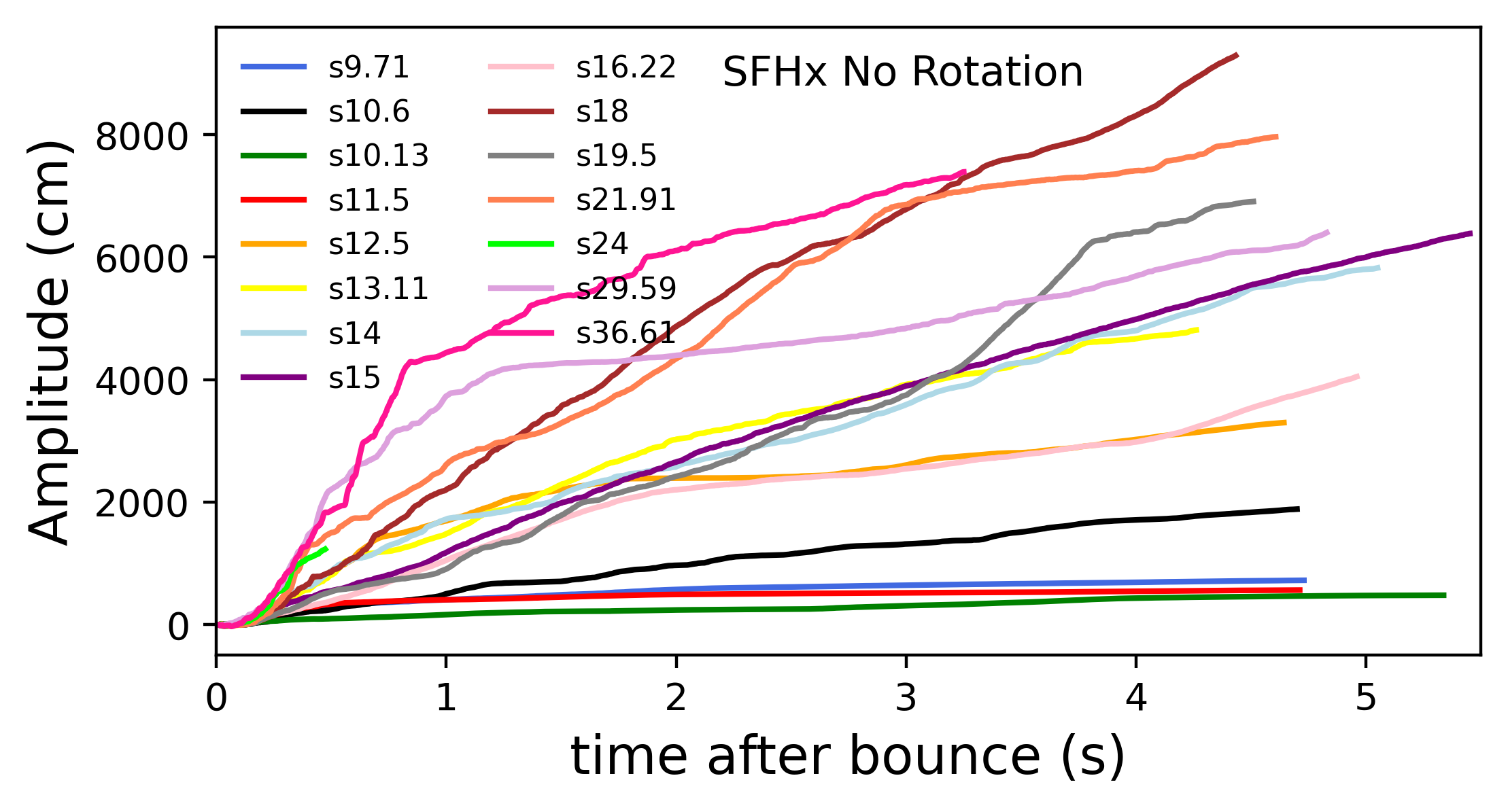}
\includegraphics[width=\columnwidth]{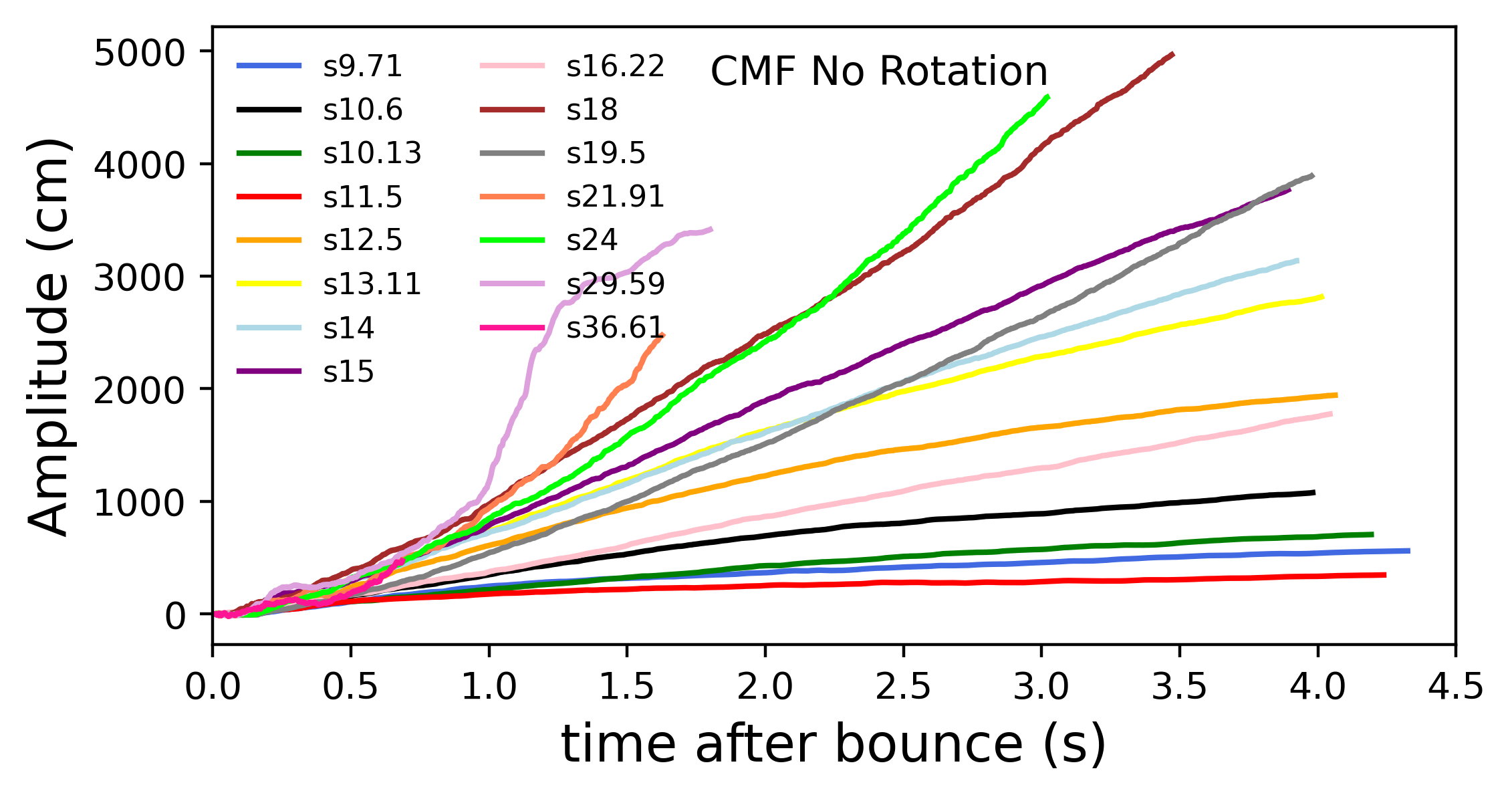}
\caption{ The gravitational-wave signals from the asymmetric emission of neutrinos for all non-rotating models. The top panel is the SFHo models, the middle panel is the SFHx models, and the bottom panel is the CMF models. The low frequency gravitational-wave emission due to the neutrinos is significantly higher amplitude than for matter, improving detection prospects. The amplitude is lower for the CMF models due to the lack of shock revival. }
\label{fig:neu_memory}
\end{figure}

\begin{figure*}
\includegraphics[width=\columnwidth]{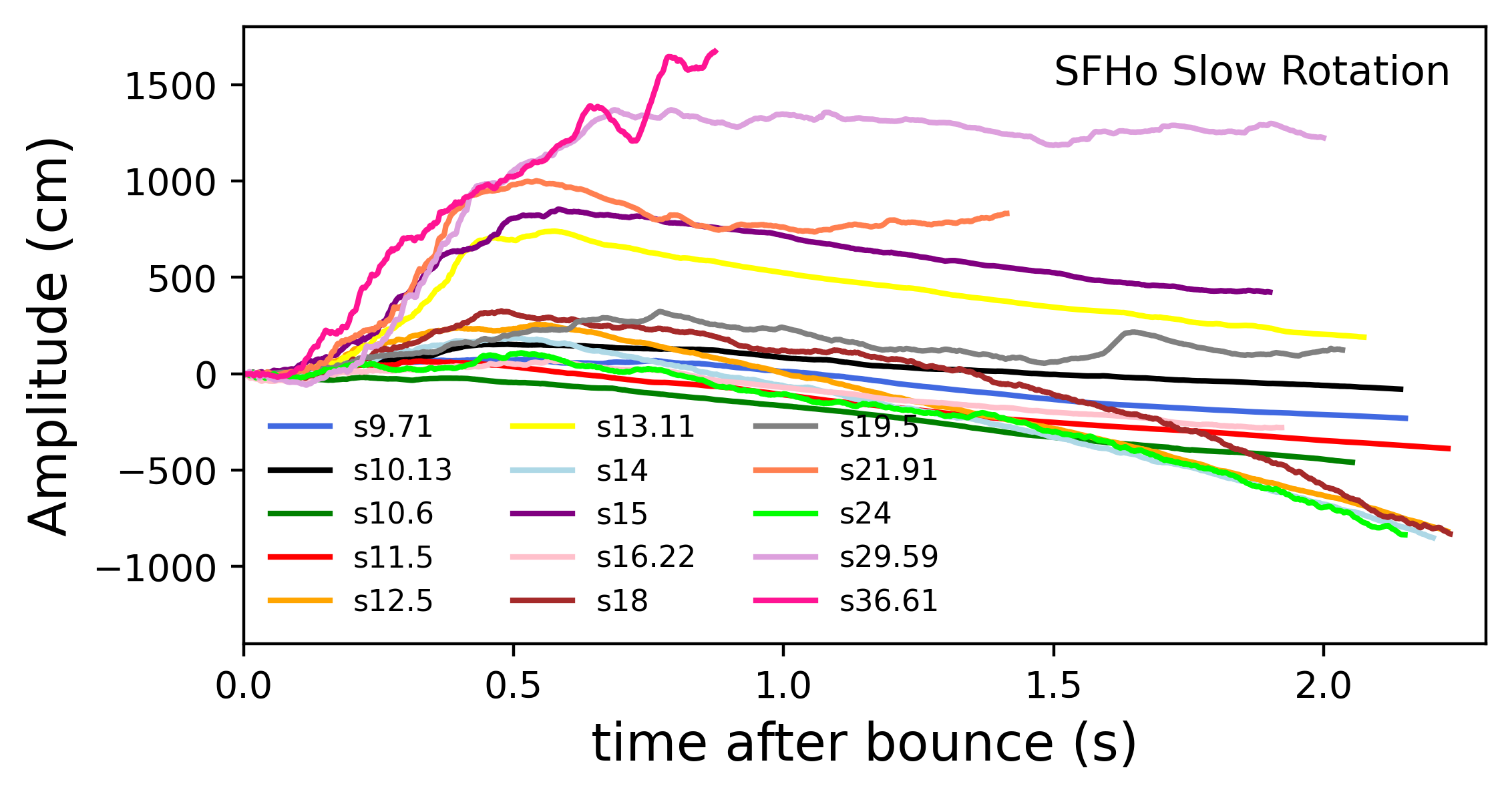}
\includegraphics[width=\columnwidth]{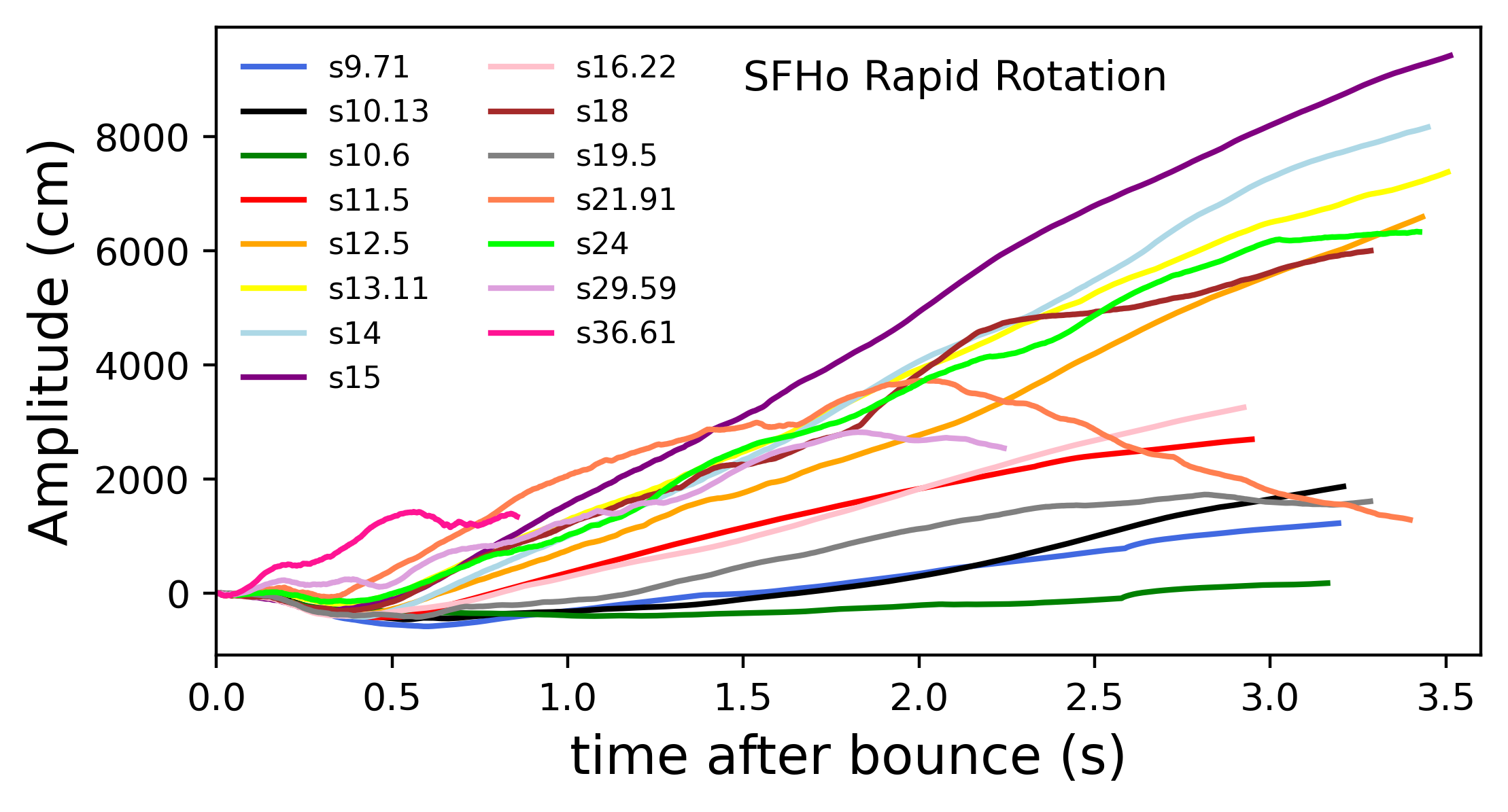}
\includegraphics[width=\columnwidth]{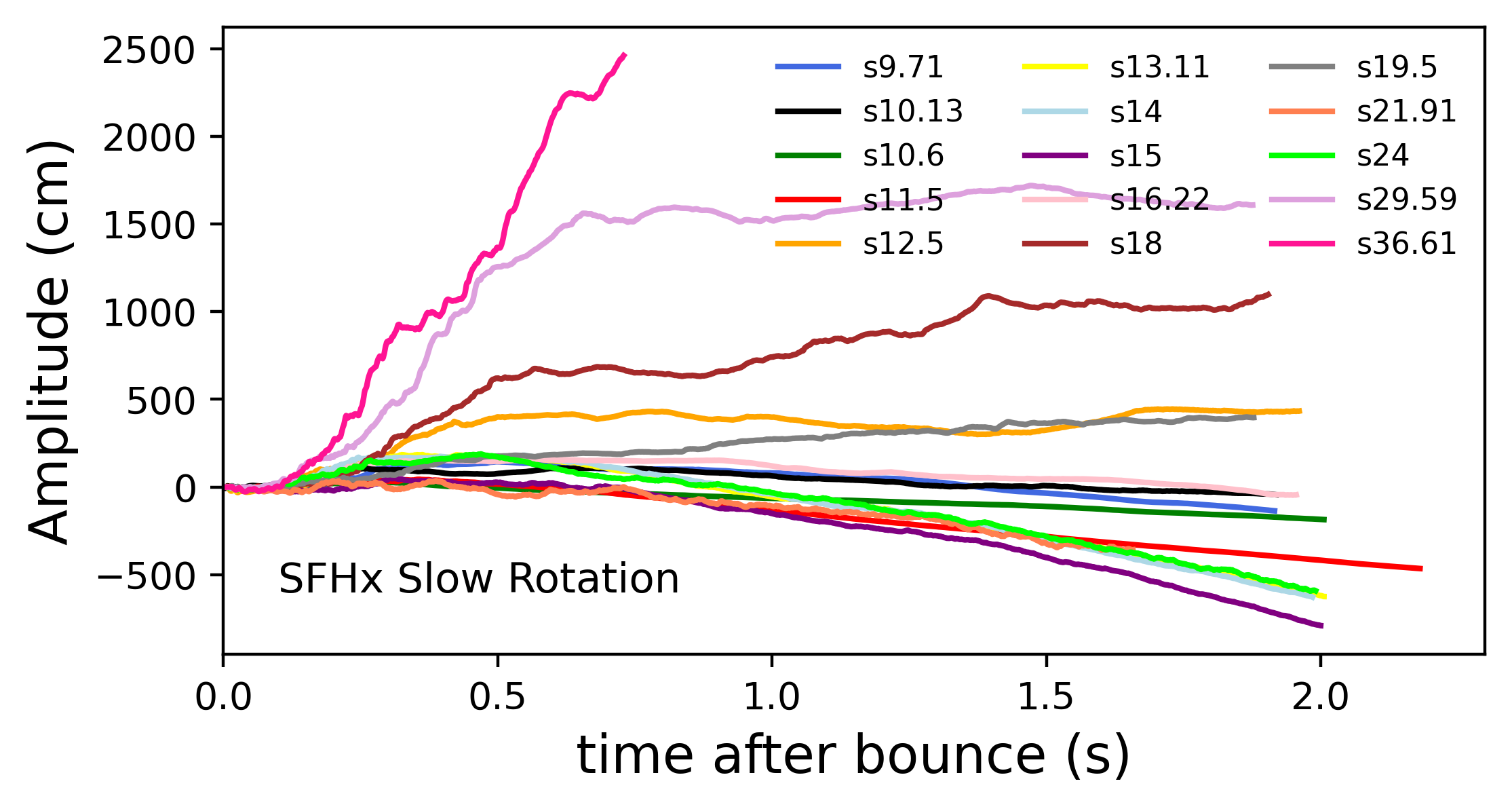}
\includegraphics[width=\columnwidth]{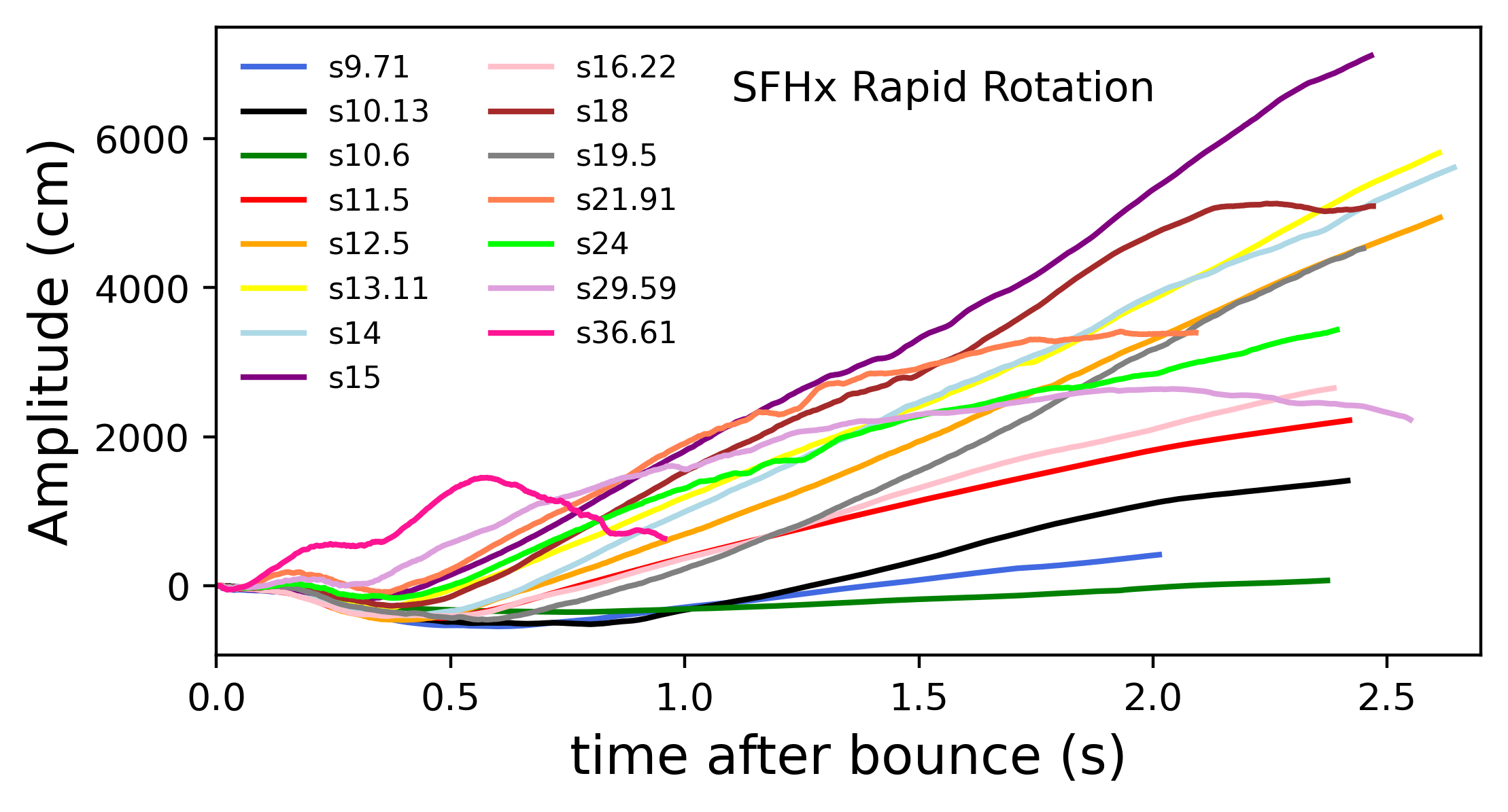}
\caption{ The gravitational-wave signals from the asymmetric emission of neutrinos for the slowly rotating models (left), and the rapidly rotating models (right). The top row is for models with the SFHo EoS, and the bottom row is models with the SFHx EoS. The slowly rotating models have less amplitude at lower frequencies than the rapid or non-rotating models. }
\label{fig:neu_memory_rot}
\end{figure*}

\begin{figure*}
\includegraphics[width=\textwidth]{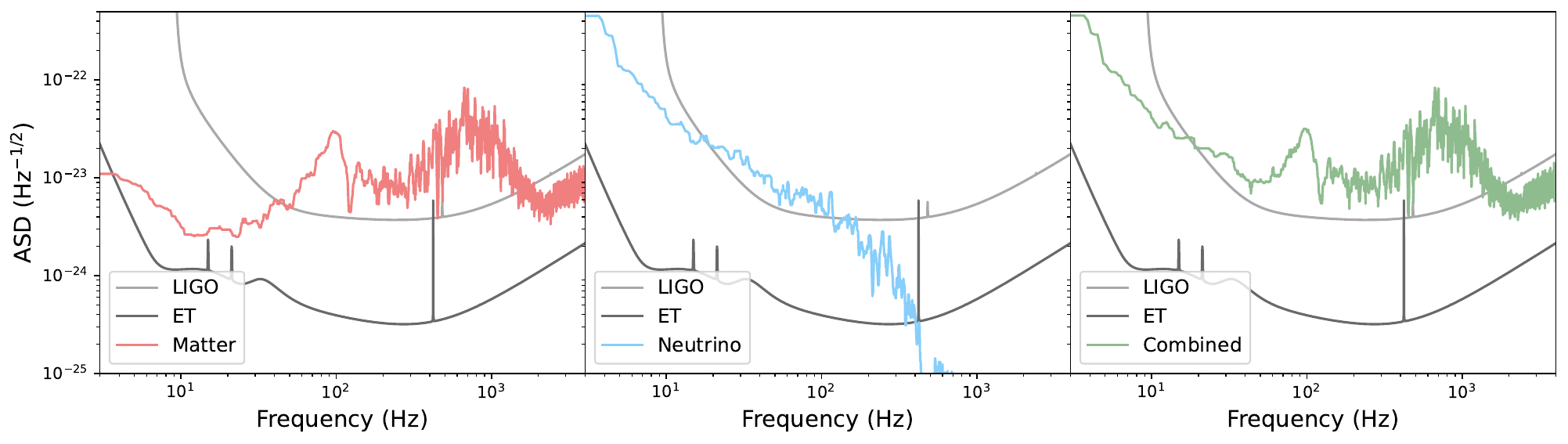}
\caption{The amplitude spectral density (ASD) of a representative gravitational-wave signal, the non-rotating s21.91 SFHx model, at 100\,kpc, and the noise curves of LIGO and Einstein Telescope (ET). The matter only curve has a peak at $\sim 100$\,Hz due to prompt-convection and another at $\sim 800$\,Hz due to the high frequency modes. The gravitational waves due to neutrino memory significantly improve detectability at frequencies that are too low for LIGO but may be achievable for ET.  
 }
\label{fig:neu_asd}
\end{figure*}

Gravitational waves due to anisotropic neutrino emission from all species are calculated using the Epstein formula \citep{epstein_78, kotake_07, mueller_14}, which in 2D gives the gravitational wave strain as,
\begin{equation}
h_{\nu} = \frac{2G}{c^4R} \int^t_0 L_{\nu}(t') \alpha_{\nu} (t') dt'
\end{equation}
where $L_{\nu}$ is the total angle-integrated neutrino energy flux, and $\alpha_{\nu}$ is the anisotropy parameter,
\begin{equation}
\alpha_{\nu} = \frac{1}{L_{\nu}} \int \pi \sin \theta (2|\cos\theta|-1) \frac{dL_{\nu}}{d\Omega} d \Omega. 
\end{equation}
The gravitational-wave memory due to the neutrinos is shown in Figure \ref{fig:neu_memory} for the non-rotating models, and Figure \ref{fig:neu_memory_rot} for the slow and rapidly rotating models. Gravitational-wave amplitudes for the neutrino memory are extreme, reaching over 10,000\,cm for the non-rotating s36.61 SFHo model. The amplitude is most likely an overestimate due to the simulations being 2D, however amplitudes beyond 1000\,cm have been observed in 3D models \citep{choi_24}. We see some correlation between mass and gravitational-wave amplitude, with the larger masses progenitors producing significantly stronger neutrino memory effects. The lowest amplitudes are generally seen in the s10.6 and s11.5 models while the largest are found in s36.61 and s18. An example of the amplitude spectral density of the memory is shown in Figure \ref{fig:neu_asd}. The neutrino memory will significantly increase detection prospects for next generation observatories with increased lower frequency sensitivity. 

For the non-rotating models, there are some differences in the neutrino memory between the different EoS. The CMF models have lower amplitude neutrino memory, mainly due to the lower neutrino luminosities and the lack of shock revival. The slowly rotating models reach smaller final amplitudes, but this is in part due to the shorter duration. The slowly rotating SFHx models show a wider range in amplitudes, from 44\,cm to 2459\,cm, in comparison to the slowly rotating SFHo models, which range from 80\,cm to 1670\,cm. The rapidly rotating models have a maximum amplitude almost as large as the non-rotating models, ranging from 175\,cm to 9412\,cm for SFHo and 71\,cm to 5809\,cm for SFHx. Even beyond 5\,s duration, the neutrino memory is still growing in amplitude by the end of the simulation time. 

%%%%%%%%%%%%%%%%%%%%%%%%%%%%%%%%%%%%%%%%%%%%%%%%%%%%%%%%%%
%%%%%%%%%%%%%%%%%%%%%%%%%%%%%%%%%%%%%%%%%%%%%%%%%%%%%%%%%%
\section{Detection and Parameter Estimation }
\label{sec:detection}

\begin{figure*}
\includegraphics[width=\columnwidth]{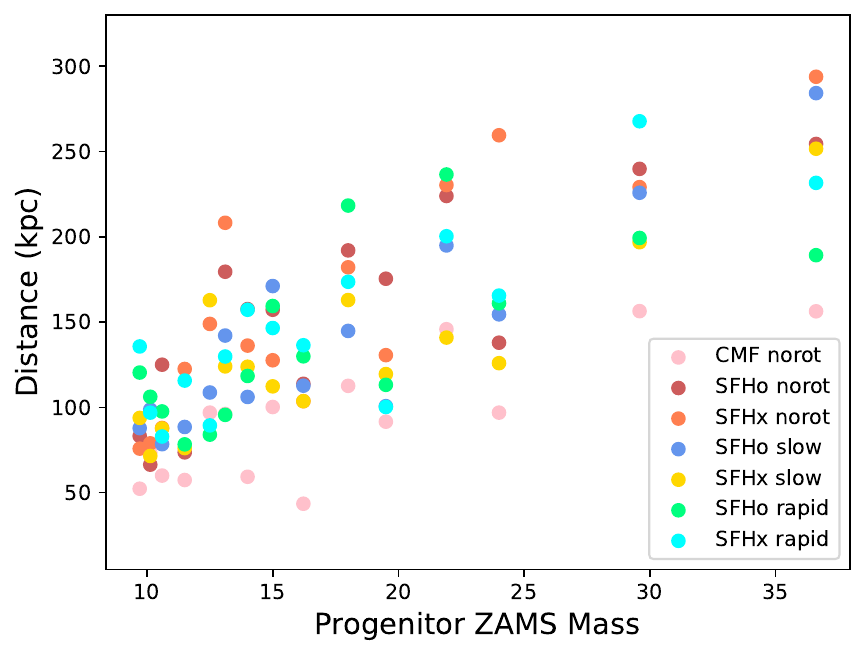}
\includegraphics[width=\columnwidth]{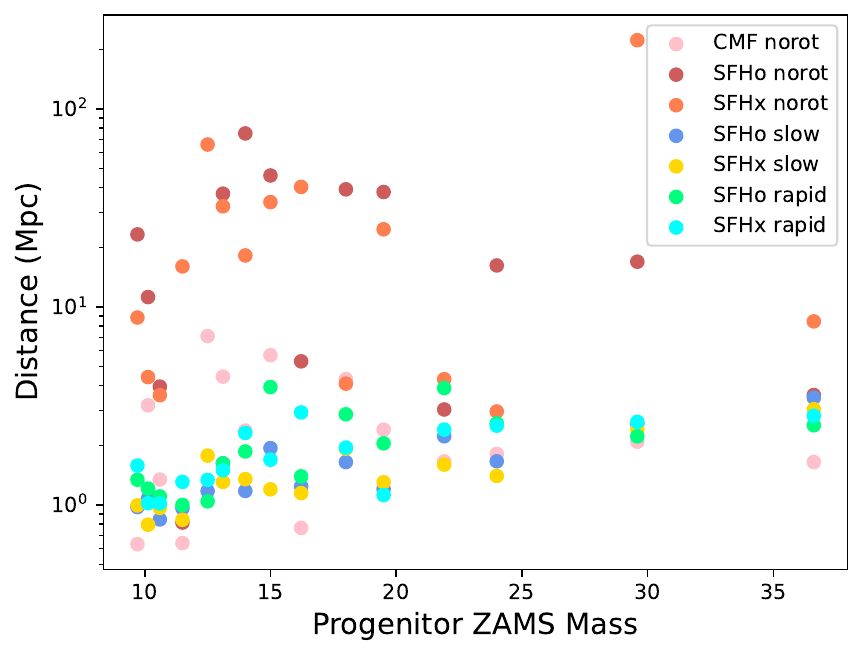}
\caption{The maximum detectable distance, which we define as signal-to-noise ratio 8, for all models in a single LIGO detector (right), and a single ET detector (left). There is a general trend of larger distances for larger progenitor stars, however there is still a lot of variety due to the EoS and rotation rate. In ET, the distances are significantly improved by the improved lower frequency sensitivity. 
 }
\label{fig:distance}
\end{figure*}

In this section, we discuss the impacts of the progenitor, rotation rate, and EoS on the prospects for detection. First, we calculate the maximum detectable distance, which we define as the distance required for a signal-to-noise ratio (SNR) of 8, at an optimal sky position \citep{powell_21}. The maximum distances will be overestimates, due to the 2D nature of our simulations, however comparisons between models with different parameters are still fruitful. 

First, we calculate the maximum detection distances for a single LIGO detector, using a low frequency cut off of 10\,Hz, and the noise curve from \citet{ligo_noise_curve}. The results are shown in Figure \ref{fig:distance}. Detection distances range from around $\sim50$\,kpc to almost 300\,kpc. However, gravitational-wave emission from 2D simulations is expected to be about a factor of several larger than in 3D \citep{andresen_17} and thus these models likely correspond to a rough LIGO maximum detection distance of $\sim5-30$\,kpc, depending on model parameters. This could be further improved if a detection is made by a network of gravitational-wave observatories. 

The results show that the various CCSN parameters have a significant impact on the detectability. 
There is a general trend of larger detection distances for higher mass progenitor stars, which is expected due to the more energetic explosions in higher mass models. 
The lowest detection distances are observed in the CMF models, due to the lack of shock revival. The non-exploding models may have some additional gravitational-wave amplitude in 3D, due to spiral SASI effects that are not present in 2D. Some of the highest distances are observed in the SFHx non-rotating models. 
The rotation does not appear to have a significant impact on the detection distances. This is partially due to shorter simulation times in comparison with the non-rotating models. However, their gravitational-wave energy is likely to change with the incorporation of 3D effects. Our previous rapidly rotating 3D models have shown early powerful explosions, that lead to significantly higher amplitude gravitational waves \citep{powell_23, powell_24}. 

In the same Figure, we also show the detection distances for the Einstein Telescope (ET), calculated with an improved lower frequency cut-off of 1\,Hz, and the noise curve from \citep{et_paper}.  
The maximum detection distances are significantly larger for ET, with the most extreme model reaching 222\,Mpc. 
For the slow and rapidly rotating models, this is mainly due to the increased sensitivity of ET at frequencies similar to the LIGO band. However, for the majority of the non-rotating models, a significant fraction of the increased detectability  is due to the better sensitivity at low frequencies for ET. The amplitude of the neutrino and matter memory below 10\,Hz is so large that it drives huge increases in detectability. This shows the importance of gravitational waves from asymmetric neutrino emission being included in detection studies for next generation observatories, as this aspect of the gravitational-wave emission has usually been neglected in detectability studies \citep{giudice_26}. The gravitational-wave memory for CCSNe is also a promising source for space based deci-Hz observatories \citep{jani_20, berti_26}.  

\begin{figure}
\includegraphics[width=\columnwidth]{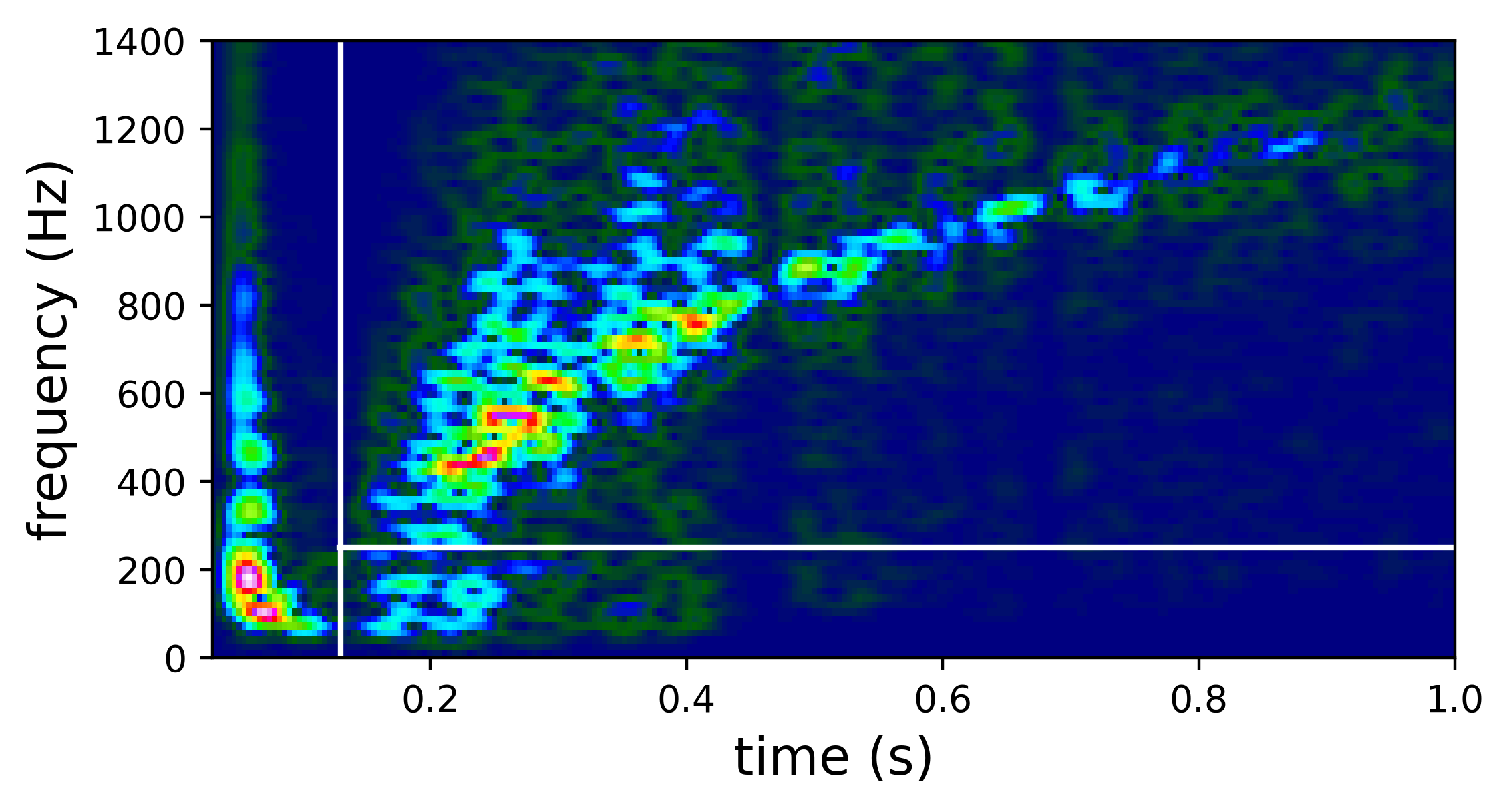}
\includegraphics[width=\columnwidth]{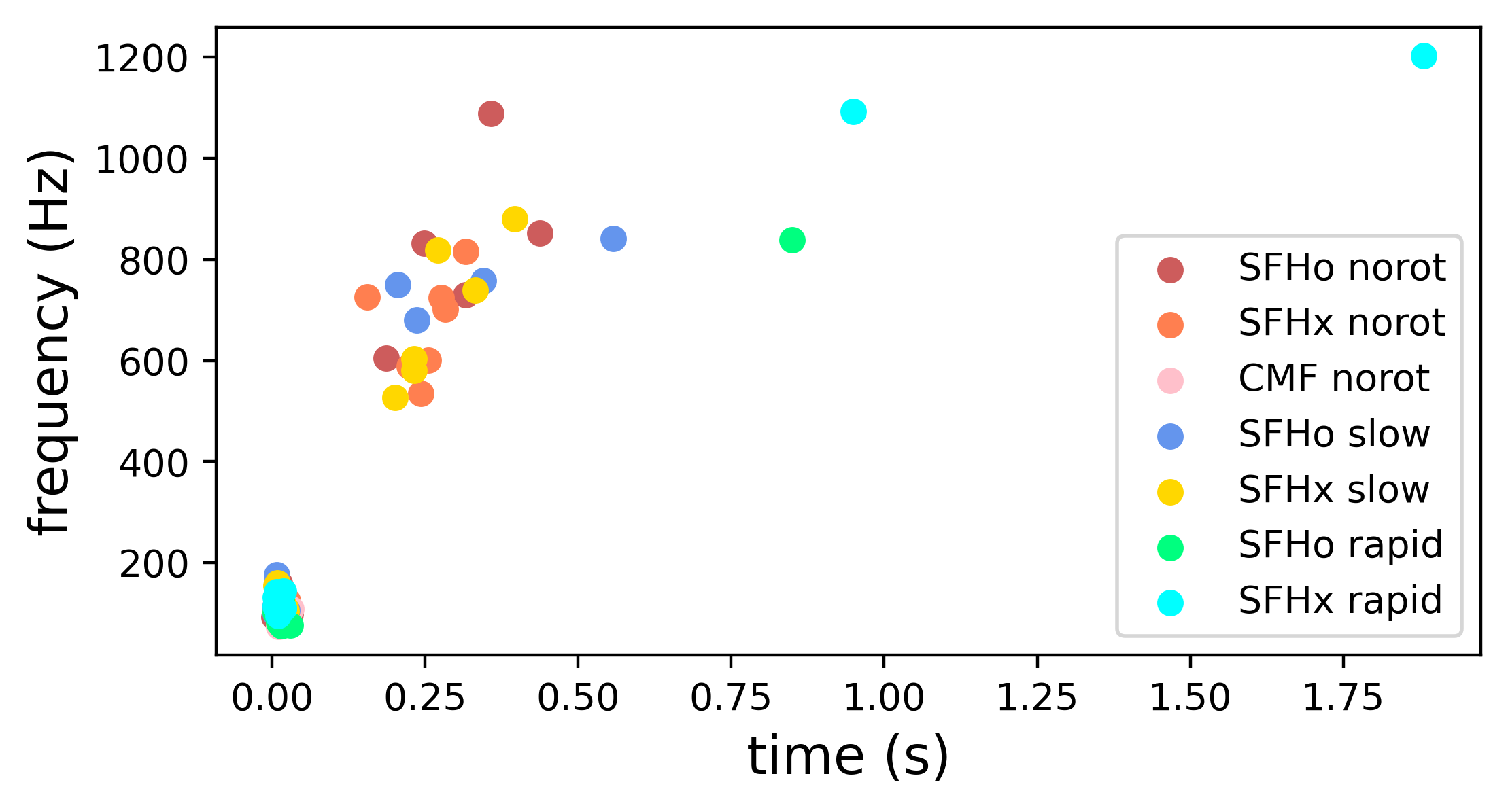}
\caption{The top panel is a spectrogram of the slowly rotating s14 SFHo model. The lines show how we divide the signal up to calculate the SNR from the different signal components. We use the first 0.1\,s to calculate the SNR from prompt convection. After 0.1\,s, we use a 250\,Hz cut to determine the SNR from the SASI and from the higher frequency modes. The bottom panel shows the time-frequency bins where the energy is highest for all models. A large number of the models are dominated by prompt convection at lower frequencies, and the others have higher amplitudes in the high frequency mode.}
\label{fig:freq_peaks}
\end{figure}

\begin{figure}
\includegraphics[width=\columnwidth]{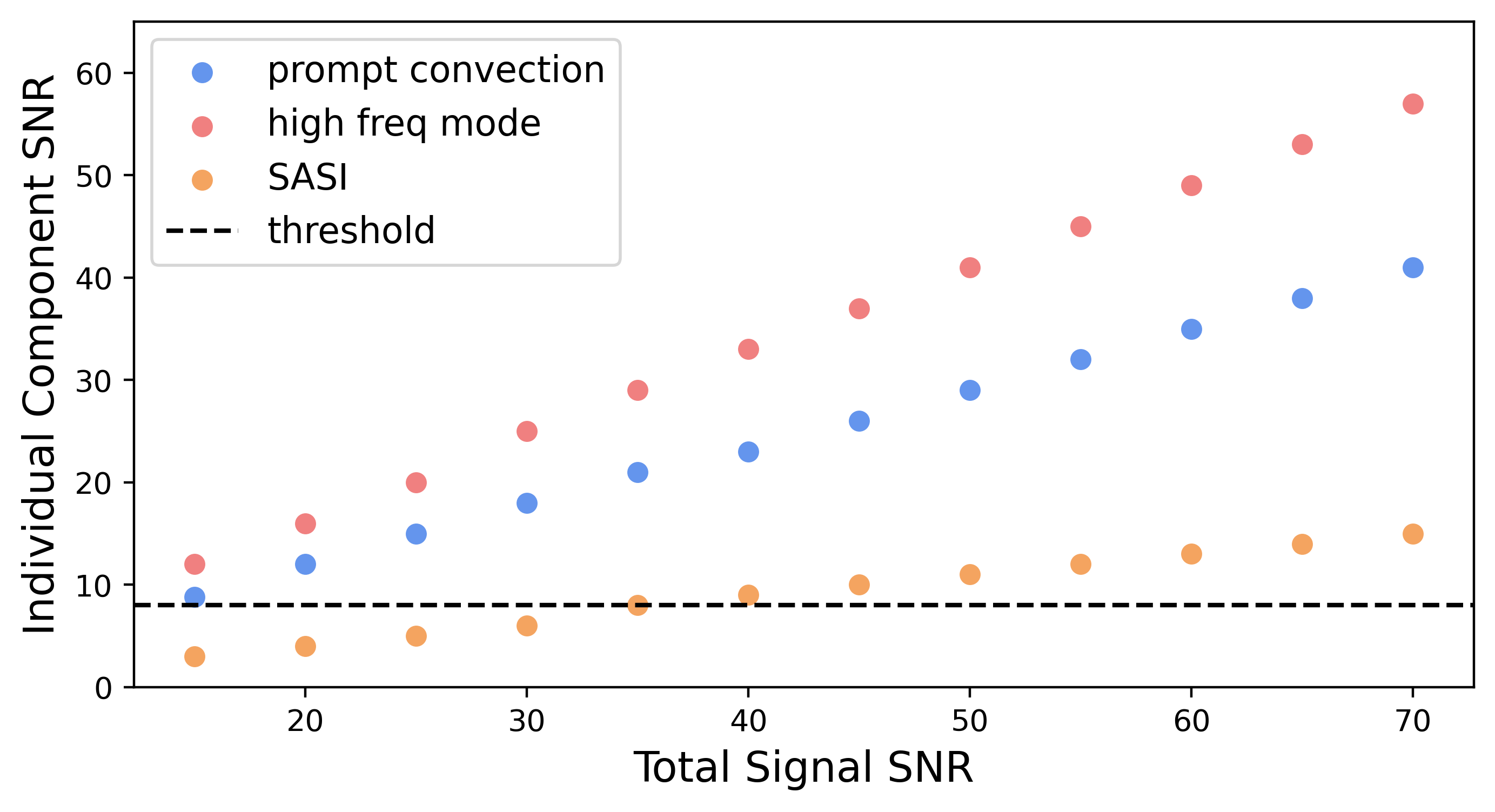}
\includegraphics[width=\columnwidth]{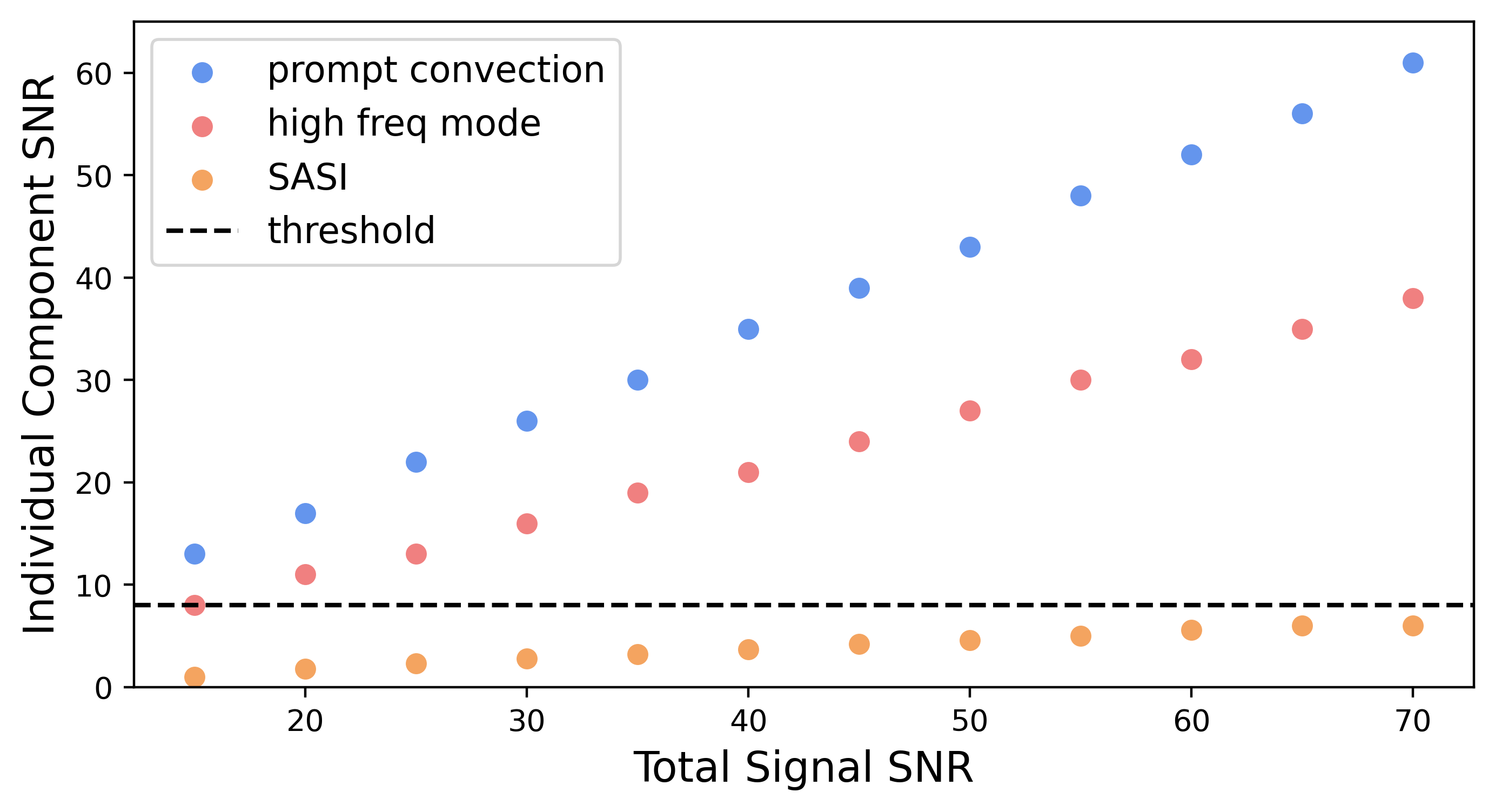}
\caption{ The top panel shows the SNR for different signal components for model CMF s15, and the bottom panel shows the same for the non-rotating model SFHo s10.6. The threshold line shows SNR 8, which we are considering as the minimum needed for detection. In the non-exploding CMF model, the SASI becomes detectable above SNR 35. For the s10.6 model, which rapidly explodes, there is no detectable SASI even at the highest SNR values. }
\label{fig:snr_regions}
\end{figure} 

After a detection is made, the next goal will be to infer the properties of the gravitational-wave source. To determine the prospects for this, we break the signal down into different components to determine which aspects of the signal would be visible to gravitational-wave observatories. In Figure \ref{fig:freq_peaks}, we show how we define the different components of the gravitational-wave signal. We consider the first 0.1\,s to be the emission due to prompt convection. For the rest of the signal, we split it by a frequency of 250\,Hz. We consider below 250\,Hz to be the gravitational-wave emission due to the SASI, and above 250\,Hz to be the gravitational-wave emission from the PNS high frequency modes. In reality, the SASI emission can extend beyond frequencies of 250\,Hz \citep[][see also Figures \ref{fig:sasi_norot} and \ref{fig:sasi_srot}]{powell_21}, however, most of the gravitational-wave energy in our models here comes from the first 1\,s, and the emission due to SASI does not go above 250\,Hz on that timescale in these 2D models. 

\begin{figure}
\includegraphics[width=\columnwidth]{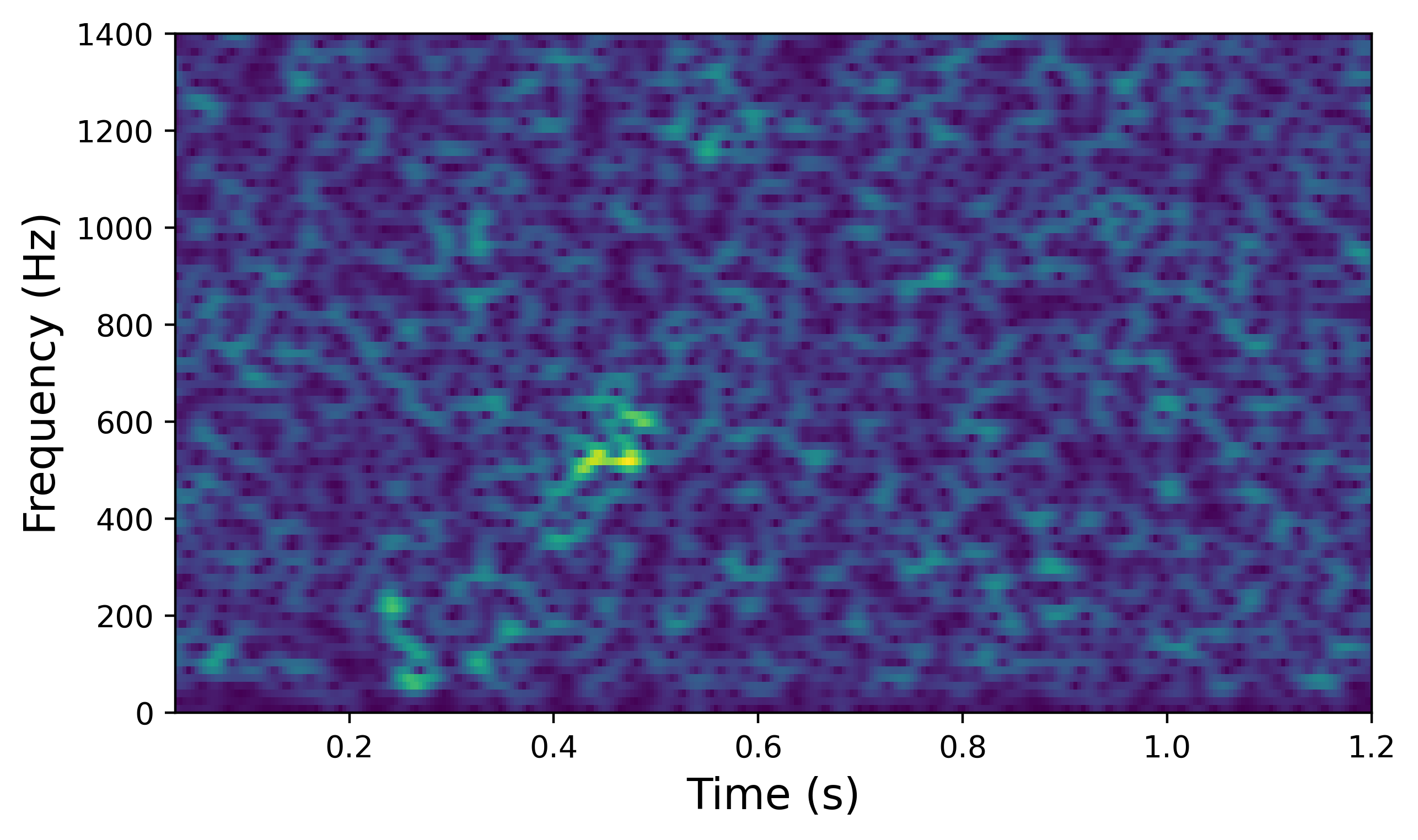}
\includegraphics[width=\columnwidth]{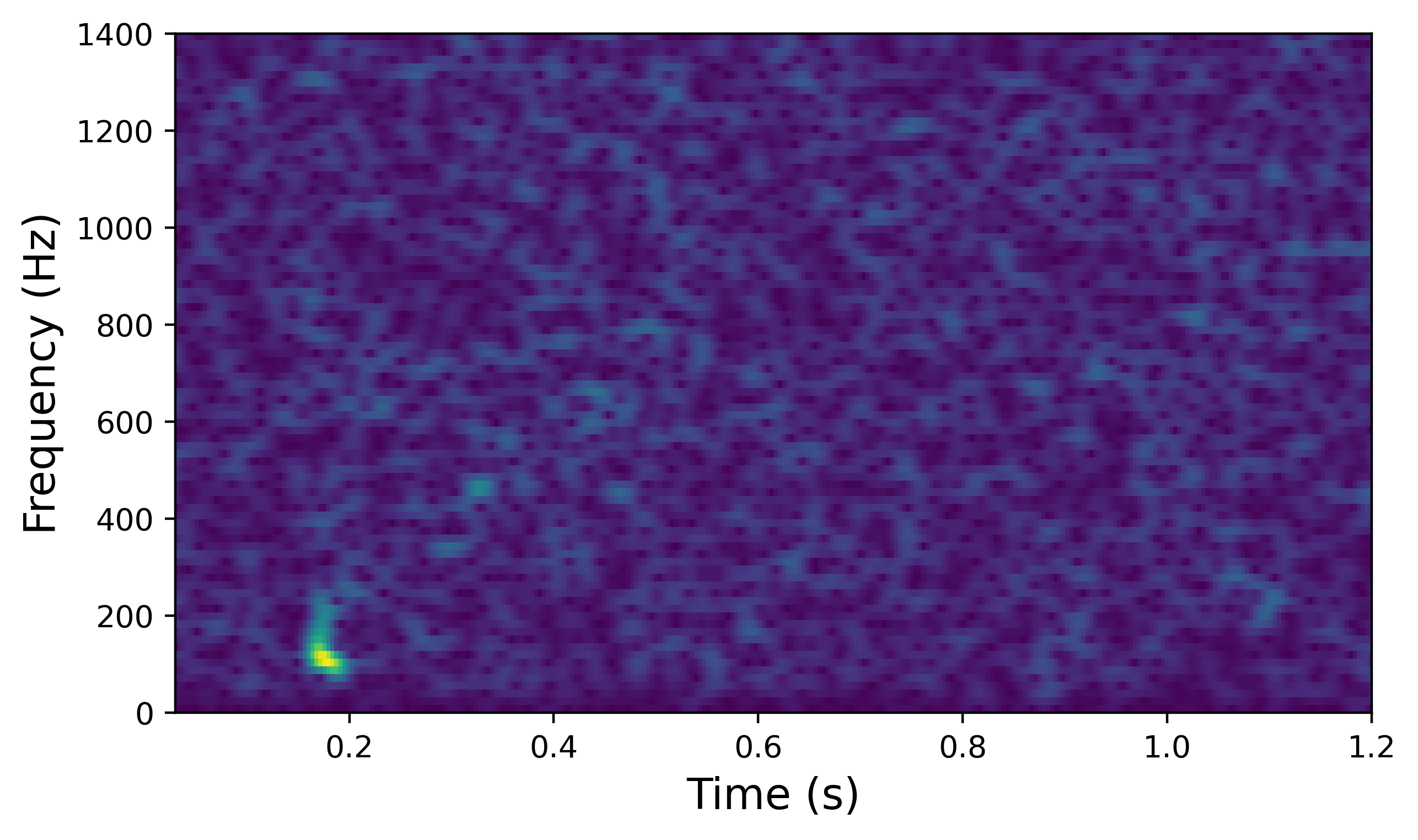}
\includegraphics[width=\columnwidth]{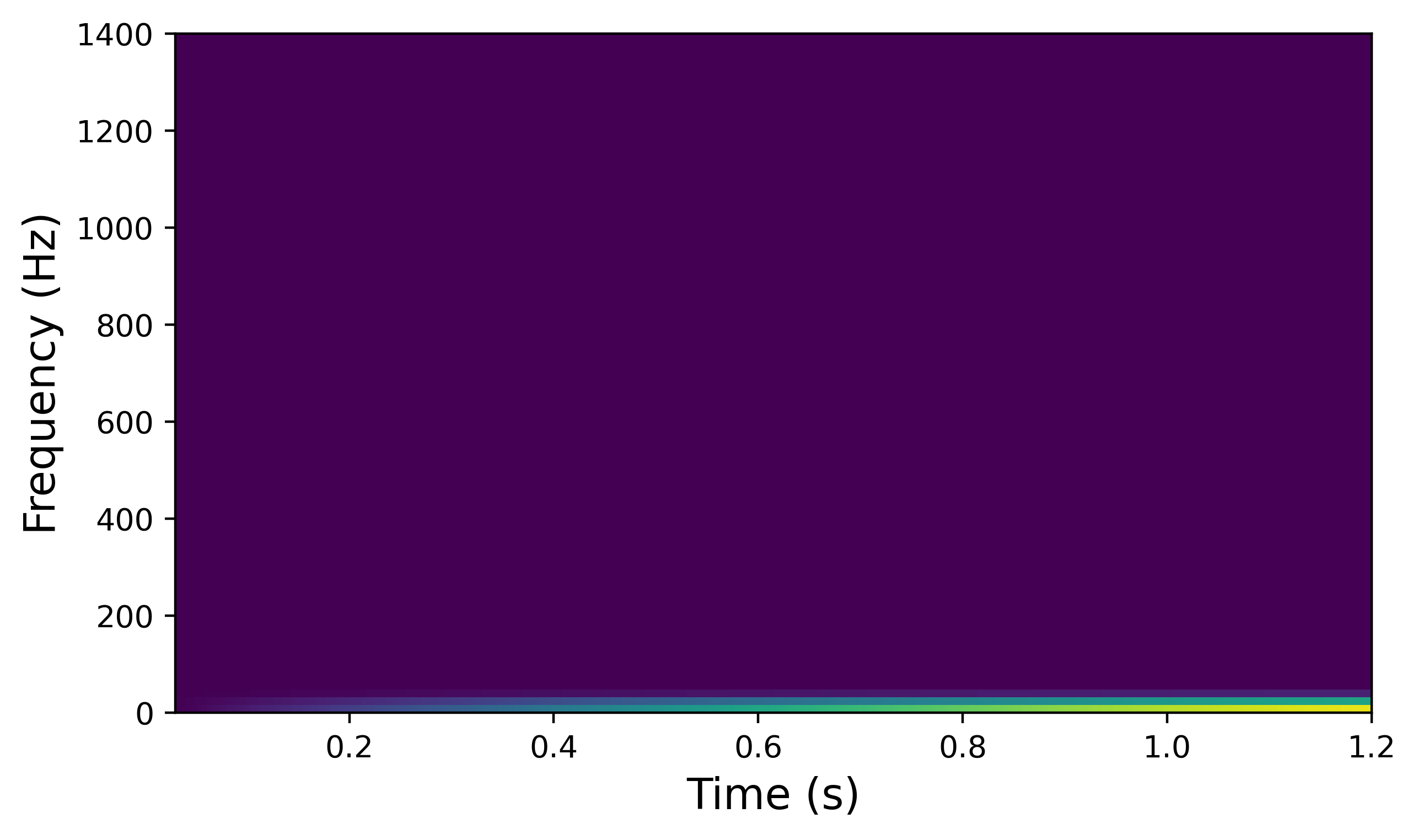}
\caption{ Three gravitational-wave signals at SNR=20. The top panel is a high frequency mode dominated signal, SFHx s12.5, in LIGO noise. The middle panel is a signal dominated by prompt convection, s10.13 SFHo, in LIGO noise. The bottom is a memory dominated signal, SFHx s29.59, in ET noise. The source properties will alter the frequency at which the signal is observed. }
\label{fig:sigs_in_noise}
\end{figure}

The results for two representative models are shown in Figure \ref{fig:snr_regions}. We use the non-rotating s10.6 SFHo model as an example that undergoes shock revival quickly, and is dominated by strong prompt convection. We also use the CMF s15 model as an example of a model that has stronger SASI due to no shock revival. We calculate the SNR for the three different signal components. For the CMF s15 model, the loudest component of the signal is the high frequency mode. At an SNR of 35, all three parts of the gravitational-wave emission are loud enough that they could likely be detected. At SNR=15, the prompt convection is starting to fall below the detectable line, so it is likely that only the highest amplitude parts of the high frequency mode would be observable. For the s10.6 SFHo model, there is not enough SASI for any low frequency emission to be detected at even the highest SNR values. The signal is dominated by prompt convection, so at SNR=15, only the prompt convection component would be observable. This shows that the gravitational-wave frequency, and the features of a low SNR detection, may be considerably different for different progenitor stars, and will be impacted by properties like the EoS and rotation. If only the gravitational waves from prompt convection are detected, then the signal duration, frequency and morphology would be similar to what we see for intermediate mass binary black hole mergers \citep{GW231123}. If a detection is suspected to come from prompt convection in a CCSN, then a model selection study similar to \citet{cuceu_26} would likely be necessary to ensure it is not a high mass binary black hole merger. For the ET observatory, the neutrino memory is by far the loudest signal feature, and may be the only part of the signal that is observed if the source is at a large distance and the SNR is low. We further illustrate this with some example SNR 20 signals in simulated gravitational-wave detector noise in Figure \ref{fig:sigs_in_noise}. 

%%%%%%%%%%%%%%%%%%%%%%%%%%%%%%%%%%%%%%%%%%%%%%%%%%%%%%%%%%
%%%%%%%%%%%%%%%%%%%%%%%%%%%%%%%%%%%%%%%%%%%%%%%%%%%%%%%%%%
\section{Stochasticity }
\label{sec:perturbations}
 
In this section, we investigate the variation of the gravitational-wave signal due to only the stochastic nature of CCSNe. All of our models include random initial perturbations in radial velocity. These perturbations are needed to seed asymmetries and power a full explosion. For four masses s14, s15, s18 and s19.5 we repeat the simulations another four times each with only the random seed for the perturbations changed. We do this for the non-rotating models only, for only the SFHo and SFHx EoS. We label the models a to e, where label e denotes the first 1\,s of the original simulation.  

The stochasticity results in some changes to the explosion dynamics which impact the gravitational-wave signals. The shock revival times are shown in Table \ref{tab:perturb}. For the models with the SFHo EoS, the explosion times of the s14 models vary by $\sim 150$\,ms. The explosion energies vary from $1.3\times 10^{50}$\,erg to $3.3\times 10^{50}$\,erg, and the final PNS masses vary between $1.67\,\mathrm{M}_{\odot}$ and $1.89\,\mathrm{M}_{\odot}$.
The s15 models show a similar variety in shock revival times, with a maximum difference of 0.159\,ms. The explosion energy varies from $1.80\times10^{50}$\,erg  to $4.50\times10^{50}$\,erg, and the final PNS mass varies from $1.78\,\mathrm{M}_{\odot}$ to $1.83\,\mathrm{M}_{\odot}$. 
Four of the s18 models undergo shock revival at similar times and one fails to explode. The explosion energies varied from $1.7\times10^{50}$\,erg to $3.8\times10^{50}$\,erg, and the final PNS masses varied from $1.82\,\mathrm{M}_{\odot}$ to $1.98\,\mathrm{M}_{\odot}$, with the largest mass found in the non-exploding model. 
For the s19.5 models, there was a smaller difference between shock revival times, with a maximum difference of 58\,ms. However, the explosion energies varied widely from $0.5\times10^{50}$\,erg to $1.5\times10^{50}$\,erg. The final PNS masses varied from $1.60\,\mathrm{M}_{\odot}$ to $1.64\,\mathrm{M}_{\odot}$. 

For the models with the SFHx EoS, the s14 models all undergo shock revival at a similar time with a maximum difference of 40\,ms. The explosion energy varies from $0.50\times10^{50}$\,erg to $1.46\times10^{50}$\,erg, and the final PNS mass is very similar for all models at $\sim 1.70\,\mathrm{M}_{\odot}$. Model s14b has a slightly smaller final PNS mass of $1.68\,\mathrm{M}_{\odot}$, and is also the model with the lowest explosion energy. 
Two of the s15 models, s15a and s15e, do not undergo shock revival, while the others explode with a maximum time difference of 86\,ms. The three exploding models have a large range of explosion energies at $3.2\times10^{50}$\,erg for s15b, $1.8\times10^{50}$\,erg for s15c, $3.1\times10^{50}$\,erg for s15d. The two non-exploding models have PNS masses of $1.91\,\mathrm{M}_{\odot}$, and the three exploding models have a similar final PNS mass of $\sim 1.8\,\mathrm{M}_{\odot}$.  
For the s18 models, four explode at a similar time, whilst s18a is $\sim200$\,ms later than the others. The lowest explosion energy is s18a at $1.4\times10^{50}$\,erg, and the largest is s18c at $5.8\times10^{50}$\,erg, as model s18a has less time for the energy to grow. Model s18a has the largest PNS mass at $1.9\,\mathrm{M}_{\odot}$, and the smallest is model s18b at $1.82\,\mathrm{M}_{\odot}$. 
The s19 models all undergo shock revival on a very similar timescale, with the smallest difference of only 20\,ms. The final explosion energies vary from $1.43\times10^{50}$\,erg to $1.04\times10^{50}$\,erg. The final PNS masses are also very similar, and vary from $1.610\,\mathrm{M}_{\odot}$ to $1.622\,\mathrm{M}_{\odot}$. 

\begin{figure}
\includegraphics[width=\columnwidth]{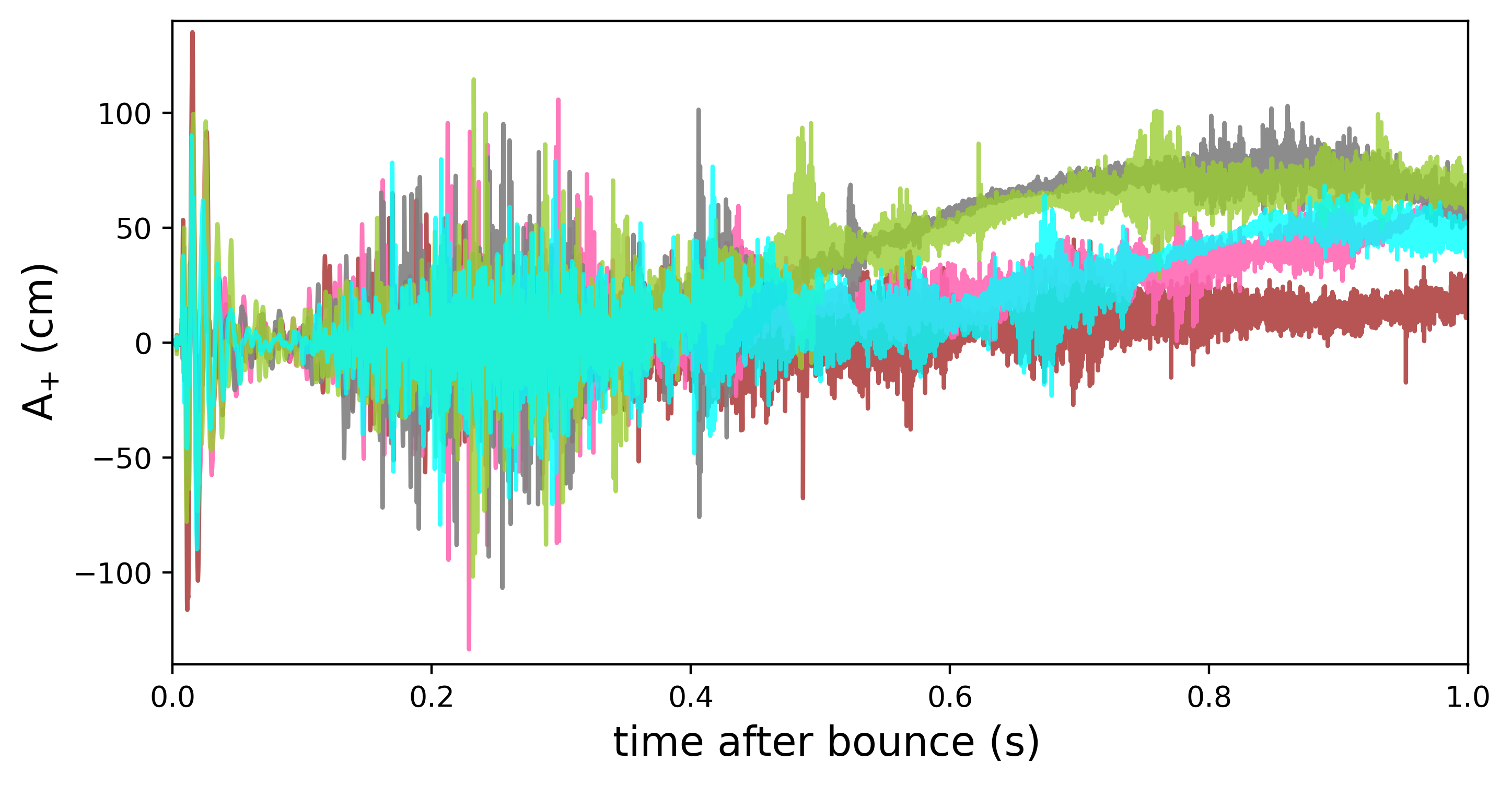}
\includegraphics[width=\columnwidth]{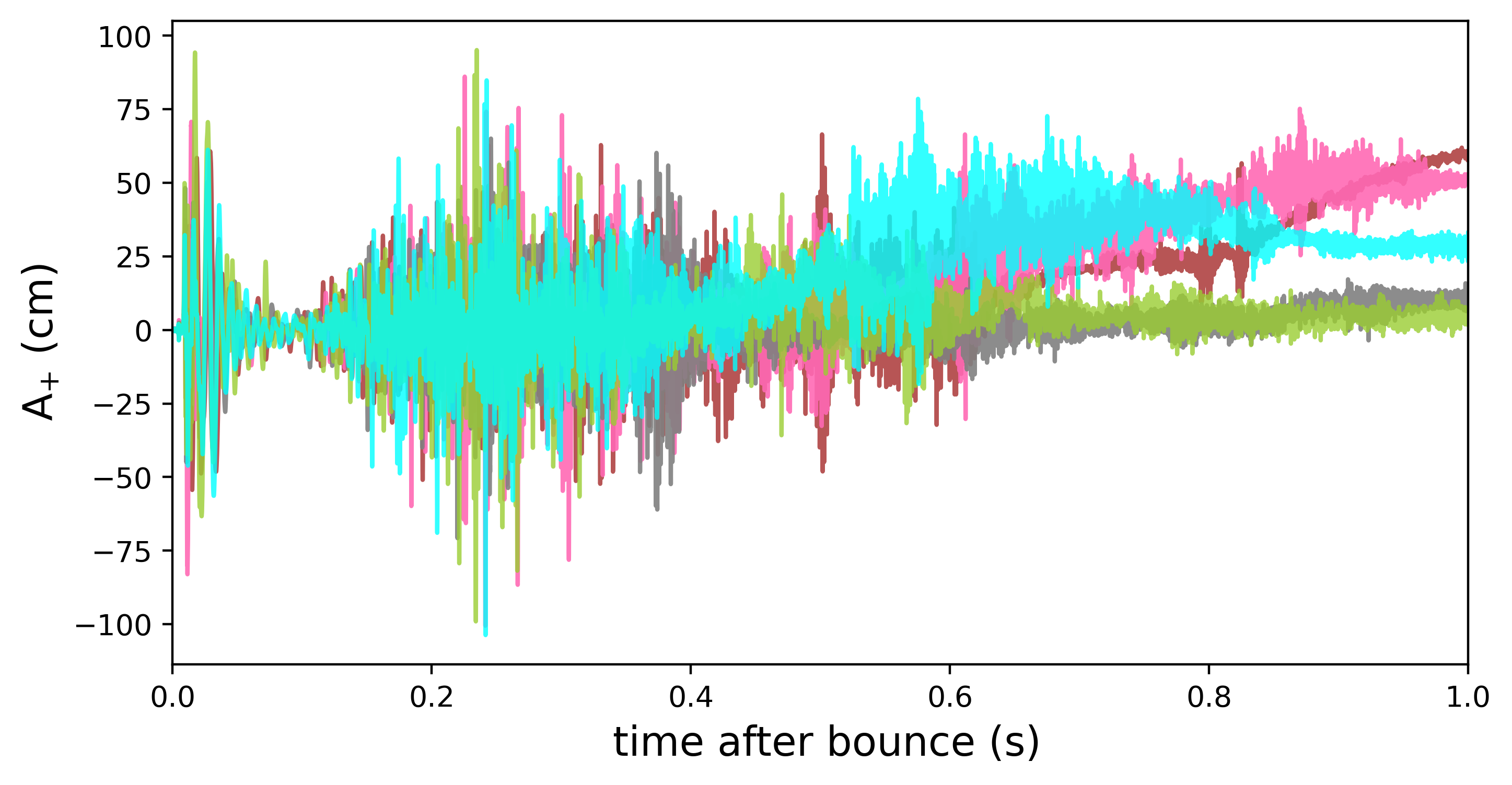}
\caption{ The top panel shows the plus polarisations of gravitational waves from the non-rotating s18 SFHx models
with different initial perturbations, and the bottom panel shows the same, but for the s14 SFHo models. Different colours represent the 5 simulations which differ only in their random radial velocity perturbations. The largest differences occur in the prompt convection. }
\label{fig:randoms}
\end{figure}

Some examples of the gravitational-wave time series are shown in Figure \ref{fig:randoms}. For the SFHo models, there is a significant variation in the amplitude of gravitational waves sourced by prompt convection, varying between 50\,cm and 94\,cm, 42\,cm and 59\,cm, 52\,cm to 85\,cm, and 52\,cm and 110\,cm, for the s14, s15, s18, and s19.5 models respectively.
Likewise for the SFHx EoS, the prompt convection signal amplitude varies between 39\,cm and 134\,cm for s14, 38\,cm to 86\,cm for s15, 49\,cm to 135\,cm for s18, and 49\,cm to 84\,cm for s19.5. The average value of the prompt convection for each model is almost always higher for the SFHx models. The exception is model s19.5, where the average is 70.4\,cm for SFHo and 68.2\,cm for SFHx. As we only have 5 simulations for each model, that one outlier may just be due to the small number statistics. 
The amplitude of the high frequency mode also has a large variations between some simulations. The largest difference between high frequency mode amplitude comes from the models where some undergo shock revival and others do not. For SFHx s15, the difference between the high frequency mode amplitude in exploding and non-exploding models is as high as 90\,cm. 

\begin{figure}
\includegraphics[width=\columnwidth]{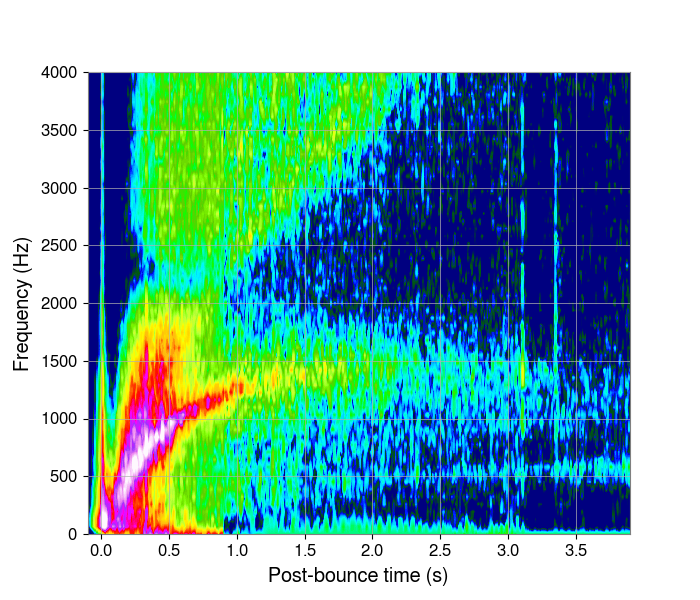}
\caption{Stacked spectrograms for the s15 SFHx models with different initial perturbations. }
\label{fig:s15_stacked}
\end{figure}

\begin{figure}
\includegraphics[width=\columnwidth]{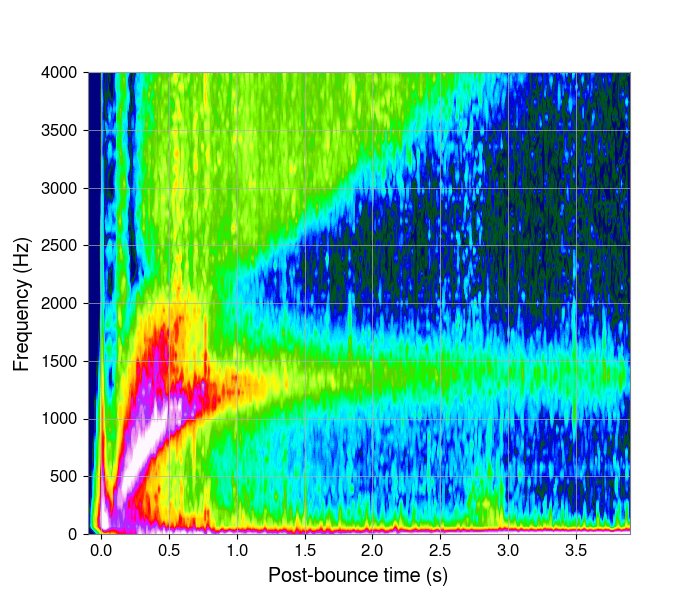}
\includegraphics[width=\columnwidth]{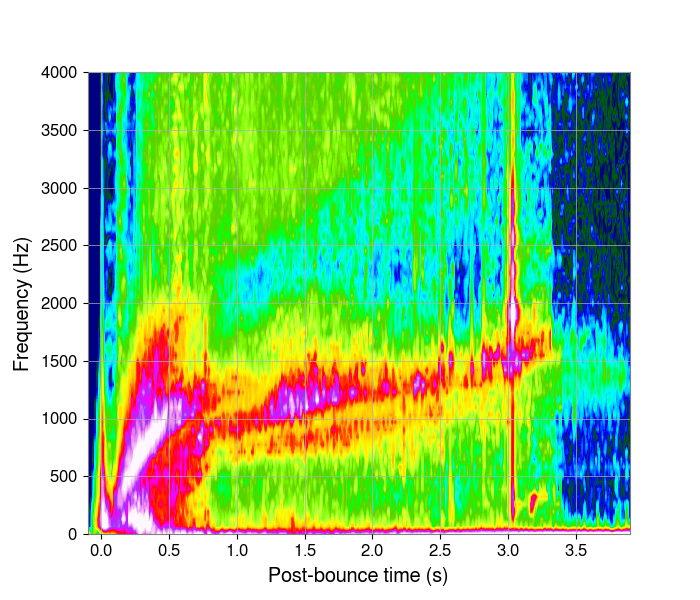}
\caption{The top panel shows the stacked spectrograms for all non-rotating and slowly rotating models. The bottom panel shows a stacked spectrogram for all non, slow and rapidly rotating models. }
%\ref{eqn:stacked}.}
\label{fig:stack_all}
\end{figure}

On the other hand, it is also useful to consider the robust features of the signal which persist across all model variations, e.g., to inform searches for supernova signals. An approach to visualize these commonalities consists in stacking the spectrograms. For the
s15 models with the SFHx EoS, we 
compute stacked spectrogram
$P_\mathrm{stacked}(t,f)$ from the
amplitudes $A_i(t,f)$ for the models with different initial perturbations,
\begin{equation}
P_\mathrm{stacked}(t,f)=\frac{1}{N(t)}\left(\sum_i|A_i(t,f)|^6\right)^{3},
\end{equation}
which has the dimension of power.
The higher weighting of the amplitudes ($|A_i(t,f)|^6$) ensures that outlier spots are smoothed away less than by simply adding power. The weighted average is computed over the $N(t)$ models with data at time $t$.
Spectrograms are computed using Wilson-Daubechies-Meyer wavelets
\citep{daubechies_91}.
The stacked spectrum in
Figure~\ref{fig:s15_stacked} indicates that the f/g-mode track is sharply defined across models. 
For comparison, in Figure \ref{fig:stack_all}, we show the stacked spectrograms for all 137 models. The width of the modes is clearly smaller for the stacked spectrograms with different perturbations, which shows that the time-frequency evolution is only weakly varied by stochasticity, and is instead more closely tuned to the actual astrophysical parameters of the star.

% The "e" is the original one
\begin{table*}
\centering
\begin{tabular}{ c  c  c  c  c  c c}
\hline\hline
Model & Revival (s) & Energy (erg) & SNR @ 10\,kpc & Revival (s) & GW Energy (erg) & SNR @ 10\,kpc \\ 
& SFHo & SFHo & SFHo & SFHx & SFHx & SFHx \\ \hline
s14a & 0.373 & 2.22e47 & 116 & 0.267 & 1.95e47 & 155 \\
s14b & 0.271 & 2.88e47 & 128 & 0.227 & 1.90e47 & 119 \\
s14c & 0.240 & 1.67e47 & 112 & 0.250 & 2.63e47 & 173 \\
s14d & 0.263 & 1.95e47 & 132 & 0.285 & 2.07e47 & 130 \\ 
s14e & 0.219 & 2.58e47 & 128 & 0.282 & 1.73e47 & 110 \\ \hline
s15a & 0.322 & 3.36e47 & 141 & N/A   & 9.47e46 & 84 \\
s15b & 0.330 & 2.92e47 & 139 & 0.283 & 3.42e47 & 163 \\
s15c & 0.377 & 3.36e47 & 136 & 0.369 & 3.33e47 & 150 \\ 
s15d & 0.269 & 3.82e47 & 145 & 0.300 & 3.23e47 & 141 \\
s15e & 0.428 & 3.70e47 & 127 & N/A   & 1.23e47 & 103 \\ \hline
s18a & 0.249 & 3.20e47 & 153 & 0.488 & 2.76e47 & 160 \\
s18b & 0.201 & 2.95e47 & 144 & 0.226 & 2.96e47 & 150 \\
s18c & 0.225 & 2.01e47 & 130 & 0.262 & 3.28e47 & 153 \\ 
s18d & N/A   & 1.37e47 & 108 & 0.261 & 3.35e47 & 170 \\
s18e & 0.267 & 3.35e47 & 153 & 0.230 & 2.30e47 & 147 \\ \hline
s19a & 0.200 & 9.96e46 & 103 & 0.178 & 1.50e47 & 105 \\ 
s19b & 0.165 & 1.16e47 & 91  & 0.164 & 1.13e47 & 99 \\
s19c & 0.171 & 1.10e47 & 94  & 0.185 & 9.59e46 & 88 \\
s19d & 0.223 & 1.65e47 & 97  & 0.169 & 9.68e46 & 109 \\
s19e & 0.166 & 1.17e47 & 141 & 0.173 & 9.50e46 & 105 \\ \hline\hline
\end{tabular}
\caption{The shock revival time, the gravitational-wave energy calculated for a model at 10\,kpc, and the signal to noise ratio for each model at 10\,kpc in a single advanced LIGO detector at design sensitivity. All calculated using the first 1\,s of the waveform. }
\label{tab:perturb}
\end{table*}

Although the time-frequency morphology is fairly similar between models with different perturbations, we also want to investigate how much the potentially large variations in amplitude impacts detectability. 
To determine this, we calculate the gravitational-wave energy E using the following equation,
\begin{equation}
E = \frac{c^3}{4G} D^2 \left( \left(\frac{dh_p}{dt}\right)^2 + \left(\frac{dh_c}{dt}\right)^2   \right) dt
\end{equation}
where D is the distance, $h_p$ is our gravitational-wave signal, and $h_c$ is 0 because the models are 2D. Additionally, we calculate the SNR of the signals at a distance of 10\,kpc and assume an optimal sky position and orientation. 

The results are shown in Table \ref{tab:perturb}. The values for the gravitational-wave energy and SNR are larger than what we would expect from 3D models \citep[cf.\ ][]{andresen_17}. However, the variation between models is not likely to be representative of what we would see in 3D. The biggest difference in SNR is found in the s15 SFHx models, which is due to the large differences in high frequency mode amplitude between the exploding and non-exploding models. The second largest difference in SNR is found between the s14 SFHx models. In this case, and for s19.5 SFHo, the differences in SNR are driven more by the differences in the amplitude of gravitational waves from prompt convection. The variation in gravitational-wave energy across models ranges from 13\% for s15 SFHo, to a huge 62\% for s15 SFHx. This variation in SNR due to CCSN stochasticity is something that should be considered carefully in future studies for realistic detection estimates for next generation observatories.

\begin{figure}
\includegraphics[width=\columnwidth]{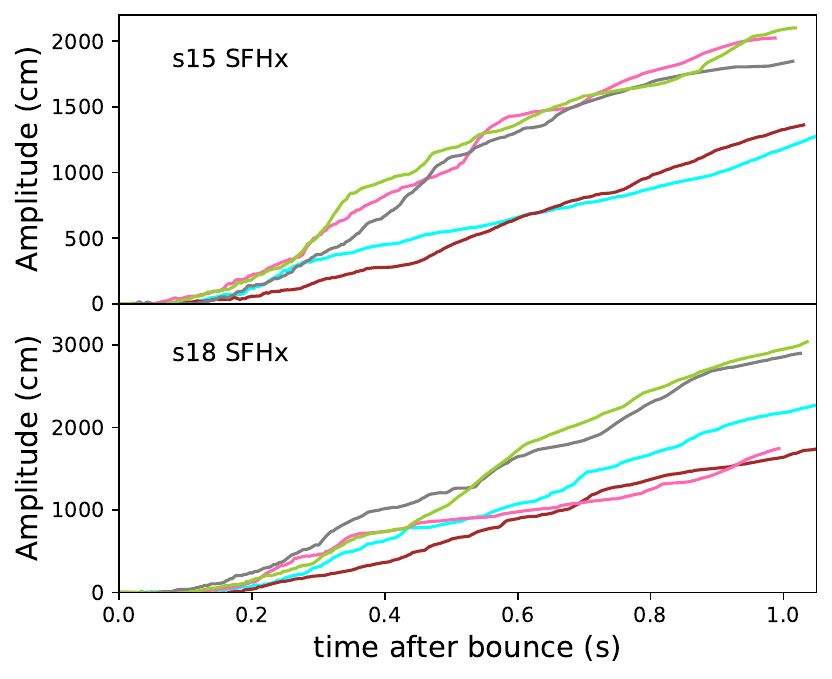}
\caption{The neutrino memory for models with different seed perturbations. The top panel is model s15 SFHx, and the bottom panel is model s18 SFHx. In some cases, the memory signal has doubled in amplitude, resulting in significant detection differences for ET. }
\label{fig:random_mem}
\end{figure}

\begin{figure}
\includegraphics[width=\columnwidth]{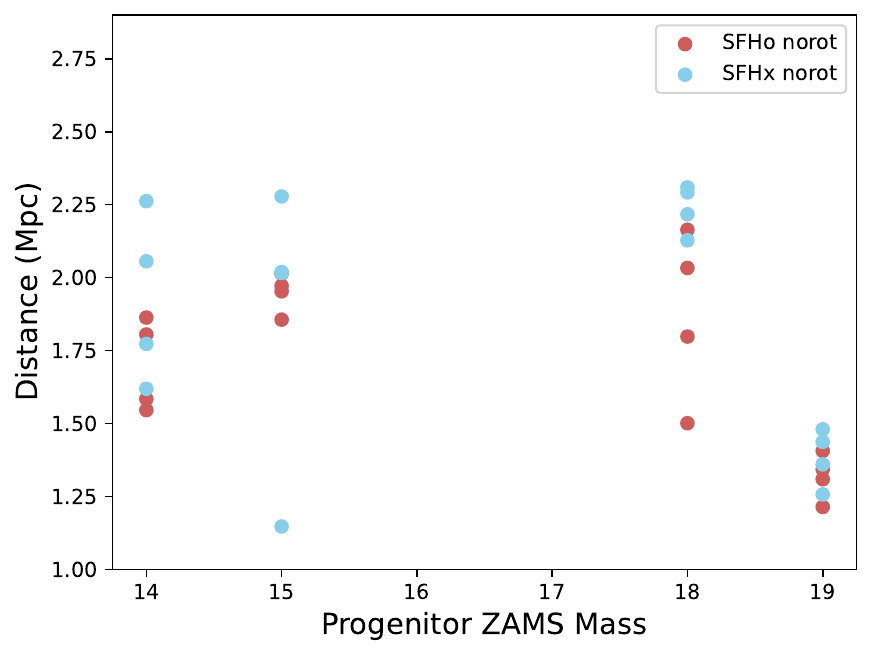}
\caption{The maximum detectable distances in ET for identical models with different seed perturbations. The largest differences are caused by successful or failed explosions. The distances are smaller than in Figure \ref{fig:distance} due to only the first second of the signal being used here. }
\label{fig:dist_perturb}
\end{figure}

We also calculate the gravitational waves due to neutrino memory for all of the models with different seed perturbations. Some examples are shown in Figure \ref{fig:random_mem}. The biggest difference is found between successful and failed shock revival, such as in model s15 SFHx. Models like s18 SFHx, which have a wide variety in explosion energies, also show significant differences between the gravitational-wave amplitudes from the asymmetric neutrino emission. Using only the first 1\,s of the signals, we show how this impacts the detection distance for ET in Figure \ref{fig:dist_perturb}. The maximum distance varies by a couple of Mpc. As the amplitude of the neutrino memory signal is still growing in most cases, this difference would likely grow if the simulations were run for longer.  

%%%%%%%%%%%%%%%%%%%%%%%%%%%%%%%%%%%%%%%%%%%%%%%%%%%%%%%%
%%%%%%%%%%%%%%%%%%%%%%%%%%%%%%%%%%%%%%%%%%%%%%%%%%%%%%%%
\section{Conclusions}
\label{sec:conclusion}

In recent years, 3D simulations of CCSNe have advanced significantly, leading to a good understanding of the morphology of the gravitational-wave signal. However, 3D simulations are still very computationally expensive, and only a small number extend beyond the first 1\,s after core-bounce. The computational expense also makes it difficult to systematically explore the progenitor parameter space in 3D. Therefore, in this study, we perform 137 simulations in 2D, to explore a range of masses, rotation rates and EoS for long durations of up to 5.5\,s. The 2D nature of our simulations also allows us to repeat models with only the initial seed perturbations changed, to investigate the impact of stochasticity on CCSN gravitational-wave emission. 

We observed large differences in the explosion dynamics between different CCSN progenitor parameters. For the CMF EoS, the majority of models did not undergo shock revival. This leads to long duration SASI activity that is visible in the gravitational-wave emission even at 5\,s post bounce, and weaker gravitational-wave energy produced by the higher frequency modes. We see differences in gravitational-wave amplitudes for different EoS during the prompt-convection phase, with CMF models having the lowest amplitude. The SFHx EoS produces the highest amplitude prompt convection for non-rotating models, and the SFHo EoS produces the highest ampltide prompt convection for slow and rapidly rotating models.  
After the prompt-convection phase, the non-rotating SFHx and SFHo models have much higher gravitational-wave amplitudes than CMF, due to rapid shock revival. Non-rotating models, with all EoS, show a secondary high frequency mode that reaches $\sim 4000$\,Hz at $\sim 2.5$\,s after bounce. Even without shock revival, we find that some CMF models still have strong enough production of low frequency gravitational waves from asymmetric neutrino emission to enhance detection prospects for observatories with improved sensitivity below 10\,Hz.

Large differences in the gravitational-wave emission were also observed between the non-rotating and the rapidly rotating models. Later shock revival was observed in the rotating models, leading to lower final gravitational-wave energies, and longer duration SASI. As well as the core-bounce signal, the rapidly rotating models had additional visible mid-frequency modes in the gravitational-wave emission. The rapidly rotating models were also a poorer fit for universal relations describing the relationship between the gravitational-wave frequency and the mass and radius of the PNS. The long duration of our simulations allowed us to test the accuracy of universal relations at late times in the signal. We then used the fits to our models to produce an updated universal relation that reduces the residuals at later times in the gravitational-wave signal. 

For the first time, we examine how much the gravitational-wave signals are impacted by the stochastic nature of CCSNe, by repeating simulations with only the random seed perturbations changed. In models close to the threshold between failed and successful explosions, the stochasticity can have a dramatic impact on the detectability of the CCSN, changing the SNR by as much as 62\%. For the least variable models, the SNR of the gravitational-wave emission varied by 13\%. The stochastic nature of CCSNe is an important aspect that should be considered carefully when predicting the future prospects for CCSN gravitational-wave detections. 

We investigated the detectability of the gravitational-wave emission across the CCSN parameter space. We find a rough correlation between larger progenitor masses and larger detection distances. Models that do not undergo shock revival are the most difficult to detect, even with long term emission due to SASI in the detectors most sensitive frequency band. We found that rotation lowers the detection horizon, due to later shock revival times. However, this is not consistent with early explosion times found in rapidly rotating 3D models. In next generation gravitational-wave observatories, better lower frequency sensitivity will result in significantly improved CCSN detection prospects due to the gravitational waves from the asymmetric emission of neutrinos. We showed that for low SNR detections, the observed gravitational-wave frequency may be very different if the model is dominated by the high frequency mode or dominated by strong prompt convection. In the non-exploding models, we find the SASI mode would only be observable for models with SNR larger than 35.

%%%%%%%%%%%%%%%%%%%%%%%%%%%%%%%%%%%%%%%%%%%%%%%%%%%%%%%%
\section*{Acknowledgements}

The authors are supported by the Australian Research Council's (ARC) Centre of Excellence for Gravitational Wave Discovery (OzGrav) through project number CE230100016. BM acknowledges support from the ARC through Discovery Projects DP240101786 and DP260104967. JP acknowledges support from the ARC through LIEF Project LE260100008. The authors acknowledge computer time allocations from Astronomy Australia Limited's ASTAC scheme, the National Computational Merit Allocation Scheme (NCMAS), and from an Australasian Leadership Computing Grant. Some of this work was performed on the Gadi supercomputer with the assistance of resources and services from the National Computational Infrastructure (NCI), which is supported by the Australian Government, and through support by an Australasian Leadership Computing Grant.  Some of this work was performed on the OzSTAR national facility at Swinburne University of Technology. The OzSTAR program receives funding in part from the Astronomy National Collaborative Research Infrastructure Strategy (NCRIS) allocation provided by the Australian Government, and from the Victorian Higher Education State Investment Fund (VHESIF) provided by the Victorian Government. 

%%%%%%%%%%%%%%%%%%%%%%%%%%%%%%%%%%%%%%%%%%%%%%%%%%
\section*{Data Availability}

The data from our simulations will be made available upon reasonable request to the authors.



\bibliographystyle{mnras}
\bibliography{example} % if your bibtex file is called example.bib

%\begin{thebibliography}{99}
%\end{thebibliography}

%%%%%%%%%%%%%%%%%%%%%%%%%%%%%%%%%%%%%%%%%%%%%%%%%%

%%%%%%%%%%%%%%%%% APPENDICES %%%%%%%%%%%%%%%%%%%%%

% \appendix

% \section{Some extra material}

% If you want to present additional material which would interrupt the flow of the main paper,
% it can be placed in an Appendix which appears after the list of references.

%%%%%%%%%%%%%%%%%%%%%%%%%%%%%%%%%%%%%%%%%%%%%%%%%%

% Don't change these lines
\bsp	% typesetting comment
\label{lastpage}
\end{document}